\documentclass[12pt]{article}
\usepackage{amsmath}
\usepackage{graphics}
\usepackage{xcolor}
\usepackage{amsfonts}
\usepackage{times}
\usepackage{bm}
\usepackage{natbib}
\usepackage{xr}
\usepackage{booktabs}
\RequirePackage{amsmath,amsfonts,amssymb,amsthm}
\usepackage{tikz}
\usepackage[flushleft]{threeparttable}
\usepackage{longtable}
\usetikzlibrary{shapes,decorations,arrows,calc,arrows.meta,fit,positioning}
\tikzset{
    -Latex,auto,node distance =1 cm and 1 cm,semithick,
    state/.style ={ellipse, draw, minimum width = 0.7 cm, fill = yellow!25},
    point/.style = {circle, draw, inner sep=0.04cm,fill,node contents={}},
    bidirected/.style={Latex-Latex,dashed},
    el/.style = {inner sep=2pt, align=left, sloped}
}
\usepackage[ruled,vlined]{algorithm2e}

\makeatletter
\renewcommand{\algocf@captiontext}[2]{#1\algocf@typo. \AlCapFnt{}#2} 
\def\@algocf@capt@plain{top}
\renewcommand{\algocf@makecaption}[2]{%
  \addtolength{\hsize}{\algomargin}%
  \sbox\@tempboxa{\algocf@captiontext{#1}{#2}}%
  \ifdim\wd\@tempboxa >\hsize
    \hskip .5\algomargin%
    \parbox[t]{\hsize}{\algocf@captiontext{#1}{#2}}
  \else%
    \global\@minipagefalse%
    \hbox to\hsize{\box\@tempboxa}
  \fi%
  \addtolength{\hsize}{-\algomargin}%
}
\makeatother

\def\T{{ \mathrm{\scriptscriptstyle T} }}

\def\dr{{ \mathrm{bc} }}
\def\pr{{ \mathrm{pr} }}
\def\var{{ \mathrm{var} }}
\def\complement{{c}}

\def\E{{E}}

\makeatletter
\newcommand*{\addFileDependency}[1]{
  \typeout{(#1)}
  \@addtofilelist{#1}
  \IfFileExists{#1}{}{\typeout{No file #1.}}
}
\makeatother

\newtheorem{lemma}{Lemma}\newtheorem{theorem}{Theorem}
\newtheorem{assumption}{Assumption}\newtheorem{remark}{Remark}

\newcommand{\F}{\mathcal{F}}
\newcommand{\R}{\mathbb{R}}
\newcommand{\bone}{\mathbf{1}}
\newcommand{\cov}{\mathrm{cov}}
\newcommand{\N}{\mathcal{N}}
\newcommand{\eff}{\mathrm{eff}}
\newcommand{\pool}{\mathrm{pool}}
\newcommand{\tap}{\mathrm{tap}}
\newcommand{\baci}{{ \mathrm{BACI} }}
\newcommand{\bacif}{{ \mathrm{BACI:F} }}
\newcommand{\paci}{{ \mathrm{PACI} }}
\newcommand{\pci}{{ \mathrm{PCI} }}
\newcommand{\plim}{\mathrm{plim}}
\newcommand{\fix}{\mathrm{fix}}

\begin{document}

\def\spacingset#1{\renewcommand{\baselinestretch}%
{#1}\small\normalsize} \spacingset{1}

\newcommand{\blind}{1}
\if1\blind
{
\date{}
  \title{\bf Pretest estimation in combining probability and non-probability samples}
  \author{Chenyin Gao\thanks{ Department of Statistics, North Carolina State University, North Carolina
27695, U.S.A. Email: cgao6@ncsu.edu}\and Shu Yang \thanks{
   Department of Statistics, North Carolina State University, North Carolina
   27695, U.S.A. Email: syang24@ncsu.edu}
    }
  \maketitle
} \fi

\if0\blind
{
  \bigskip
  \bigskip
  \bigskip
  \begin{center}
    {\LARGE\bf  Pretest estimation in combining probability and non-probability samples}
\end{center}
  \medskip
} \fi

\bigskip

\begin{abstract}
Multiple heterogeneous data sources are becoming increasingly available for statistical analyses in the era of big data. As an important example
in finite-population inference, we develop a unified framework of the test-and-pool approach to general parameter estimation by combining
gold-standard probability and non-probability samples.
We focus on the case when the study variable is observed in both datasets for estimating the target parameters, and each contains other auxiliary variables. Utilizing
the probability design, we conduct a pretest procedure to determine
the comparability of the non-probability data with the probability data and decide whether
or not to leverage the non-probability data in a pooled analysis. 
When the probability and non-probability
data are comparable, our approach combines both data for efficient
estimation. Otherwise, we retain only the probability data for estimation.
We also characterize the asymptotic distribution of the proposed test-and-pool estimator under a local alternative and provide a data-adaptive procedure to
select the critical tuning parameters that target the smallest
mean square error of the test-and-pool estimator. Lastly, to deal with the non-regularity
of the test-and-pool estimator, we construct a robust confidence interval that
has a good finite-sample coverage property.
\end{abstract}
\noindent%
\vfill

\newpage
\spacingset{1.5} 

\section{Introduction}\label{sec:introduction}
It has been widely accepted that probability sampling, where
each selected sample is treated as a representative sample to the
target population, is the best vehicle for finite-population inference.
Since the sampling mechanism is known based on survey design, each
weight-calibrated sample can be used to obtain consistent estimators for the target population; see \cite{sarndal2003model}, \cite{cochran2007sampling} and \cite{fuller2009sampling}
for textbook discussions. However, complex and ambitious surveys are
facing more and more hurdles and concerns recently, such as costly intervention strategies and lower participation rates. \cite{baker2013summary}
address some of the current challenges in using probability samples
for finite-population inference. On the other hand, higher demands of small area estimation and other more factors have led researchers to seek out alternative data
collection with less program budget \citep{williams2018trends, kalton2019developments}. In particular, lots of attention
has been drawn to the studies of non-probability samples.

Non-probability samples are sets of selected objects where the
sampling mechanism is unknown. First of all, non-probability samples are readily
available from many data sources, such as satellite information \citep{mcroberts2010advances},
mobile sensor data \citep{palmer2013new}, and web survey panels \citep{tourangeau2013science}. In addition, these non-representative samples are far more cost-effective
compared to probability samples and have the potential of providing
estimates in near real-time, unlike the traditional inferences derived
from probability samples \citep{rao2020making}. Based on these big and easy-accessible
data, a wealth of literature has been proposed which enunciates the
bright future while properly utilizing such amount of data (e.g.,
\citealp{couper2013sky}, \citealp{citro2014multiple}, \citealp{tam2015big},
and \citealp{pfeffermann2015methodological}).

However, the naive use of such data cannot ensure the statistical validity of the resulting estimators because such non-probability samples
are often selected without sophisticated supervision. Therefore, the acquisition
of large whereas highly unrepresentative data is likely to produce
erroneous conclusions. \cite{couper2000web} and \cite{elliott2017inference} present more recent examples where non-probability samples can often lead to estimates
with significant selection biases. To overcome these challenges,
it is essential to establish appropriate statistical tools to draw
valid inferences when integrating data from the probability and non-probability samples. Various data integration methods have been proposed in the literature to leverage the unique strengths of the probability and non-probability samples; see \cite{yang2020statistical} for a review, and the existing methods for data integration can be categorized into three
types including the inverse propensity score adjustment \citep{rosenbaum1983central, elliott2007bayesian},
calibration weighting \citep{deville1992calibration,kott2006using},
and mass imputation \citep{rivers2007sampling,kim2019sampling,yang2018integration}.

But most of the works assume that the non-probability sample is comparable
to the probability sample in terms of estimating the finite-population parameters, which may not be satisfied in many applications due to the unknown sampling mechanism of the non-probability samples. Thus, the non-probability samples with unknown sampling mechanisms may bias the estimators for the target parameters. To resolve this issue, \cite{robbins2019blending} propose a pretest to gauge the statistical adequacy of integrating the probability and non-probability samples in an application. The pretesting procedure has been broadly practiced in econometrics and medicine, and its implications are of considerable interests (e.g., \cite{wallace1977pretest,toyoda1979pre,baltagi2003fixed,yang2020elastic}). Essentially,
the final value of the estimator depends on the outcome of a random testing event and therefore is a stochastic mixture of two different estimators. Despite the long history of the application
of the pretest, few literature investigates the theoretical properties of the underlying non-smooth distribution for the pretest estimators.

In this paper, we establish a general statistical framework for the test-and-pool analysis of the probability and non-probability samples by constructing a test
to gauge the comparability of the non-probability data and decide whether or not
to use non-probability data in a pooled analysis. In addition, we consider the null, fixed, and local alternative hypotheses for the pre-testing, representing different levels of comparability of the non-probability data with the probability data. In particular, the non-probability sample is perfectly comparable under the null hypothesis, whereas it is starkly incomparable under the fixed alternative. Therefore, the fixed alternative cannot adequately capture the finite-sample behavior of the pre-testing estimator, under which the test statistic will diverge to infinity as the sample size increases. Toward this end, we establish the asymptotic distribution of the proposed estimator under local alternatives, which provides a better approximation of the finite-sample behavior of the pretest estimator when the idealistic assumption required for the non-probability data is weakly violated. Also, we provide a data-adaptive
procedure to select the optimal values of the tuning parameters achieving the smallest mean square error of the pretest estimator. Lastly,
we construct a robust confidence interval 
accounting for the non-regularity of the estimator, which has a valid coverage
property.

The rest of the paper is organized as follows. Section \ref{sec:Basic-setup}
lays out the basic setup and presents an efficient estimator for combing the non-probability sample and the probability sample. Section
\ref{sec: TAP} proposes a test statistic and the test-and-pool estimator. In Section \ref{sec:Asymptotic-properties-of}, we present the asymptotic properties of the test-and-pool estimator, an adaptive inference procedure, and lastly a data-adaptive selection scheme of the tuning parameters. Section \ref{sec:Simulation} presents a simulation
study to evaluate the performance of our test-and-pool estimator. Section \ref{sec:Application} provides a real-data illustration. All proofs are given in the Appendix.

\section{Basic setup\label{sec:Basic-setup} }

\subsection{Notation: two data sources\label{subsec:notation}}

Let $\F_{N}=\{V_{i}=(X_{i}^{\T},Y_{i})^{\T}:i\in U\}$
with $U=\{1,\ldots,N\}$ denote a finite population of size $N$, where $X_{i}$ is a vector of covariates and $Y_{i}$ is the study variable. We assume
that $F_{N}$ is a random sample from a superpopulation model $\zeta$ and our objective is to estimate the finite-population parameter $\mu_g\in\R^l$, defined as the solution to
\begin{equation}
\frac{1}{N}\sum_{i=1}^{N}S(V_{i};\mu)=0,\label{eq:general-par}
\end{equation}
where $S(V_i;\mu)$ is a $l$-dimensional estimating function. The class of parameters is fairly general. For example, if $S(V;\mu)=Y-\mu$,
$\mu_{g}=\overline{Y}_{N}=N^{-1}\sum_{i=1}^{N}Y_{i}$ is the population
mean of $Y_{i}$. If $S(V;\mu)=\bone(Y<c)-\mu$ for some constant
$c$, where $\bone(\cdot)$ is an indicator function, $\mu_{g}=N^{-1}\sum_{i=1}^{N}\bone(Y_{i}<c)$
is the population proportion of $Y_{i}$ less than $c$. If $S(V;\mu)=X(Y-X^{\T}\mu),$
$\mu_{g}=(\sum_{i=1}^{N}X_{i}X_{i}^{\T})^{-1}(\sum_{i=1}^{N}X_{i}Y_{i})$
is the coefficient of the finite-population regression projection
of $Y_{i}$ onto $X_{i}$.

Suppose that there are two data sources, one from a probability sample, referred
to as Sample A, and the other from a non-probability sample, referred to as Sample
B. Assume Sample A to be independent of Sample B, and the observed units can be envisioned as being generated through two phases of sampling \citep{chen2019doubly}. Firstly, a superpopulation model $\zeta$ generates the finite population $\F_N$. Then, the probability (or non-probability) sample is selected from it using some known (or unknown) sampling schemes. Hence, the considered total variance of estimators is based on the randomness induced by both the superpopulation model and the sampling mechanisms; see Table \ref{tab:notations} for the notations of probability order, expectation and (co-)variance. For example, $\E_{\mathrm{p}}(\cdot\mid \F_N)$ is the average over all possible samples under the probability design for particular finite population $\F_N$, and $\E(\cdot)$ is the average over all possible samples from all possible finite populations.

\begin{table}[!htbp]
 \caption{Notation and definitions}
 \vspace{0.15cm}
     \centering
     \resizebox{\textwidth}{!}{\begin{tabular}{llll}
    \toprule
       Randomness  & order notation & expectation & (co-)variance \\
     \midrule
       probability design  & $o_{\mathrm{p}}(1), O_{\mathrm{p}}(1)$ & $\E_{\mathrm{p}}\left(\cdot \mid \F_{N}\right)$ & $\var_{\mathrm{p}}\left(\cdot \mid \F_{N}\right),
       \cov_{\mathrm{p}}\left(\cdot \mid \F_{N}\right)$ \\
       non-probability design  & $o_{\mathrm{np}}(1), O_{\mathrm{np}}(1)$ & $\E_{\mathrm{np}}\left(\cdot \mid \F_{N}\right)$ & $\var_{\mathrm{np}}\left(\cdot \mid \F_{N}\right),
       \cov_{\mathrm{np}}\left(\cdot \mid \F_{N}\right)$ \\
       $\zeta$ model  & $o_{\zeta}(1), O_{\zeta}(1)$ & $\E_{\zeta}\left(\cdot \right)$ & $\var_{\zeta}\left(\cdot \right),
       \cov_{\zeta}\left(\cdot \right)$ \\
       total variance   & $o_{\zeta\text{-}\mathrm{p}\text{-}\mathrm{np}}(1), O_{\zeta\text{-}\mathrm{p}\text{-}\mathrm{np}}(1)$ & $\E\left(\cdot \right)$ & $\var\left(\cdot\right),\cov\left(\cdot\right)$\\
       \bottomrule
     \end{tabular}}
        \label{tab:notations}
\end{table}

Thus far, our focus has been on the setting where the covariates $X$ and the study variable $Y$ are available in both the probability and non-probability samples, which has also been considered in \cite{elliott2007use} and \cite{elliot2013combining}. The sampling indicators are denoted
by $\delta_{A,i}$ and $\delta_{B,i}$, respectively; e.g., $\delta_{A,i}=1$
if unit $i$ is selected into Sample A and zero otherwise. Sample
A contains observations $\mathcal{O}_{A}=\{(d_{i}=\pi_{A,i}^{-1},X_{i},Y_{i}):i\in\mathcal{A}\}$
with sample size $n_{A},$ where $\pi_{A,i}$ is the known first-order inclusion probability for Sample A, and Sample B contains observations $\mathcal{O}_{B}=\{(X_{i},Y_{i}):i\in\mathcal{B}\}$
with sample size $n_{B}$. The unknown propensity score for being selected into Sample B is denoted by $\pi_{B,i}$. Here, $\mathcal{A}$
and $\mathcal{B}$ denote the indexes of units in Samples A and B with total sample size $n = n_A + n_B$ and negligible sampling fractions, i.e., $n/N=o(1)$. Let the limits of the fractions of Sample A and B be $f_A = \lim_{n \rightarrow \infty} n_A/n$ and $f_B = \lim_{n \rightarrow \infty} n_B/n$ with $0<f_A,f_B<1$. 


\subsection{Assumptions and separate estimators}

As observing $(X_i, Y_i)$ for all units $i$ in $U$ is usually not feasible in practice, we can estimate the population estimating
equation (\ref{eq:general-par}) by the design-weighted
sample analog 
under the probability sampling design
\begin{equation}
\frac{1}{N}\sum_{i=1}^{N}\frac{\delta_{A,i}}{\pi_{A,i}}S(V_{i};\mu)=0,\label{eq:ipw}
\end{equation}
yielding a design-weighted Z-estimator $\widehat{\mu}_{A}$ \citep{van2000asymptotic}. When $S(V;\mu)$ is a score function, the resulting estimator will be a pseudo maximum likelihood estimator \citep{skinner1992pseudo}. For example,
for estimating $\overline{Y}_{N}$, we have $S(V;\mu) = Y - \mu$, which leads to  $\widehat{\mu}_{A}=(\sum_{i=1}^N\delta_{A,i}\pi_{A,i}^{-1})^{-1}\allowbreak\sum_{i=1}^N \delta_{A,i} \pi_{A,i}^{-1} Y_i$. We now make the following assumption for the design-weighted Z-estimator.
\begin{assumption}
(Design consistency and central limit theorem) \label{asmp:Sampling-design}
Let $\widehat{\mu}_{A}$
be the corresponding design-weighted Z-estimator of $\mu_g$, which satisfies that $\text{var}_\mathrm{p}(\widehat{\mu}_{A}\mid \F_N)=O_\zeta(n_{A}^{-1})$
and $\{\text{var}_\mathrm{p}(\widehat{\mu}_{A})\}^{-1/2}$ $\times(\widehat{\mu}_{A}-\mu_g)\mid \F_{N}\rightarrow \N(0,1)$
in distribution as $n_{A}\rightarrow\infty$. \end{assumption} 
Under the typical regularity conditions \citep{fuller2009sampling}, Assumption
\ref{asmp:Sampling-design} holds for many common sampling designs
such as probability proportional to size and stratified simple random sampling. Under Assumption \ref{asmp:Sampling-design}, $\widehat{\mu}_{A}$
is design-consistent and does not rely on any modeling assumptions.
This explains why the probability sampling has been the gold standard approach for finite-population
inference, and we make this assumption throughout this article.

Let $f(Y\mid X)$ be the conditional density function of $Y$ given $X$ in the superpopulation model $\zeta$, and let $f(X)$ and $f(X\mid\delta_{B}=1)$ be the density function of $X$ in the finite population and the non-probability sample, respectively. To correct
for the selection bias of the non-probability sample, most of the existing literature
considers the following assumptions \citep[e.g.,][]{rivers2007sampling,vavreck20082006,chen2019doubly}.
\begin{assumption} (Common support and ignorability of sampling) \label{asmp:MAR} (i)
The vector of covariates $X$ has a compact and convex support, with
its density bounded and bounded away from zero. Also, there exist positive constants
$C_{l}$ and $C_{u}$ such that $C_{l}\leq f(X)/f(X\mid\delta_{B}=1)\leq C_{u}$
almost surely. (ii) Conditional on $X$, the density of $Y$ in the non-probability sample follows the superpopulation model; i.e., $f(Y\mid X,\delta_{B}=1)=f(Y\mid X)$. (iii) The sample inclusion indicator $\delta_{B,i}$ and $\delta_{B,j}$ are independent given $X_i$ and $X_j$ for $i\neq j$.
\end{assumption} Assumption \ref{asmp:MAR} (i) and (ii) constitute
the strong sampling ignorability condition \citep{rosenbaum1983central}.
Assumption \ref{asmp:MAR} (i) implies that the support of $X$ in
the non-probability sample is the same as that in the finite population, and it can
also be formulated as a positivity assumption that $\pr(\delta_{B}=1\mid X)>0$
for all $X$. This assumption does not hold if certain units would never be included in the non-probability sample. Assumption \ref{asmp:MAR} (ii) is equivalent to the ignorability of the sampling mechanism for the non-probability sample conditional on the covariates $X$, i.e., $\pr(\delta_{B}=1\mid X,Y)=\pr(\delta_{B}=1\mid X)$ \citep{little1982models}. This assumption holds if the set of covariates contain all the outcome predictors that affect the possibility of being selected into the non-probability sample. Assumption \ref{asmp:MAR} (iii) is a critical condition to employ the weak law of large numbers under the non-probability sampling design \citep{chen2019doubly}. Under Assumption \ref{asmp:MAR}, the non-probability sample can be used to produce consistent estimators. However, this assumption may be unrealistic if the non-probability data collection
suffers from uncontrolled selection biases \citep{bethlehem2016solving}, measurement
errors \citep{couper2000web}, or other error-prone issues. Thus,
we consider Assumption \ref{asmp:MAR} as an idealistic assumption,
which may be violated and require pretesting.

Under Assumptions \ref{asmp:Sampling-design} and \ref{asmp:MAR}, let $\Phi_A(V,\delta_A;\mu)$ and $\Phi_B(V,\delta_A,\delta_B;\mu)$ be two $l$-dimensional estimating functions for the target parameter $\mu_g$ when using the probability sample and the combined samples, respectively. In practice, $\Phi_A(\cdot)$ and $\Phi_B(\cdot)$ may depend on unknown nuisance functions, and solving $\E\{\Phi_A(V,\delta_{A};\mu)\}=0$ and $\E\{\Phi_B(V,\delta_{A},\delta_{B};\mu)\}=0$ is not feasible. By replacing the nuisance functions with their estimated counterparts, and the expectations with the empirical averages, we obtain $\widehat{\mu}_A$ and $\widehat{\mu}_B$ by solving 
\begin{equation}
\frac{1}{N}\sum_{i=1}^{N}\widehat{\Phi}_{A}(V_{i},\delta_{A,i};\mu)=0,\quad \frac{1}{N}\sum_{i=1}^{N}\widehat{\Phi}_{B}(V_{i},\delta_{A,i},\delta_{B,i};\mu)=0,
\label{eq:SA+SB}
\end{equation}
respectively, where $\{\widehat{\Phi}_{A}(\cdot)$, $\widehat{\Phi}_{B}(\cdot)\}$ are 
the estimated version of $\{{\Phi}_{A}(\cdot)$, ${\Phi}_{B}(\cdot)\}$.
\begin{remark}
For estimating the finite population means, that is, $\mu_{g}=\overline{Y}_{N}$, $\Phi_{A}(\cdot)$ and $\Phi_{B}(\cdot)$ are commonly chosen as
\begin{align}
 & \Phi_{A}(V,\delta_{A};\mu)=\frac{\delta_{A}}{\pi_{A}}(Y-\mu),\label{eq:S_A_1}\\
 & \Phi_{B}(V,\delta_{A},\delta_{B};\mu)=
\frac{\delta_{B}}{\pi_{B}\left(X\right)}\left\{Y-m\left(X\right)\right\}+\frac{\delta_{A}}{\pi_{A}} m\left(X\right)-\mu, \label{eq:S_B_1}
\end{align}
where $\pi_B(X)=\pr(\delta_B=1\mid X)$ and $m(X)=\E(Y\mid X,\delta_B=1)$. To obtain the estimators $\widehat{\mu}_A$ and $\widehat{\mu}_B$, parametric models $\pi_B(X;\alpha)$ and $m(X;\beta)$ can be posited for the nuisance functions $\pi_B(X)$ and $m(X)$, respectively. 

In addition, researchers might be interested in estimating the individual-level outcomes rather than the population-level outcomes. In this case, $\Phi_A(\cdot)$ and $\Phi_B(\cdot)$ can be specified for estimating the outcome model $m(X;\beta)$ as:
\begin{align*}
    &\Phi_A(V,\delta_A;\beta)=
    \frac{\delta_A}{\pi_A}\frac{\partial m(X;\beta)}{\partial\beta}\{Y-m(X;\beta)\}\\
    &\Phi_B(V,\delta_A,\delta_B;\beta)=\left(\frac{\delta_A}{\pi_A}+\frac{\delta_B}{\pi_B(X)}\right) \frac{\partial m(X;\beta)}{\partial \beta}\{Y-m(X;\beta)\}.
\end{align*}
\label{rmk:estimating_example}
\end{remark}
Next, we adopt the model-design-based framework for inference, which incorporates the randomness over the two phases of sampling \citep{kalton1983models, molina2001modelling,binder2003design,xu2013pseudo}. 
The asymptotic properties for $\widehat{\mu}_{A}$ and $\widehat{\mu}_{B}$
can be derived using the standard M-estimation theory under suitable moment conditions.
\begin{lemma}
\label{lemma:mu_A_B_H0}Suppose Assumptions \ref{asmp:Sampling-design},
\ref{asmp:MAR} and additional regularity conditions \ref{asmp:regularity} hold. Then, we have 
\begin{equation}
n^{1/2}\left(\begin{array}{c}
\widehat{\mu}_{A}-\mu_{g}\\
\widehat{\mu}_{B}-\mu_{g}
\end{array}\right){\rightarrow}\N\left\{ \left(\begin{array}{c}
{0}_{l\times1}\\
{0}_{l\times1}
\end{array}\right),\left(\begin{array}{cc}
V_{A} & \Gamma\\
\Gamma^{\T} & V_{B}
\end{array}\right)\right\} ,\label{eq:m_est}
\end{equation}
where $V_{A}$ , $V_{B}$, and $\Gamma$ are defined explicitly in the Appendix . 
\end{lemma}
In Lemma \ref{lemma:mu_A_B_H0}, we extend the conditional normality to unconditional as in \cite{schenker1988asymptotic}, which implies that the asymptotic (co-)variances terms $V_A,V_B$ and $\Gamma$ refer to all the sources of uncertainty over the two phases.

\subsection{Efficient estimator}\label{subsec:Efficient-estimator-combining} 

Under Assumptions \ref{asmp:Sampling-design}
and \ref{asmp:MAR}, both $\widehat{\mu}_{A}$ and $\widehat{\mu}_{B}$
are consistent, and it is appealing to combine $\widehat{\mu}_{A}$
with $\widehat{\mu}_{B}$ to achieve efficient estimation. We consider
a class of linear combinations of the functions in (\ref{eq:SA+SB}):
\begin{equation}
\sum_{i=1}^{N}\{\widehat{\Phi}_{A}(V_{i},\delta_{A,i};\mu)+\Lambda \widehat{\Phi}_{B}(V_{i},\delta_{A,i},\delta_{B,i};\mu)\}=0,\label{eq:general comb}
\end{equation}
where $\Lambda$ is the linear coefficient that gauges how much information
of the non-probability sample should be integrated with the probability sample. Equation (\ref{eq:general comb})
leads to a class of composite estimators which is a weighted average of $\widehat{\mu}_{A}$ and $\widehat{\mu}_{B}$ with $\Lambda$-indexed weight $\omega_A$ and $\omega_B$. When $\Lambda=0$, (\ref{eq:general comb}) provides the design-consistent
estimator $\widehat{\mu}_{A}$. The optimal choice $\Lambda_{\eff}$ can be empirically tuned to minimize the asymptotic variance of the composite estimator, leading to the efficient estimator $\widehat{\mu}_{\text{eff}}$. However, the major concern
for $\widehat{\mu}_{\eff}$ is the possible
bias due to the violation of Assumption \ref{asmp:MAR} (ii) for the
non-probability sample. When it is violated, it is reasonable to choose $\Lambda=0$
and prevent any bias associated with the non-probability sample. 

\section{Test-and-pool estimator\label{sec: TAP}}

Motivated by the above reasoning, we develop a strategy that pretests the comparability of the non-probability sample with the probability sample first and then decides whether or not we should combine them for efficient estimation. 
We formulate the hypothesis test in Section \ref{subsec:hypothesis and test}, and construct the test-and-pool estimator in Section \ref{subsec:Elastic}.
\subsection{Hypothesis and test \label{subsec:hypothesis and test}}
We formalize the null hypothesis $H_{0}$ when
Assumption \ref{asmp:MAR} holds, and the fixed and local alternatives $H_a$ and $H_{a,n}$
when Assumption \ref{asmp:MAR} is violated. To be specific, we consider
\begin{align}
 & H_{0}:\E\{\Phi_{B}(V, \delta_A,\delta_B;\mu_{g,0})\}=0,\label{eq:H_0}\\
 & H_a:\E\{\Phi_{B}(V, \delta_A,\delta_B;\mu_{g,0})\}=\eta_{\fix},\label{eq:H_a}\\
 & H_{a,n}:\E\{\Phi_{B}(V, \delta_A,\delta_B;\mu_{g,0})\}=n_{B}^{-1/2}\eta,\label{eq:H_n}
\end{align}
where $\mu_{g,0}=\E_\zeta(\mu_g)$, $\mu_g=\mu_{g,0}+O_\zeta(N^{-1/2})$, and $\eta_{\fix}$, $\eta$ are two fixed parameters. The fixed alternative $H_a$ is commonly considered in the standard hypothesis testing framework. However, it enforces the bias of the estimating function $\Phi_B(\cdot)$ to be fixed and indicates a strong violation of Assumption 2.2, under which the test statistic $T$ will diverge to infinity with the sample size. Moreover, the inference under the fixed alternative can not capture the finite-sample behavior of the test well and lacks uniform validity. On the contrary, the local alternative provides a useful tool to study the finite-sample distribution of non-regular estimators when the signal of violation is weak, i.e., in the $n_B^{-1/2}$ neighborhood of zero. In such cases, we allow the existence of a set of unmeasured covariates whose association with either the possibility of being selected into Sample B or the outcome is small. Also, the local alternative $H_{a,n}$ is more general in the sense that it reduces to $H_a$ with $\eta=\pm\infty$, and has been
widely employed to illustrate the non-regularity settings, such as 
weak instrumental variables regression \citep{staiger1994instrumental}, regression
estimators of weakly identified parameters \citep{cheng2008robust}
and test errors in classification \citep{laber2011adaptivejasa}. We will mainly exploit the local alternative to show the inherent non-regularity
of the pretest estimator.

Under the null hypothesis \eqref{eq:H_0}, $\widehat{\mu}_{B}$ is consistent, and hence, it is reasonable to combine $\widehat{\mu}_{A}$
and $\widehat{\mu}_{B}$ for efficient estimation. However, when the
null hypothesis is violated as in \eqref{eq:H_n}, the efficient estimator
is biased. Lemma \ref{lemma:mu_A_B_Hn} presents the asymptotic properties
of the separate and efficient estimators under $H_{a,n}$. 
\begin{lemma}
\label{lemma:mu_A_B_Hn}Suppose Assumptions \ref{asmp:Sampling-design}, \ref{asmp:MAR} (i) and (iii), and all the regularity conditions in Lemma \ref{lemma:mu_A_B_H0}
hold. Then, under the local alternative $H_{a,n}$, the asymptotic
distributions for $\widehat{\mu}_{A}$ and $\widehat{\mu}_{B}$ are
\begin{equation}
\begin{split}
    &n^{1/2}\left(\begin{array}{c}
\widehat{\mu}_{A}-\mu_{g}\\
\widehat{\mu}_{B}-\mu_{g}
\end{array}\right){\rightarrow}N\left\{ \left(\begin{array}{c}
{0}_{l\times1}\\
-f_{B}^{-1/2}\E\left\{ \partial \Phi_{B}(V,\delta_A,\delta_B;\mu_{g,0})/\partial\mu\right\} ^{-1}\eta
\end{array}\right),\left(\begin{array}{cc}
V_{A} & \Gamma\\
\Gamma^{\T} & V_{B}
\end{array}\right)\right\}. 
\end{split}
\label{eq:normalAB}
\end{equation}
The asymptotic distribution of the efficient estimator $\widehat{\mu}_\eff$ is
\[
n^{1/2}(\widehat{\mu}_{\eff}-\mu_g){\rightarrow}\N\left\{ b_{\eff}(\eta),V_{\eff}\right\} ,
\]
where $b_{\eff}(\eta)=-f_{B}^{-1/2}\omega_{B}(\Lambda_\eff)\E\left\{ \partial \Phi_{B}(V,\delta_A,\delta_B; \mu_{g,0})/\partial\mu\right\} ^{-1}\eta$.
The exact form of $\omega_{B}(\Lambda_\eff)$ and $V_{\eff}$ are  presented in Lemma \ref{lem:mu_eff}. 
\end{lemma}
By Lemma \ref{lemma:mu_A_B_Hn}, among the three estimators $\widehat{\mu}_{A}$,
$\widehat{\mu}_{B}$ and $\widehat{\mu}_{\eff}$, when $H_{0}$ holds,
$\widehat{\mu}_{\text{eff}}$ is optimal because it is consistent and the most
efficient; while when $H_{0}$ is violated, $\widehat{\mu}_{A}$ is
optimal because it is consistent but the other two estimators are not.

We now use pretesting to guide choosing the estimators. To test $H_{0}$,
the key insight is that $\widehat{\mu}_{A}$ is always consistent
for $\mu_{g}$ by Assumption \ref{asmp:Sampling-design}, and if $H_{0}$ holds, $\widehat{\Phi}_{B,n}(\widehat{\mu}_A) = n_B^{1/2}N^{-1} \sum_{i=1}^{N}\allowbreak\widehat{\Phi}_{B}(V_{i},\delta_{A,i},\delta_{B,i};\widehat{\mu}_{A})$
should behave as a mean-zero random vector asymptotically. Thus, we construct the
test statistic $T$ as 
\begin{equation}
    T=\left\{ \widehat{\Phi}_{B,n}(\widehat{\mu}_A)\right\} ^{\T}\widehat{\Sigma}_{T}^{-1}\left\{\widehat{\Phi}_{B,n}(\widehat{\mu}_A)\right\} ,
    \label{eq:testT}
\end{equation}
where $\Sigma_{T}$ is the asymptotic variance of ${\Phi}_{B,n}(\widehat{\mu}_A,\widehat{\tau})$,
and $\widehat{\Sigma}_{T}$ is a consistent estimator of $\Sigma_{T}$. The exact form of $\Sigma_{T}$ in \eqref{eq:sigma_T} involves $V_{A}$, $V_{B}$, and $\Gamma$. Thus, $\widehat{\Sigma}_{T}$
can be obtained by replacing the unknown components in the expression
of $\Sigma_{T}$ with their estimated counterparts, and the expectations
with the empirical averages. In addition, we can consider the replication-based
method for variance estimation in Algorithm \ref{alg:replication} adapted from \citep{mashreghi2014bootstrap}. 

Lemma \ref{lemma:test} serves as the foundation for our data-driven
pooling step in Section \ref{subsec:Elastic}. 
\begin{lemma}
\label{lemma:test}Suppose Assumptions \ref{asmp:Sampling-design}, \ref{asmp:MAR} (i) and (iii), and all the regularity conditions in Lemma \ref{lemma:mu_A_B_H0}
hold. Under $H_{0},$ the test statistic $T
{\rightarrow}\chi_{l}^{2}$
, i.e., a chi-square distribution with degree of freedom $l$. Under
$H_{a,n}$, $T
{\rightarrow}\chi_{l}^{2}(\eta^{\T}\Sigma_{T}^{-1}\eta/2)$
with non-central parameter $\eta^{\T}\Sigma_{T}^{-1}\eta/2$
as $n\rightarrow\infty$. 
\end{lemma}

\subsection{Data-driven pooling\label{subsec:Elastic}}

If $T$ is large, it indicates that $H_{0}$ may be violated and thus
it is desirable to retain only the probability sample for estimation. If $T$
is small, it indicates that $H_{0}$ may be accepted and suggests combining
the probability and non-probability samples for efficient estimation. This strategy leads
to the test-and-pool estimator $\widehat{\mu}_{\tap}$ as the solution to 
\begin{equation}
\sum_{i=1}^{N}\{\widehat{\Phi}_{A}(V_{i},\delta_{A,i};\mu)+\bone(T<c_{\gamma})\Lambda \widehat{\Phi}_{B}(V_{i},\delta_{A,i},\delta_{B,i};\mu)\}=0,\label{eq:tap}
\end{equation}
where $c_{\gamma}$ is the $(1-\gamma)$ critical value of $\chi_{l}^{2}$.
In (\ref{eq:tap}), we can fix $\Lambda$ to be the optimal form $\Lambda_{\eff}$
leading to an efficient estimator under $H_{0}$ in Section \ref{subsec:Efficient-estimator-combining}.
Alternatively, we view $c_{\gamma}$ and $\Lambda$ jointly as tuning
parameters that determine how much information from the non-probability sample
can be borrowed in pooling. Larger $c_{\gamma}$ and $\Lambda$ borrow
more information from the non-probability sample, leading to more efficient but
more error-prone estimators, and vice versa. We will use a data-adaptive
rule to select $(\Lambda, c_{\gamma})$ that minimizes the mean squared
error of $\widehat{\mu}_{\tap}$. 

\begin{remark}
Compare to the t-test-based pooling estimator in \cite{mosteller1948pooling}, our proposed method is more general in the sense that (a) the auxiliary covariates are used to provide a more informative model of $\mu_{g}$; (b) our test statistic $T$ is motivated by the estimating function, which can be more robust to model misspecification, and (c) a data-adaptive selection of $(\Lambda,c_{\gamma})$ is adopted for minimizing the post-integration mean squared error.
\end{remark}

\section{Asymptotic properties of the test-and-pool estimator \label{sec:Asymptotic-properties-of}}
In this section, we characterize the asymptotic properties of $\widehat{\mu}_{\tap}$. Before proceeding further, we introduce more notations.
Let $I_{l\times l}$ be a $l\times l$ identify matrix, $F_{l}(\cdot;\eta)$
be the cumulative distribution function for $\chi_{l}^{2}$
with non-central parameter $\eta$, and $F_{l}(\cdot)=F_{l}(\cdot;0).$
Denote $V_{\text{A-eff}}=V_{A}-V_{\eff}$ and $V_{\text{B-eff}}=V_{B}-V_{\eff}$, which are both positive-definite.

\subsection{Asymptotic distribution\label{subsec:Asym_dist}}

By construction, the estimator $\widehat{\mu}_{\tap}$ is a pretest
estimator that first constructs $T$ for pretesting $H_{0}$ and then
forms the test-based weights for combining $\widehat{\mu}_{A}$ and
$\widehat{\mu}_{B}$. It is challenging to derive the asymptotic distribution
of $\widehat{\mu}_{\tap}$ because it is involved with the test statistic
$T$ and two asymptotically dependent components $\widehat{\mu}_{A}$
and $\widehat{\mu}_{B}$. In order to formally characterize the asymptotic
distribution of $\widehat{\mu}_{\tap}$, we decompose the asymptotic
representation of $\widehat{\mu}_{\tap}$ by two orthogonal components,
one is affected by the testing and the other is not. 

 First, by Lemma \ref{lemma:mu_A_B_Hn}, let $n^{1/2}(\widehat{\mu}_{A}-\mu_{g}){\rightarrow}Z_{1}$
and $n^{1/2}(\widehat{\mu}_{B}-\mu_{g}){\rightarrow}Z_{2}$,
where $Z_{1}$ and $Z_{2}$ are multivariate normal random vectors
as in (\ref{eq:normalAB}). 

Second, by Lemma \ref{lemma:test}, asymptotically, we write $T$
as a quadratic form $W_{2}^{T}W_{2}$ with $W_{2}=-f_{B}^{1/2}\Sigma_{T}^{-1/2}\E\left\{ \partial \Phi_{B}(\mu_{g,0},\tau_{0})/\partial\mu\right\} ^{-1}(Z_{1}-Z_{2})$.
We then find another standardized $l-$variate normal vector $W_{1}=f_{B}^{1/2}\Sigma_{S}^{-1/2}\{(\Gamma^{\T}-V_{B})(\Gamma-V_{A})^{-1}Z_{1}+Z_{2}\}$
that is orthogonal to $W_{2}$, where $\cov(W_1,W_2)=0_{l\times l}$, $\E(W_{1})=\mu_{1},\var(W_{1})=I_{l\times l}$
and $\E(W_{2})=\mu_{2},\var(W_{2})=I_{l\times l}$, $\Sigma_{S}$
is introduced for the purpose of standardization. 

Third, $\widehat{\mu}_{\tap}$ can be asymptotically represented by
two components involving $W_{1}$ and $W_{2}$, respectively, one
component is affected by the test constraint and the other component
is not. Following the above steps, Theorem \ref{thm:mu_TAP} characterizes
the asymptotic distribution of $\widehat{\mu}_{\tap}$. 
\begin{theorem}
\label{thm:mu_TAP}Suppose the assumptions in Lemma \ref{lemma:mu_A_B_Hn}
hold except that Assumption \ref{asmp:MAR} (ii) may be violated as dictated
by $H_{a,n}$ in (\ref{eq:H_n}). Let $W_{1}$ and $W_{2}$ to be
independent normal random vectors with mean $\mu_{1}$ and $\mu_{2}$
(given below, which vary by hypothesis) and variance matrices $I_{l\times l}$ {.}
The test-and-pool estimator $\widehat{\mu}_{\tap}$ follows the following asymptotic
distribution
\[
n^{1/2}(\widehat{\mu}_{\tap}-\mu_{g}){\rightarrow}\begin{cases}
-V_{\eff}^{1/2}W_{1}+(\omega_{A}V_{\text{A}-\eff}^{1/2}-\omega_{B}V_{\text{B}-\eff}^{1/2})W_{[0,c_{\gamma}]}^{t} & w.p.\ \xi,\\
-V_{\eff}^{1/2}W_{1}+V_{\text{A}-\eff}^{1/2}W_{[c_{\gamma},\infty]}^{t} & w.p.\ 1-\xi,
\end{cases}
\]
where 
$W_{[a,b]}^{t}$ is the truncated normal distribution $W_{2}\mid(a\leq W_{2}^{\T}W_{2}\leq b)$
and $\xi=F_{l}(c_{\gamma};\mu_{2}^{\T}\mu_{2}/2)$. 

(a) Under $H_{0}$, $\mu_{1}=\mu_{2}=0,\xi=F_{l}(c_{\gamma};0)=\gamma$. 

(b) Under $H_{a,n}$,
$\mu_{1}=-\Sigma_{S}^{-1/2}\E\left\{ \partial \Phi_{B}(\mu_{g,0},\tau_{0})/\partial\mu\right\} ^{-1}\eta$,
$\mu_{2}=-\Sigma_{T}^{-1/2}\eta$ and $\xi=F_{1}(c_{\gamma};\allowbreak\mu_{2}^{T}\mu_{2}/2)$. 

\end{theorem}

Theorem \ref{thm:mu_TAP} reveals that the asymptotic distribution
of $\widehat{\mu}_{\tap}$ depends on the local parameter $\eta$
and thus characterizes the non-regularity of the pretest estimator.
When $H_{0}$ is violated weakly (a small perturbation in the true
data generating model), the asymptotic distribution of $\widehat{\mu}_{\tap}$
can change abruptly depending on $\eta.$ The non-regularity of $\widehat{\mu}_{\tap}$
also poses challenges for inference as shown in Section \ref{sec:ACI}.
Based on Theorem \ref{thm:mu_TAP}, we derive the asymptotic biases
and mean squared errors of $\widehat{\mu}_{\tap}$ under $H_{0}$ and $H_{a,n}$,
which serve as the stepping stone to a data-driven procedure to select
the tuning parameters $\Lambda$ and $c_{\gamma}$.

\subsection{Asymptotic bias and mean squared error\label{subsec:bias_mse}}

Based on the Theorem \ref{thm:mu_TAP}, the asymptotic distribution
of $\widehat{\mu}_{\tap}$ involves elliptical truncated normal distributions
\citep{tallis1963elliptical,barr1999mean}. To understand the asymptotic behavior of our proposed estimator,
it is crucial to comprehend the essential properties of elliptical truncated
multivariate normal distributions. We derive the moment generating
function and subsequently the mean square error  of the estimator $\widehat{\mu}_{\tap}$. The exact form of mean squared error given by $\text{mse}(\text{\ensuremath{\Lambda}},c_{\gamma};\eta)$
in (\ref{eq:mse_general}), albeit complicated, reveals that the amount
of information borrowed from the non-probability sample (controlled by $\Lambda$
and $c_{\gamma}$) should tailor to the strength of violation of $H_{0}$
(dictated by local parameter $\eta$). For illustration, we consider
a toy example in the supplemental material.

We search for the optimal values $(\Lambda^{*},c_{\gamma}^{*})$ that
minimize $\text{mse}(\Lambda,c_{\gamma};\widehat{\eta})$ using standard
numerical optimization algorithm \citep{nelder1965simplex}, where
$\widehat{\eta}=\Phi_{B,n}(\widehat{\mu}_A, \widehat{\tau})$. Note that the decision of rejecting $H_0$ or not is subject to the hypothesis testing errors, namely the Type I error and Type II error. That is, the test statistic $T$ can be larger than $c_\gamma$ even when $H_0$ holds; similarly, it can be small when $H_{a,n}$ holds. However, the data-adaptive tuning procedure aims at minimizing the mean squared error of the estimator $\widehat{\mu}_{\tap}$, which implicitly restricts these two testing errors to be small.

\subsection{Adaptive inference\label{sec:ACI}}

Standard approaches to inference, e.g., the nonparametric bootstrap,
require the estimators to be regular \citep{shao1994bootstrap}. In
non-regular settings, researchers have proposed alternative approaches
such as the $m$-out-$n$ bootstrap or subsampling. However, these
approaches critically rely on a proper choice of $m$ or the subsample
size; otherwise, the small sample performances can be poor. The non-regularity
is induced because the asymptotic distribution of the estimator $\widehat{\mu}_{\tap}$
depends on the local parameter, thus, it does not converge uniformly
over the parameter space. \cite{laber2011adaptivejasa}
propose adaptive confidence intervals for test errors in the classification
problems. Following this idea, we construct the bound-based adaptive
confidence interval (BACI) for the estimator $\widehat{\mu}_{\tap}$ that guarantees good
coverage properties. To avoid the non-regularity, our general strategy
is to derive two smooth functionals that bound the estimator $\widehat{\mu}_{\tap}$.
Because these two functionals are regular, standard approaches to inference
can be adopted and valid confidence intervals follow.

To be concrete, we construct a bound-based adaptive confidence interval for $a^{\T}\mu_{g},$ where
$a\in\R^{l}$ is fixed. By Theorem \ref{thm:mu_TAP}, we can reparametrize
the asymptotic distribution of $a^{\T}n^{1/2}(\widehat{\mu}_{\tap}-\mu_{g})$
as 
\begin{equation}
a^{\T}n^{1/2}(\widehat{\mu}_{\tap}-\mu_{g}){\rightarrow} R_{n}+a^{\T}\omega_{B}(V_{\text{B-eff}}^{1/2}+V_{\text{A-eff}}^{1/2})U_{n},\label{eq:tap_one_smoother}
\end{equation}
where 
\begin{align*}
R_{n} & =-a^{\T}V_{\eff}^{1/2}W_{1}+a^{\T}(\omega_{A}V_{\text{A-eff}}^{1/2}-\omega_{B}V_{\text{B-eff}}^{1/2})W_{2}+a^{\T}\omega_{B}(V_{\text{B-eff}}^{1/2}+V_{\text{A-eff}}^{1/2})\mu_{[c_{\gamma},\infty)}^{t},\\
U_{n} & =W_{[c_{\gamma},\infty)}^{t}-\mu_{[c_{\gamma},\infty)}^{t},
\end{align*}
and $\mu_{[c_{\gamma},\infty)}^{t}=\mu_{2}\bone_{\mu_{2}^{\T}\mu_{2}>c_{\gamma}}$.
By construction, $R_{n}$ is regular and asymptotically normal, but
$U_{n}$ is nonsmooth. Nonsmoothness and nonregularity are interrelated.
To illustrate, if $\mu_{2}=0$, $U_{n}$ follows a standard truncated
normal distribution with truncated probability $\pr(W_{2}^{\T}W_{2}\leq c_{\gamma}\mid\mu_{2}=0)$;
whereas, if $|\mu_{2}|\rightarrow\infty$, $\pr(W_{2}^{\T}W_{2}\leq c_{\gamma}\mid\mu_{2})$
diminishes to zero, implying that $U_{n}$ follows a standard normal
distribution. Thus, the limiting distribution of $a^\T n^{1/2}(\widehat{\mu}_{\tap}-\mu_{g})$
is not uniform over local parameter $\mu_{2}$ (or equivalently $\eta$).

\begin{figure}[!tb]
    \centering
    \includegraphics[width = .75\linewidth]{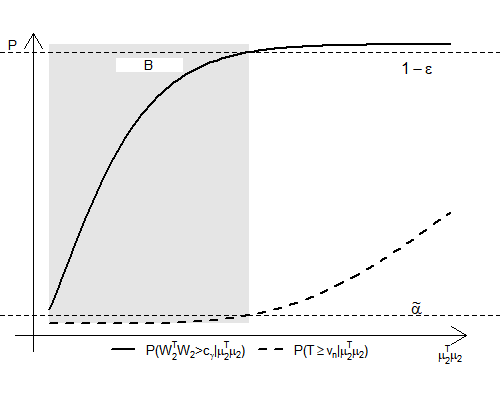}
    \caption{Illustration of the nonregular zone $\mathbb{B}$ (shaded) and two power functions:
the solid and dash lines are $\pr(W_{2}^{\T}W_{2}>c_{\gamma}\mid\mu_{2}^{\T}\mu_{2})$
and $\pr(T\protect\geq v_{n}\mid\mu_{2}^{\T}\mu_{2})$ as functions
of $\mu_{2}^{\T}\mu_{2}$, respectively \label{fig:ht}}
\end{figure}

Our goal is to form the least conservative smooth upper and lower
bounds. An important observation is that if $|\mu_{2}|$ is sufficiently
large, we may treat $U_{n}$ as regular. Thus, we define $\mathbb{B}$
as the nonregular zone for $\mu_{2}^{\T}\mu_{2}$ such that $\max_{\mu_{2}^{\T}\mu_{2}\in\mathbb{B}}\pr(W_{2}^{\T}W_{2}\geq c_{\gamma}\mid\mu_{2})\leq1-\varepsilon$
for small $\epsilon>0$ and $\mathbb{B}^{\complement}$ the regular
zone. When $\mu_{2}^{\T}\mu_{2}\in\mathbb{B}^{\complement}$,
standard inference can apply, and bounds are only needed when $\mu_{2}^{\T}\mu_{2}\in\mathbb{B}$
to avoid the inference procedure to be overly conservative. We then
require another test procedure to test $\mu_{2}^{\T}\mu_{2}\in\mathbb{B}$
against $\mu_{2}^{\T}\mu_{2}\in\mathbb{B}^{\complement}$.
Toward this end, we use $T\geq v_{n}$, where $v_{n}$ is chosen such
that $\max_{\mu_{2}^{\T}\mu_{2}\in\mathbb{B}}\pr(T\geq v_{n}\mid\mu_{2})=\tilde{\alpha}$
for a pre-specified $\tilde{\alpha}$. Figure \ref{fig:ht} illustrates
the regular and nonregular zones and the test. If $T\geq \nu_n$, we conclude the regularity of the estimator $\widehat{\mu}_{\tap}$ and construct a normal confidence interval, but if $T<\nu_n$, we construct the least favorable confidence interval by taking the union for all $\mu_2\in\mathbb{R}^l$. In practice, $v_{n}$
can be determined by the double bootstrapping satisfying the regularity
condition that $\lim_{n\rightarrow\infty}v_{n}/n=0$; see Section \ref{subsec:double-bootstrap} of the supplemental material for more details.

Accordingly, $U_{n}$ can be decomposed into two components $U_{n}=(W_{[c_{\gamma},\infty)}^{t}-\mu_{[c_{\gamma},\infty)}^{t})\bone_{T\geq\upsilon_{n}}+(W_{[c_{\gamma},\infty)}^{t}-\mu_{[c_{\gamma},\infty)}^{t})\bone_{T<v_{n}}$
and only regularize (i.e., deriving bounds for) the latter component.
Continuing with (\ref{eq:tap_one_smoother}), we can take the supremum
over all $\mu_{2}$ in the nonregular zone to construct the upper
bound $U(a)$,
\begin{equation}
\begin{aligned}
 {U}(a) =& R_{n}+a^{\T}\omega_{B}(V_{\text{B-eff}}^{1/2}+V_{\text{A-eff}}^{1/2})(W_{[c_{\gamma},\infty)}^{t}-\mu_{[c_{\gamma},\infty)}^{t})\bone_{T\geq\upsilon_{n}}\\
 & +\sup_{\mu_{2}\in  \R^{l}}\left\{ a^{\T}\omega_{B}(V_{\text{B-eff}}^{1/2}+V_{\text{A-eff}}^{1/2})(W_{[c_{\gamma},\infty)}^{t}-\mu_{[c_{\gamma},\infty)}^{t})\right\}\bone_{T<v_{n}} \label{eq:supp}
\end{aligned}
\end{equation}
The lower bound ${L}(a)$ for $a^{\T}n^{1/2}(\widehat{\mu}_{\tap}-\mu_{g})$
can be computed in an analogous way by replacing $\sup$ with $\inf$
in (\ref{eq:supp}). Taking the supremum and the infimum of $\mu_{2}$
over $\R^{l}$ renders the two bounds ${U}(a)$ and
${L}(a)$ smooth and regular. The limiting distribution of ${U}(a)$ is 
\begin{align}
{U}(a){\rightarrow} & R+a^{\T}\omega_{B}(V_{\text{B-eff}}^{1/2}+V_{\text{A-eff}}^{1/2})(W_{[c_{\gamma},\infty)}^{t}-\mu_{[c_{\gamma},\infty)}^{t})\bone_{\mu_{2}^{\T}\mu_{2}\in\mathbb{B}^{\complement}}\nonumber \\
 & +\sup_{\mu_{2}\in \R^{l}}\left\{ a^{\T}\omega_{B}(V_{\text{B-eff}}^{1/2}+V_{\text{A-eff}}^{1/2})(W_{[c_{\gamma},\infty)}^{t}-\mu_{[c_{\gamma},\infty)}^{t})\right\}\bone_{\mu_{2}^{\T}\mu_{2}\in\mathbb{B}} .\label{eq:upper_bound}
\end{align}
Similarly, the limiting distribution of ${L}(a)$ is (\ref{eq:upper_bound})
by replacing $\sup$ with $\inf$. Based on the limiting distribution
of ${U}(a)$ and ${L}(a)$, if $\pr(\mu_{2}^{\T}\mu_{2}\in\mathbb{B})=0$,
${U}(a)$ and ${L}(a)$ have approximately the same
limiting distributions as $a^{\T}n^{1/2}(\widehat{\mu}_{\tap}-\mu_{g})$.
However, if $\pr(\mu_{2}^{\T}\mu_{2}\in\mathbb{B})\neq0$, ${U}(a)$
is stochastically larger and ${L}(a)$ is stochastically smaller
than $a^{\T}n^{1/2}(\widehat{\mu}_{\tap}-\mu_{g})$.

Based on the regular bounds ${U}(a)$ and ${L}(a)$,
we construct the $(1-\alpha)\times100\%$ bound-based adaptive confidence interval of $a^{\T}\mu_{g}$
as 
\begin{equation}
\mathbb{C}_{\mu_g,1-\alpha}^{\baci}(a)=\left[a^{\T}\widehat{\mu}_{\tap}-{\widehat{U}}_{1-\alpha/2}(a)/\surd{n},a^{\T}\widehat{\mu}_{\tap}-{\widehat{L}}_{\alpha/2}(a)/\surd{n}\right],\label{eq:ci_tap1}
\end{equation}
where ${\widehat{U}}_{d}(a)$ and ${\widehat{L}}_{d}(a)$
approximate the $d$-th quantiles of the distribution of ${U}(a)$
and ${L}(a),$ respectively, which can be obtained by the
nonparametric bootstrap method. 
\begin{theorem}
\label{thm:ACI}Assume the conditions in Theorem \ref{thm:mu_TAP}
hold true. Furthermore, assume matrices $\Sigma_{T}$, $\Sigma_{S}$ in Lemma \ref{lemma:mu_A_B_Hn} and their consistent estimates $\widehat{\Sigma}_T, \widehat{\Sigma}_S$ are strictly positive-definite, and the sequence
$v_{n}$ satisfies $v_{n}\rightarrow\infty$ and $v_{n}/n\rightarrow0$
with probability one. The asymptotic coverage rate of
(\ref{eq:ci_tap1}) satisfies 
\begin{equation}
\pr\left\{a^{\T}\mu_{g}\in \mathbb{C}_{\mu_g,1-\alpha}^{\baci}(a)\right\}\geq1-\alpha.\label{eq:cvg}
\end{equation}
In particular, if Assumption \ref{asmp:MAR} is strongly violated
with $\pr(\mu_{2}^{\T}\mu_{2}\in\mathbb{B}^{\complement})=1$,
the inequality in (\ref{eq:cvg}) becomes equality. 
\end{theorem}

\begin{remark}
\label{rem:1} We discuss an alternative approach to construct valid
confidence intervals for the non-regular estimators using projection
sets \citep{robins2004optimal} (referred to as projection-based adaptive
confidence intervals (PACI), $\mathbb{C}_{\mu_g,1-\alpha}^{\paci}(a)$). The basic idea is as follows.
For a given $\mu_{2}$, the limiting distribution of $\widehat{\mu}_{\tap}$
is known and a regular $(1-\tilde{\alpha}_{1})\times100\%$ confidence interval $\mathbb{C}_{\mu_{g},1-\tilde{\alpha}_{1}}(a;\mu_{2})$
of $a^{\T}\mu_{g}$ can be formed through the standard procedure.
Since $\mu_{2}$ is unknown, a $(1-\alpha)\times100\%$ projection
confidence interval of $\mu_{g}$ can be conservatively constructed
as the union of all $\mathbb{C}_{\mu_{g},1-\tilde{\alpha}_{1}}(a;\mu_{2})$
over $\mu_{2}$ in its $(1-\tilde{\alpha}_{2})\times100\%$ confidence
region, where $\alpha=\tilde{\alpha}_{1}+\tilde{\alpha}_{2}$. Such
strategy may be overly conservative, and in that way, the projection-based adaptive confidence interval then introduces a pretest in order to mitigate the conservatism. If the
pretest rejects $H_{0}:\mu_{2}^{\T}\mu_{2}\in\mathbb{B}$,
$\mathbb{C}_{\mu_{g},1-\tilde{\alpha}_{1}}(a;\widehat{\mu}_{2})$
is used; otherwise, the union of $\mathbb{C}_{\mu_{g},1-\tilde{\alpha}_{1}}(a;\mu_{2})$ is used. The technical details for
the $\mathbb{C}_{\mu_g,1-\alpha}^{\paci}(a)$ are presented in the supplemental material. Our simulation
study later shows that the $\mathbb{C}_{\mu_g,1-\alpha}^{\paci}(a)$ is more conservative than the proposed
$\mathbb{C}_{\mu_g,1-\alpha}^{\baci}(a)$. 
\end{remark}

\section{Simulation study\label{sec:Simulation}}

In this section, we evaluate the finite-sample performances of the
proposed estimator $\widehat{\mu}_{\tap}$ and $\mathbb{C}_{\mu_g,1-\alpha}^{\baci}(a)$. First, we generate the finite population $\F_N$ with size $N=10^5$. For each subject $i$, generate $X_i = (1, X_{1,i},X_{2,i})^\T$, where $X_{1,i}\sim \N(0,1)$ and $X_{2,i}\sim \N(1,1)$, and generate $Y_i$ by $Y_{i}=1+X_{1,i}+X_{2,i}+u_{i}+u_{i}^2+\varepsilon_{i}$, where $u_{i}\sim\N(0,1)$ and $\epsilon_{i}\sim\N(0,1)$. Generate samples from the finite population $\F_N$ by Bernoulli sampling with specified inclusion probabilities
\[
\begin{aligned}
&\log\left(\frac{\pi_{A,i}}{1-\pi_{A,i}}\right)\mid X_{i}=\nu_A + .2X_{1,i}+.1X_{2,i},\\
&\log\left(\frac{\pi_{B,i}}{1-\pi_{B,i}}\right)\mid X_{i}=\nu_B + .1X_{1,i}+.2X_{2,i}+.5n_B^{-1/2}bu_{i},
\end{aligned}
\]
where $\nu_A$ and $\nu_B$ are adaptively chosen to ensure the target sample sizes $n_A\approx 600$ and $n_B \approx 5000$. We assume that $(X_i, Y_i)$ are observed but $u_i$ is unobserved, and we vary $b$ in $\{0,10,100\}$ to represent the scenarios where $H_0$ holds, is slightly violated  or strongly violated , respectively. 

We compare the estimator $\widehat{\mu}_{\tap}$ with other estimators: (a) $\widehat{\mu}_{A}$: the solution to $\sum_{i=1}^{N}\Phi_{A}(V_{i},\delta_{A,i};\allowbreak\mu)=0$ with $\Phi_{A}(V_{i},\delta_{A,i};\mu)$ defined in (\ref{eq:S_A_1}). (b) $\overline{\mu}_{B}$: the naive sample mean $\overline{\mu}_{B}
=(\sum_{i=1}^N \delta_{B,i})^{-1}\allowbreak\sum_{i=1}^N \delta_{B,i}Y_i$. (c) $\widehat{\mu}_{\dr}$: the solution to $\sum_{i=1}^{N}\Phi_{B}(V_{i},\delta_{A,i},\delta_{B,i};\mu, \alpha,\beta)=0$ with $\Phi_{B}(V_{i},\delta_{A,i},\delta_{B,i};\allowbreak\mu, \alpha,\beta)$ defined in (\ref{eq:S_B_1}), where $(\alpha,\beta)$ are estimated by using the maximum pseudo-likelihood estimator $\widehat{\alpha}$ and the ordinary least square estimator $\widehat{\beta}$ \citep{haziza2006nonresponse}; see Equations \eqref{eq:Phi_tau_alpha_1} and \eqref{eq:Phi_tau_beta_2}. (d) $\widehat{\mu}_{\text{eff}}$: the solution to (\ref{eq:general comb})
with the optimal choice $\Lambda_\eff$ specified in (\ref{eq:lambda_eff}) and the consistent estimators $(\widehat{\alpha},\widehat{\beta})$ obtained from (c). (e) $\widehat{\mu}_{\text{eff}:B}$: $\widehat{\mu}_{\text{eff}}$, where $\alpha$ is estimated in the same manner as (c) but $\beta$ is estimated solely based on the non-probability sample; see Equation \eqref{eq:Phi_tau_beta_1}. (f) $\widehat{\mu}_{\text{eff}:\rm KH}$: $\widehat{\mu}_{\text{eff}}$, where $(\alpha,\beta)$ are estimated simultaneously by adopting the methods proposed by \cite{kim2014doubly}; see Equations \eqref{eq:Phi_tau_alpha_KH} and \eqref{eq:Phi_tau_beta_KH}. (g) $\widehat{\mu}_{\tap}$, $\widehat{\mu}_{\tap:B}$, $\widehat{\mu}_{\tap:\rm KH}$: the solution to (\ref{eq:tap}),
where $(\Lambda,c_{\gamma})$ are chosen by our data-adaptive procedure
with $(\widehat{\alpha},\widehat{\beta})$ obtained from (d), (e), (f), respectively. (h) $\widehat{\mu}_{\text{Bayes}:1}$, $\widehat{\mu}_{\text{Bayes}:2}$, $\widehat{\mu}_{\text{Bayes}:3}$: the Bayesian approaches for combining the non-probability sample with the probability sample assuming different informative priors \citep{sakshaug2019supplementing}.

For all estimators, we specify the model $\pi_B(X;\alpha)$ to be a
logistic regression model with $X_{i}$ and the outcome mean model $m(X;\beta)$
to be a linear regression model with $X_{i}$. For non-regular estimators
$\widehat{\mu}_{\tap}$, $\widehat{\mu}_{\tap:B}$ and $\widehat{\mu}_{\text{eff}:\rm KH}$,
we construct the $\mathbb{C}^{\baci}_{\mu_g,1-\alpha}(a)$ in (\ref{eq:ci_tap1}) with a data-adaptiv choice of $\nu_n$, the $\mathbb{C}^{\baci}_{\mu_g,1-\alpha}(a)$ with a fixed $v_{n}=\log\log n\{\mathbb{C}^{\bacif}_{\mu_g,1-\alpha}(a)\}$ ($\text{BACI}_F$), and the $\mathbb{C}^{\paci}_{\mu_g,1-\alpha}(a)$. For any confidence intervals requiring the nonparametric
bootstrap, the bootstrap size is $2000$. For the Bayesian estimators, the point estimates are obtained by the Markov chain Monte Carlo sampling with size $2000$ after additional $500$ burn-in samples.

\begin{table}[!tb]
\caption{Simulation results for bias $(\times 10^{-3})$, variance (var) $(\times 10^{-3})$ and mean squared error (MSE) $(\times 10^{-3})$ of $\widehat{\mu}_{A},\overline{\mu}_{B},\widehat{\mu}_{\dr},\widehat{\mu}_{\text{eff}}, \widehat{\mu}_{\text{Bayes}}$
and $\widehat{\mu}_{\tap}$ when $H_0$ holds, is slightly violated or strongly violated \label{tab:Summary-stat}}  
\vspace{0.15cm}
\centering
\resizebox{.8\textwidth}{!}{\begin{tabular}{llccccccccc}
    \toprule
    \multicolumn{1}{l}{$H_0$} &       & \multicolumn{3}{c}{holds} & \multicolumn{3}{c}{slightly violated} & \multicolumn{3}{c}{strongly violated} \\
          &       & \multicolumn{1}{l}{bias} & \multicolumn{1}{l}{var} & \multicolumn{1}{l}{MSE} & \multicolumn{1}{l}{bias} & \multicolumn{1}{l}{var} & \multicolumn{1}{l}{MSE} & \multicolumn{1}{l}{bias} & \multicolumn{1}{l}{var} & \multicolumn{1}{l}{MSE} \\
    \midrule
    \multicolumn{1}{l}{{Regular}} &
    $\widehat{\mu}_A$
    & {-4.1} & {10.4} & {10.4} & {-4.1} & {10.4} & {10.4} & {-4.1} & {10.4} & {10.4} \\
          & $\overline{\mu}_B$ & 284.1  & 1.2   & 81.9  & 355.3  & 1.2   & 127.4  & 1318.8  & 2.0   & 1741.4  \\
          & $\widehat{\mu}_\dr$ & -0.4  & 4.2   & 4.2   & 71.0  & 4.3   & 9.3   & 1048.0  & 5.0   & 1103.2  \\
          & $\widehat{\mu}_\eff$ & -0.9  & 4.1   & 4.1   & 62.3  & 4.2   & 8.1   & 851.5  & 6.6   & 731.7  \\
          & $\widehat{\mu}_{\eff:B}$ & -0.9  & 4.1   & 4.1   & 62.3  & 4.2   & 8.1   & 851.7  & 6.6   & 732.1  \\
          & $\widehat{\mu}_{\eff:\text{KH}}$ & -0.9  & 4.1   & 4.1   & 62.3  & 4.2   & 8.1   & 851.5  & 6.7   & 731.7  \\
    \midrule
    \multicolumn{1}{l}{{Bayes}} & $\widehat{\mu}_{\text{Bayes:1}}$ & -3.7  & 14.1  & 14.1  & 1.0   & 14.0  & 14.0  & -4.3  & 14.1  & 14.1  \\
          & $\widehat{\mu}_{\text{Bayes:2}}$ & -4.1  & 10.8  & 10.8  & {17.1} & {11.1} & {11.4} & {7.0} & {13.8} & {13.8} \\
          & $\widehat{\mu}_{\text{Bayes:3}}$ & {-2.4} & {8.9} & {8.9} & 51.2  & 9.0   & 11.6  & 614.0  & 10.8  & 387.9  \\
    \midrule
    \multicolumn{1}{l}{{TAP}} & $\widehat{\mu}_{\tap}$ & -4.8  & 7.6   & 7.6   & 10.1  & 9.3   & 9.4   & -4.1  & 10.4  & 10.4  \\
          & $\widehat{\mu}_{\tap:B}$ & -4.8  & 7.6   & 7.6   & 10.1  & 9.3   & 9.4   & -4.1  & 10.4  & 10.4  \\
          & $\widehat{\mu}_{\tap:\text{KH}}$ & {-4.8} & {7.6} & {7.6} & {10.1} & {9.3} & {9.4} & {-4.1} & {10.4} & {10.4}\\
    \bottomrule
    \end{tabular}%
    }
\end{table}%

Table \ref{tab:Summary-stat} reports the bias, variance and mean squared error of each estimator over $2000$ simulated datasets. The benchmark estimators $\widehat{\mu}_{A}$ have small biases across all
scenarios, guaranteed by the probability sampling design. On the other hand, the non-probability-only estimators $\overline{\mu}_{B}$ exhibit high biases in all cases, mainly due to the effect of selection bias. When the
impact of the unmeasured confounder $b$ increases, the pooled estimators
$\widehat{\mu}_{\text{eff}},\widehat{\mu}_{\text{eff}:B}$ and $\widehat{\mu}_{\text{eff}:\rm KH}$
are becoming more biased. Additionally, the Bayesian methods, particularly $\widehat{\mu}_{\text{Bayes}:2}$, perform reasonably well when $H_0$ holds or is slightly violated, but it tends to have large biases when $H_0$ is strongly violated. Whereas the proposed estimators $\widehat{\mu}_{\tap},\widehat{\mu}_{\tap:B}$
and $\widehat{\mu}_{\tap:\rm KH}$ have small biases regardless of the
strength of the unmeasured confounder. When $H_{0}$ is slightly violated,
our proposed estimators have slightly larger biases but smaller mean squared errors than
$\widehat{\mu}_{A}$ by integrating the non-probability sample. When $H_{0}$
is strongly violated, the proposed estimators perform similarly to $\widehat{\mu}_{A}$
with the protection of pretesting.

\begin{table}[tb]
    \caption{Simulation results for coverage rates (CR) $(\times 10^{-2})$ and widths $(\times 10^{-3})$ for 95\% confidence intervals when $H_0$ holds, is slightly violated or strongly violated\label{tab:CIs}}
    \vspace{0.15cm}
    \centering
    {\begin{tabular}{llcccccc}
    \toprule
\multicolumn{1}{l}{$H_0$} &       & \multicolumn{2}{c}{holds} & \multicolumn{2}{c}{slightly violated} & \multicolumn{2}{c}{strongly violated} \\
          & CIs   & CR    & \multicolumn{1}{l}{width} & \multicolumn{1}{l}{CR} & \multicolumn{1}{l}{width} & \multicolumn{1}{l}{CR} & \multicolumn{1}{l}{width} \\
    \midrule
    $\widehat{\mu}_A$ & Wald & {95.2 } & {404.1} & {95.3} & {404.1} & {95.2} & {404.0} \\
    $\overline{\mu}_B$ &       & 0.0   & 135.5  & 0.0   & 138.8  & 0.0   & 173.7  \\
    $\widehat{\mu}_\dr$ &       & 95.9  & 262.8  & 81.8  & 264.4  & 0.0   & 282.4  \\
    $\widehat{\mu}_\eff$ &       & 95.9  & 259.5  & 85.1  & 260.9  & 0.0   & 273.6  \\
    \midrule
    $\widehat{\mu}_{\text{Bayes:1}}$ & \textsc{hpdi} & 98.3  & 463.0  & 97.5  & 461.5  & 97.3  & 462.8  \\
    $\widehat{\mu}_{\text{Bayes:2}}$ &       & 97.8  & 404.2  & 97.4  & 409.8  &  {97.5} &  {458.3} \\
    $\widehat{\mu}_{\text{Bayes:3}}$ &       &  {99.3} &  {368.2} &  {97.4} &  {370.6} & 0.0   & 407.0  \\
    \midrule
    $\widehat{\mu}_\tap$ & \textsc{paci}  & 98.4  & 558.7  & 98.4  & 535.7  & 99.2  & 541.0  \\
          & $\textsc{baci}_F$ & 94.7  & 399.1  & 95.9  & 402.3  & 94.7  & 402.6  \\
          & \textsc{baci}  & {92.1} & {363.1} & {93.3} &  {367.2} &  {94.8} &  {402.8} \\
    \bottomrule
    \end{tabular}}
\end{table}

Table \ref{tab:CIs} reports the properties of $95\%$ Wald confidence intervals for the regular estimators, the highest
posterior density intervals (HPDIs) for the Bayesian estimators, and various adaptive confidence intervals for the non-regular estimators $\widehat{\mu}_{\tap}$, where the Wald confidence intervals are constructed, and the Bayesian credible intervals are constructed based on the posterior samples after burn-in. Because the confidence intervals (and the point estimates; see Table \ref{tab:Summary-stat}) are not sensitive to the methods of estimating the nuisance parameters $(\alpha, \beta)$, we only present the confidence intervals for $\widehat{\mu}_{\eff:\rm KH}$ and $\widehat{\mu}_{\tap:\rm KH}$ for simplicity. Based on Table \ref{tab:CIs}, $\mathbb{C}_{\mu_g,1-\alpha}^{\paci}$ tend to overestimate
the uncertainty, leading to over-conservative confidence intervals.
$\mathbb{C}_{\mu_g,1-\alpha}^{\baci}$ and $\mathbb{C}_{\mu_g,1-\alpha}^{\bacif}$ are less conservative and alleviate
the over-coverage issues; thus, the empirical coverage rates are close
to the nominal level in all cases. Moreover, $\mathbb{C}_{\mu_g,1-\alpha}^{\baci}$ have narrower intervals than $\mathbb{C}_{\mu_g,1-\alpha}^{\bacif}$ by using
the double bootstrap procedure to select $v_{n}$ at the expense of
computational burden. When $H_0$ holds, the $\mathbb{C}_{\mu_g,1-\alpha}^{\baci}$ are narrower than the Wald for the probability-only estimator $\widehat{\mu}_A$, indicating the advantages of implementing the test-and-pool strategy in these cases. When $H_0$ is slightly violated, the benefit in coverage rate is not significantly observed under similar coverage rates. When $H_0$ is strongly violated, the adaptive confidence interval $\mathbb{C}_{\mu_g,1-\alpha}^{\baci}$ reduces to the Wald confidence intervals for $\widehat{\mu}_A$. Lastly, the credible intervals for the Bayesian estimators do not have satisfactory coverage properties as the model misspecification persists across scenarios, which is aligned with the Bernstein-von Mises Theorem \citep[Chapter~10.2]{van2000asymptotic}.


\section{A real-data illustration\label{sec:Application}}

To demonstrate the practical use, we apply the proposed method to a probability sample from the 2015 Current Population Survey (CPS) and a non-probability
sample from the 2015 Behavioral Risk Factor Surveillance System (BRFSS) survey. Note that the Behavioral Risk Factor Surveillance System survey itself is a probability sample and we manually discard its sampling weights to recast it as a non-probability sample for illustrating our proposed method.

To apply the proposed method, we use a two-phase sampling survey data with sizes $n_{A}=1000$ and $n_{B}=8459$. We focus on two outcome variables
of interest: employment (percentages of working and retired) and educational
attainment (high school or less as h.s.o.l, and college or
above as c.o.a.). Both datasets provide measurements on the
outcomes of interest and some common covariates including age, sex
(female or not), race (white and black), origin (Hispanic or not),
region (northeast, south, or west), and marital status (married or not). To illustrate the heterogeneity in the study populations, Table \ref{tab:unadjust-mean-of} contrasts the means of variables from
the CPS sample (design-weighted averages) and the BRFSS sample (simple averages). Based on Table \ref{tab:unadjust-mean-of}, the BRFSS sample may not be representative of the target population, and the pretesting procedures before pooling should be expected.

\begin{table}[!ht]
\caption{The covariate means by two samples: CPS sample (a probability sample) and BRFSS sample (a hypothetical non-probability sample)\label{tab:unadjust-mean-of}}
    \vspace{0.15cm}
\centering %
\resizebox{.8\textwidth}{!}{
\begin{tabular}{llllllll}
\toprule 
Data source  & \multicolumn{1}{l}{age} & \multicolumn{1}{l}{\%sex} & \multicolumn{1}{l}{\%white} & \multicolumn{1}{l}{\%black} & \multicolumn{1}{l}{\%hispanic} & \multicolumn{1}{l}{\%northeast} & \multicolumn{1}{l}{\%south}\tabularnewline
\midrule 
CPS  & 47.5  & 56.5  & 81.9  & 11.0  & 13.3  & 18.1  & 37.7\tabularnewline
BRFSS  & 48.3  & 54.2  & 83.2  & 8.4  & 8.3  & 20.0  & 27.6\tabularnewline
\midrule 
 & \multicolumn{1}{l}{\%west} & \multicolumn{1}{l}{\%married} & \multicolumn{1}{l}{\%working} & \multicolumn{1}{l}{\%retired} & \multicolumn{1}{l}{\%h.s.o.l.} & \multicolumn{1}{l}{\%c.o.a.} & \tabularnewline
\midrule 
CPS  & 24.1  & 52.5  & 58.7  & 13.6  & 39.4  & 30.3  & \tabularnewline
BRFSS  & 29.5  & 50.8  & 52.2  & 24.5  & 21.2  & 41.9  & \tabularnewline
\bottomrule
\end{tabular}}
\end{table}

\begin{table}[!ht]
\caption{Estimated population mean (EST), standard errors (SE) and confidence intervals of $\mu_{g}$ for selected
covariates when combining two datasets 
\label{tab:Estimated-population-mean}}
\vspace{0.15cm}
\centering
\resizebox{.8\textwidth}{!}{
\begin{tabular}{llllll}
\toprule 
\multicolumn{1}{l}{Outcome $Y$} &  & \%working  & \%retire  & \%h.s.o.l.  & \%c.o.a.\\
\midrule
{$\widehat{\mu}_{A}$} & \textsc{est}   & 58.7  & 13.6  & 39.4  & 30.3\\
 & \textsc{se}   & 1.51  & 1.17  & 1.60  & 1.59\\
 & Wald & (54.8,62.3)  & (11.6,16.2)  & (35.7,43.0)  & (27.2,33.7)\\
{$\widehat{\mu}_{\dr}$} & \textsc{est}   & 56.5  & 20.0  & 25.8  & 32.3\\
 & \textsc{se}  & 1.03  & 1.24  & 0.93  & 1.20\\
 & Wald  & (54.2,58.8)  & (17.9,22.4)  & (234.0,27.5)  & (30.3,34.5)\\
{$\widehat{\mu}_{\text{eff}:\rm KH}$} & \textsc{est}   & 56.6  & 17.3  & 26.4  & 32.1\\
 & \textsc{se}   & 0.80  & 0.19  & 0.87 & 0.62\\
 & Wald   & (54.3,58.9)  & (15.4,19.6)  & (24.6,28.1)  & (30.1,34.3)\\
 \midrule
 {$\widehat{\mu}_{\text{Bayes}:1}$} & \textsc{est}   & 59.8  & 14.1  & 40.5  & 30.7\\
 & \textsc{se}    & 1.97  & 1.37  & 2.00  & 1.84\\
 & \textsc{hpdi}  & (56.0, 63.6)  & (11.4,16.8)  & (36.6,44.4)  & (27.2,34.4)\\
 {$\widehat{\mu}_{\text{Bayes}:2}$} & \textsc{est}  & 59.8  & 14.0  & 40.3  & 30.9\\
 & \textsc{se}    & 2.01  & 1.33  & 1.92  & 1.84\\
 & \textsc{hpdi}  & (56.1,63.9)  & (11.4,16.4)  & (36.4,44.0)  & (27.2,34.5)\\
 {$\widehat{\mu}_{\text{Bayes}:3}$} & \textsc{est}   & 58.6  & 14.1  & 37.6  & 31.1\\
 & \textsc{se}    & 1.94  & 1.30  & 1.92  & 1.76\\
 & \textsc{hpdi}  & (54.7, 62.4)  & (11.6,16.7)  & (33.7,41.4)  & (27.7,34.7)\\
 \midrule
{$\widehat{\mu}_{\tap:\rm KH}$} & \textsc{est}   & 58.7  & 13.6  & 39.0  & 31.7\\
 & \textsc{se}  & 1.51  & 1.17  & 1.55  & 0.64\\
 & \textsc{baci}  & (54.9,62.6)  & (11.6,15.8)  & (35.8,42.6)  & (31.0,33.6)\\
 \bottomrule
\end{tabular}
}
\end{table}

Table \ref{tab:Estimated-population-mean} presents the results. For all estimators, we specify the propensity score
model to be a logistic regression model with the covariates (all variables
excluding the outcome variable) and the outcome mean model to be a
logistic regression model with the covariates. The
efficient estimator $\widehat{\mu}_{\eff}$ gains efficiency in all
estimators compared to both $\widehat{\mu}_{A}$ and $\widehat{\mu}_{\dr}$;
however, it may be subject to biases if the non-probability sample does not satisfy
the required assumptions. In the test-and-pool analysis, the pretesting rejects
the use of the non-probability sample for the employment variables "working"
and "retired " but accepts the use of the non-probability sample for the education
variables  "high school or less" and "college or above". Thus, for the employment
variables, $\widehat{\mu}_{\tap}=\widehat{\mu}_{A}$, and for the
educational attainment variables, $\widehat{\mu}_{\tap}$ gains efficiency
over $\widehat{\mu}_{A}$. The Bayesian estimators with the informative priors 2 and 3 are more efficient than the prior 1. However, they still yield larger standard errors compared to the probability-only estimator $\widehat{\mu}_A$ perhaps because the non-probability-based informative priors are biased for the model parameters for the probability sample. From the test-and-pool analysis, the employment rate
and the retirement rate are $58.7\%$ and $13.6\%$, respectively,
the percentage of the U.S. population with a high school education
or less is $39.0\%$ and the percentage of the population with a college
education or above is $31.7\%$ in 2015.

\section{Concluding remarks\label{sec:Concluding-remarks}}

When utilizing the non-probability samples, researchers often assume that the
observed covariates contain all the information needed for recovering
the sampling mechanism. However, this assumption may be violated,
and hence the integration of the probability and non-probability samples is subject to biases.
In this paper, we propose the test-and-pool estimator that firstly scrutinizes
the assumption required for combining by hypothesis testing and carefully
combines the probability and non-probability samples by a data-driven procedure to achieve
the minimum mean squared error. In theoretical development, we treat $(\Lambda, c_{\gamma})$ jointly as two tuning parameters and establish the asymptotic
distribution of the pretesting estimator without taking their uncertainties into account. The non-regularity of the pretest estimator invalidates the conventional method for generating reliable inferences. To address this issue, the proposed adaptive confidence interval has been designed to effectively handle the non-smoothness of the pretest estimator and ensure uniform validity of inferences. It is important to note, however, that this approach may result in a little gain in the precision of the confidence interval, although the point estimator might have a significant gain in the MSE compared to the estimator based only on the probability sample. Further research is required to develop a valid post-testing confidence interval that offers reduced conservatism.

Pretest estimation is the norm rather than the exception in applied
research, so the theories that we have established are highly relevant to researchers who engage in applied work. The proposed framework can be extended in the following directions. First, in this work, we study the implications of pretesting on estimation and inference under one single pretest. In practice, researchers may engage in multiple presetting. For example, in the data integration context, one can encounter multiple data sources \citep{rothwell2005subgroup,yang2020combining}, requiring pretesting of the comparability of each data source and
the benchmark. Multiple presetting alters the current asymptotic results and is an important future research topic. Second, our framework considers a fixed number of covariates; however, in reality, practitioners often collect a rich set of auxiliary variables, rendering variable selection imperative \citep{yang2020doubly}. Developing a valid statistical
framework to deal with issues arising from selective inference is a challenging but important topic for further investigation.
Third, small area estimation has received a lot of attention in the
data integration context \citep{rao2014small,kalton2019developments}.
The typical estimator in small area estimation is a weighted average of the design-based estimator and a model-based synthetic estimator. \cite{beaumont2020probability} discussed the trade-off of the efficiency gain from invoking model assumptions and the risk that these assumptions do not hold. Thus, pretesting can be potentially useful for small-area estimation, which we will investigate in the future.

\section{Acknowledgment}
Yang's research is partially supported by NIH 1R01AG066883 and 1R01ES031651.

\bibliographystyle{dcu} 
\bibliography{ci}

\setcounter{equation}{0}
\setcounter{table}{0}
\setcounter{theorem}{0}
\setcounter{lemma}{0}
\setcounter{figure}{0}
\renewcommand{\theequation}{S\arabic{equation}}
\renewcommand{\thetable}{S\arabic{table}}  
\renewcommand{\thefigure}{S\arabic{figure}}
\renewcommand{\thesection}{S\arabic{figure}}

\setcounter{algocf}{0} 
\makeatletter
\renewcommand{\thealgocf}{S\@arabic\c@algocf}
\makeatother
\renewcommand\thetheorem{{S}\arabic{theorem}}
\renewcommand\thelemma{{S}\arabic{lemma}}
\renewcommand\thecorollary{{S}\arabic{corollary}}
\renewcommand\theproposition{{\rm S}\arabic{proposition}}
\renewcommand\thedefinition{{\rm S}\arabic{definition}}
\renewcommand\theassumption{{ S}\arabic{assumption}}
\renewcommand\theremark{{S}\arabic{remark}}
\renewcommand\thestep{{\rm S}\arabic{step}}
\renewcommand\thecondition{{\rm S}\arabic{condition}}
\renewcommand\theexample{{\rm S}\arabic{example}}

\newpage
\appendix
\section{Proofs}

\subsection{Regularity conditions}
Let $\F_{N}=\{V_{i}=(X_{i}^{\T},Y_{i})^{\T}:i\in U\}$, $\Phi_{A}(V,\delta_A;\mu)$ and $\Phi_{B}(V,\delta_A,\delta_B;\mu,\tau)$ be $l-$dimensional estimating functions for the parameter $\mu_{g}\in\R^l$ when using the probability sample and the combined samples, respectively. Let $\Phi_\tau(V,\delta_A,\delta_B;\tau)$ be the $k$-dimensional estimating equations for the nuisance parameter $\tau_0 \in \R^k$. Then, we construct one stacked estimating equation system $\Phi(V,\delta_A,\delta_B;\theta)$ with $\theta=(\mu_A^\T,\mu_B^\T,\tau^\T)^\T$ and $\dim(\theta)=2l+k$. For establishing our stochastic statements, we require the following regularity conditions.

\begin{assumption}\label{asmp:regularity} 
The following regularity conditions hold.
\begin{enumerate}
\item[a)] The parameter $\theta=(\mu_A^\T,\mu_B^\T,\tau^\T)^\T$ belongs to a compact
parameter spaces $\Theta$ in $\R^{2l+k}$.
\item[b)] There exist a unique solution $\theta_0 = (\mu_{A,0}^\T,\mu_{B,0}^\T, \tau_0^\T)^\T$ lying in the interior of the compact space $\Theta$ such that
\begin{equation*}
    \mathbb{E}\{\Phi_A(V,\delta_A;\mu_{A,0})\}=
    \mathbb{E}\{\Phi_B(V,\delta_A,\delta_B;\mu_{B,0},\tau_0)\}=
    \mathbb{E}\{\Phi_\tau(V,\delta_A,\delta_B;\tau_0)\}=0.
    \end{equation*}
\item[c)] $\Phi(V,\delta_A,\delta_B;\theta)$ is integrable with respect to the joint distribution of $(V,\delta_A,\delta_B)$ for all $\theta$ in a neighborhood of $\theta_0$.
\item[d)]  The first two partial derivatives of $\E\{\Phi(V,\delta_A,\delta_B;\theta)\}$ and their empirical estimators are invertible for all $\theta$ in a neighborhood of $\theta_0$.
\item[e)] For all $j,k,l\in\{1,\cdots,2l+k\}$, there is an integrable function $B(V,\delta_A,\delta_B)$ such that
    $$|\partial \Phi_j(V,\delta_A,\delta_B;\theta)/\partial\theta_k\partial \theta_l|\leq B(V,\delta_A,\delta_B),\quad
    \mathbb{E}\left\{
    B(V,\delta_A,\delta_B)
    \right\}<\infty,
    $$
    for all $\theta$ in a neighborhood of $\theta_0$ almost surely.
    \item [f)] $\{V_i: i\in \mathcal{U}\}$ are a set of i.i.d. random variables s.t. $\mathbb{E}\{
    \left|
    \Phi(V,\delta_A,\delta_B;\theta)
    \right|^{2+\delta}\}$ is uniformly bounded for $\theta$ in a neighborhood of $\theta_0$.
    \item[g)] The sample sizes $n_A$ and $n_B$ are in the same order of magnitude, i.e., $n_A=O(n_B)$. The sampling fractions for both Sample A and B are negligible, i.e., $n/N=o(1)$, where $n=n_A+n_B$.
\item[h)] There exist $C_1$ and $C_2$ such that $0<C_1\leq N\pi_{A,i}/n_A\leq C_2$ and $0<C_1\leq N\pi_{B,i}/n_B\leq C_2$ for all $i\in\mathcal{U}$.
\end{enumerate}
\end{assumption}
Assumption \ref{asmp:regularity} a)-e) are typical finite moment conditions to ensure the consistency of the solution to the estimating functions \cite[Appendix~B]{robins1994estimation}, \cite[Section~3.2]{tsiatis2007semiparametric}, \cite[page~293]{boos2013essential} and \cite[Appendix~C]{vermeulen2015bias}. Assumption \ref{asmp:regularity} f) is required for obtaining the asymptotic normality of $\mu_g$ under superpopulation. Assumption \ref{asmp:regularity} g) states that the sampling fraction is negligible, which is helpful for subsequent variance estimation, and we can use $O(n_A^{-1/2})$, $O(n_B^{-1/2})$ and $O(n^{-1/2})$ interchangeably. Assumption \ref{asmp:regularity} h) implies that the inclusion probabilities for Samples $A$ and $B$ are in the order of $n/N$, which is necessary to establish their root-$n$ consistency.

It is noteworthy that in Assumption \ref{asmp:Sampling-design}, the asymptotic normality is ascertained for the design-weighted estimators given the finite population $\F_N$. Hereby, we extend the conditional normality to the unconditional one, which averages over all possible finite populations satisfying the Assumption \ref{asmp:regularity} (f). The following lemma plays a key role to establish the stochastic statements \cite[Theorem~1.3.6.]{fuller2009sampling}.

\begin{lemma}
    \label{lemm:uncondition}
    Under Assumption \ref{asmp:Sampling-design} and Assumption \ref{asmp:regularity} (f), let $\{\mathcal{F}_N\}$ be a sequence of finite populations and 
    $\mathcal{A}_N$ be a sample selected from the $N$th population by PR design with size $n_N$. Assume that 
    $$
    \lim_{N\rightarrow \infty}n_N=\infty,\quad
    \lim_{N\rightarrow \infty} N-n_N=\infty.
    $$
    We know that the distribution of the design-weighted estimator $\widehat{\mu}_{g}$ and finite-population estimator $\mu_g$ are both asymptotically normal distributed such that
    $$
    \widehat{\mu}_{g}\mid \mathcal{F}_N \overset{\cdot}{\sim} \mathcal{N}(\mu_g, V_1),\quad \mu_g \overset{\cdot}{\sim} \mathcal{N}(\mu_{g,0},V_2),
    $$
    where $\overset{\cdot}{\sim}$ denotes the asymptotic distribution. Then, $\widehat{\mu}_{g}-\mu_g$ is also asymptotically normal.
    \end{lemma}
By lemma \ref{lemm:uncondition}, the sampling fraction is negligible, and therefore the limiting variance of $\lim_{N\rightarrow \infty}n_N^{1/2}(\mu_g-\mu_{g,0})$ is $0$, indicating that the intermediate step of producing the finite population is of little significance.

\subsection{Proof of Lemmas \ref{lemma:mu_A_B_H0} and \ref{lemma:mu_A_B_Hn}}
In the general case, we begin to investigate the statistical properties of $$\Phi_{A,n}(\widehat{\mu}_A,\widehat{\tau}) = n^{1/2}N^{-1}\sum_{i=1}^{N}\Phi_{A}(V_{i},\delta_{A,i};\widehat{\mu}_{A},\widehat{\tau}),$$ and $$\Phi_{B,n}(\widehat{\mu}_B,\widehat{\tau}) =n^{1/2}N^{-1}\sum_{i=1}^{N}\Phi_{B}(V_{i},\delta_{A,i},\delta_{B,i};\allowbreak\widehat{\mu}_{B},\widehat{\tau}).$$ First, to simplify our notations, let
$$
\begin{aligned}
&\dot{\Phi}_{A}(V,\delta_{A};\mu,\tau)=\partial \Phi_{A}(V,\delta_{A};\mu,\tau)/\partial\mu,\\
&\dot{\Phi}_{B}(V,\delta_A,\delta_B;\mu,\tau)=\partial \Phi_{B}(V,\delta_A,\delta_B;\mu,\tau)/\partial\mu,\\
&\phi_{B,\tau}(V,\delta_A,\delta_B;\mu,\tau) = \partial \Phi_{B}(V,\delta_{A},\delta_{B};{\mu},{\tau})/\partial \tau,\\
&\phi_{\tau}(V,\delta_A,\delta_B;\tau) = \partial \Phi_{\tau}(V,\delta_{A},\delta_{B};{\tau})/\partial \tau.
\end{aligned}
$$
By the Taylor expansion of
$\Phi_{B,n}(\widehat{\mu}_B,\widehat{\tau})$ at $(\mu_g, \tau_0)$,
we have 
\begin{eqnarray}
0&=&\Phi_{B,n}(\widehat{\mu}_B,\widehat{\tau}) \nonumber \\ &=&n^{1/2}N^{-1}\sum_{i=1}^{N}\Phi_{B}(V_{i},\delta_{A,i},\delta_{B,i};\mu_{g},\tau_{0})\nonumber\\
&&+n^{1/2}N^{-1}\sum_{i=1}^{N}
\phi_{B,\tau}(V_i,\delta_{A,i},\delta_{B,i};\widehat{\mu}_{B}^*,\widehat{\tau}^*)
(\widehat{\tau}-\tau_{0})\nonumber \\
 && +n^{1/2}N^{-1}\sum_{i=1}^{N}\dot{\Phi}_{B}(V_i,\delta_{A,i},\delta_{B,i};\widehat{\mu}_{B}^*,\widehat{\tau}^*)(\widehat{\mu}_{B}-\mu_{g}),\label{eq:muB-mu}
\end{eqnarray}
for some $(\widehat{\mu}_{B}^*,\widehat{\tau}^*)$ lying between $(\widehat{\mu}_B,\widehat{\tau})$ and $(\mu_g,\tau_0)$, which leads to 
\begin{align}
  &-n^{1/2}N^{-1}\sum_{i=1}^{N}\dot{\Phi}_{B}(V_i;\widehat{\mu}_{B}^*,\widehat{\tau}^*)(\widehat{\mu}_{B}-\mu_{g})
 \label{eq:mu_B-mu_0}\\
  &=n^{1/2}N^{-1}\sum_{i=1}^{N}\Phi_{B}(V_{i},\delta_{A,i},\delta_{B,i};\mu_{g},\tau_{0})+n^{1/2}N^{-1}\sum_{i=1}^{N}
\phi_{B,\tau}(V_i,\delta_{A,i},\delta_{B,i};\widehat{\mu}_{B}^*,\widehat{\tau}^*)
 (\widehat{\tau}-\tau_{0}).\nonumber 
\end{align}
Also, under Assumption \ref{asmp:regularity} a), b) and c), by the Taylor expansion, we have 
\begin{align}
n^{1/2}(\widehat{\tau}-\tau_{0}) & =-\left\{ \frac{1}{N}\sum_{i=1}^{N}\phi_{\tau}(V_{i};\tau_{0})\right\} ^{-1}\nonumber\\
&\times \left\{ n^{1/2}N^{-1}\sum_{i=1}^{N}\Phi_{\tau}(V_{i},\delta_{A,i},\delta_{B,i};\tau_{0})\right\} +o_{\zeta\text{-}\mathrm{p}\text{-}\mathrm{np}}(1),\label{eq:tau_hat}
\end{align}
as $\widehat{\tau}\rightarrow\tau_0$. Also, under Assumption \ref{asmp:regularity} (e), we know that
\begin{equation}
\begin{aligned}
    &N^{-1}\sum_{i=1}^N
\dot{\Phi}_A(V_i;\widehat{\mu}_A^*,\widehat{\tau}^*)
\rightarrow
\E\{
\dot{\Phi}_A(V;\mu_{g,0},\tau_0)
\},
\quad
N^{-1}\sum_{i=1}^N
\phi_r(V_i;{\tau}_0)
\rightarrow
\E\left\{
\phi_\tau(V;\tau_0)
\right\},
\\
&N^{-1}\sum_{i=1}^N
\dot{\Phi}_B(V_i;\widehat{\mu}_B^*,\widehat{\tau}^*)
\rightarrow
\E\{
\dot{\Phi}_B(V;\mu_{g,0},\tau_0)
\},\\
& N^{-1}\sum_{i=1}^N 
\phi_{B,\tau}(V_i;\widehat{\mu}_B^*,\widehat{\tau}^*)\rightarrow
\E\left\{
\phi_{B,\tau}(V;{\mu}_{g,0},{\tau}_0)
\right\},
\end{aligned}
    \label{eq:expectation_avg}
\end{equation}
where the first two probability convergence can be straightforward to obtain by Weak Law of Large Numbers under Assumption \ref{asmp:regularity} f) and continuous mapping theorem as $\mu_g \rightarrow\mu_{g,0}$, $(\widehat{\mu}_A,\widehat{\tau}) \rightarrow(\mu_{g,0},\tau_0)$ by design and $(\widehat{\mu}_A^*, \widehat{\tau}^*)$ is lying between $(\widehat{\mu}_A,\widehat{\tau})$ and $({\mu}_{g,0},{\tau}_0)$. As for the third and fourth probability convergence, we first prove that $\mu_{B,0}-\mu_{g,0}=o_{\mathrm{np}\text{-}\mathrm{p}\text{-}\zeta}(1)$ under the local alternative $\E\{\Phi_B(V,\delta_A,\delta_B;\mu_{g,0},\tau_0)\}=n_B^{-1/2}\eta$ in Lemma \ref{lemma:Phi_B_expectation}.

\begin{lemma}\label{lemma:Phi_B_expectation}
Under Assumptions 
\ref{asmp:Sampling-design}, \ref{asmp:MAR} (iii) and suitable moments conditions in Assumption \ref{asmp:regularity}, we have $\mu_{B,0}-\mu_{g,0}=O_{\mathrm{np}\text{-}\mathrm{p}\text{-}\zeta}(n^{-1/2})$.
\end{lemma}
Next, we have under Assumption \ref{asmp:regularity} e), 
\begin{equation}
    \begin{aligned}
&N^{-1}\sum_{i=1}^N 
\dot{\Phi}_B(V_i;\widehat{\mu}_B^*,\widehat{\tau}^*) \\
&\cong 
N^{-1}\sum_{i=1}^N 
\dot{\Phi}_B(V_i;{\mu}_{g,0},{\tau}_0)+
N^{-1}\sum_{i=1}^N 
\frac{\partial^2 \Phi_B(V_i;{\mu}_B^{\#},{\tau}_0)}{\partial \mu\partial \mu^\T}
(\widehat{\mu}_B^*-\mu_{g,0})\\
& \cong N^{-1}\sum_{i=1}^N 
\dot{\Phi}_B(V_i;{\mu}_{g,0},{\tau}_0) + O_{\zeta\text{-}\mathrm{p}\text{-}\mathrm{np}} \{(\widehat{\mu}_B^*-\mu_{B,0}) + (\mu_{B,0} - \mu_{g,0})\}\\
& = \E\{\dot{\Phi}_B(V;{\mu}_{g,0},{\tau}_0)\} + o_{\zeta\text{-}\mathrm{p}\text{-}\mathrm{np}}(1),
\end{aligned}
\label{eq:mu_B2mu_0}
\end{equation}
where $A_n \cong B_n$ means that $A_n = B_n+o_{\zeta\text{-}\mathrm{p}\text{-}\mathrm{np}}(1)$ and ${\mu}_B^{\#}$ lies between $\widehat{\mu}_{B}^*$ and $\mu_{g,0}$. Since $\widehat{\mu}_{B}\rightarrow\mu_{B,0}, {\mu}_{g}\rightarrow\mu_{g,0}$ and $\widehat{\mu}_B^*$ lies between $\widehat{\mu}_B$ and $\mu_g$, we establish the second approximation in (\ref{eq:mu_B2mu_0}) as
$$
(\widehat{\mu}_B^* -\mu_{B,0}) + (\mu_{B,0}-\mu_{g,0})
=O_{\mathrm{np}}(n_B^{-1/2}) + 
O_{\zeta}(N^{-1/2}) = o_{\zeta\text{-}\mathrm{p}\text{-}\mathrm{np}}(1),
$$
since $n_B/N=o(1)$. The probability convergence of $N^{-1}\sum_{i=1}^N\phi_{B,\tau}(V_i;\widehat{\mu}_B^*,\widehat{\tau}^*)$ can be established similarly and hence we obtain the last two parts of (\ref{eq:expectation_avg}). By plugging (\ref{eq:tau_hat}) and (\ref{eq:expectation_avg}) into (\ref{eq:mu_B-mu_0}),
we obtain the influence function for $\widehat{\mu}_{B}$ as 
\begin{align}
  &n^{1/2}(\widehat{\mu}_{B}-\mu_{g})\cong-
 \E\left\{\dot{\Phi}_{B}(V;{\mu}_{g,0},{\tau}_0)
 \right\} ^{-1} \times\left[n^{1/2}N^{-1}\sum_{i=1}^{N}\Phi_{B}(V_{i},\delta_{A,i},\delta_{B,i};\mu_{g},\tau_{0})\right.\nonumber \\
 &\left.-
 \E\left\{\phi_{B,r}(V;{\mu}_{g,0},{\tau}_0)\right\}
 \cdot\E\left\{\phi_{\tau}(V;\tau_{0})\right\} ^{-1}\left\{ n^{1/2}N^{-1}\sum_{i=1}^{N}\Phi_{\tau}(V_{i},\delta_{A,i},\delta_{B,i};\tau_{0})\right\} \right]\nonumber \\
 &\cong n^{1/2}N^{-1}\sum_{i=1}^{N}\psi_{B}(V_{i};\mu_{g},\tau_{0}),\label{eq:mu_B_IF}
\end{align}
where $\psi_{B}(V_{i};\mu,\tau)$ is the influence function
for estimation of $\widehat{\mu}_{B}$ under $H_0$. For completeness, we define the influence function $\psi_{A}(V_{i};\mu,\tau)$
for estimator $\widehat{\mu}_{A}$ in an analogous way as 
\begin{align}
  &n^{1/2}(\widehat{\mu}_{A}-\mu_{g})\cong -n^{1/2}N^{-1}\sum_{i=1}^{N}\left\{ N^{-1}\sum_{i=1}^{N}\dot{\Phi}_{A}(V_{i},\delta_{A,i};\widehat{\mu}_A^*,\widehat{\tau}^*)\right\} ^{-1} \nonumber \\
 & \times\left\{ \Phi_{A}(V_{i},\delta_{A,i};\mu_{g},\tau_{0})+
 \phi_{A,\tau}(V_i;\widehat{\mu}_A^*,\widehat{\tau}^*)
 \cdot(\widehat{\tau}-\tau_{0})\right\}\label{eq:mu_A-mu_0}\\
 & \cong -\E\left\{\dot{\Phi}_{A}(V;{\mu}_{g,0},{\tau}_0)\right\} ^{-1} \times\left[n^{1/2}N^{-1}\sum_{i=1}^{N}\Phi_{A}(V_{i},\delta_{A,i};\mu_{g},\tau_{0})
 \right.\nonumber\\
 &\left.-\E\left\{\phi_{A,\tau}(V;{\mu}_{g,0},{\tau}_0)\right\}
 \cdot
 \E\left\{ \phi_{\tau}(V;\tau_{0})\right\} ^{-1}\left\{ n^{1/2}N^{-1}\sum_{i=1}^{N}\Phi_{\tau}(V_{i},\delta_{A,i},\delta_{B,i};\tau_{0})\right\} \right]\nonumber \\
 & \cong n^{1/2}N^{-1}\sum_{i=1}^{N}\psi_{A}(V_{i};\mu_{g},\tau_{0}),\label{eq:mu_A_IF}
\end{align}
where $\phi_{A,r}(V;\mu,\tau)=\partial \Phi_{A}(V,\delta_A;\mu,\tau)/\partial\tau$. By Lemma \ref{lemm:uncondition}, the joint asymptotic distribution for $n^{1/2}(\widehat{\mu}_{A}-\mu_{g})$
and $n^{1/2}(\widehat{\mu}_{B}-\mu_{g})$ would be 
\begin{align*}
&n^{1/2}\left(\begin{array}{c}
\widehat{\mu}_{A}-\mu_{g}\\
\widehat{\mu}_{B}-\mu_{g}
\end{array}\right) \rightarrow\\
&\N\left\{ \left(\begin{array}{c}
0_{l\times1}\\
-f_{B}^{-1/2} \left[\E\left\{\partial \Phi_{B}(V_{i};\mu_{g,0},\tau_0)/\partial\mu\right\}\right]^{-1}\eta
\end{array}\right),\left(\begin{array}{cc}
V_{A} & \Gamma\\
\Gamma^{\T} & V_{B}
\end{array}\right)\right\},
\end{align*}
where $V_A,\Gamma$ and $V_B$ are the total (co-)variance of two-phase design averaging over the finite populations:
\begin{alignat*}{2}
    V_A &=
&&nN^{-2}\E_\zeta 
\left[
\var_{\mathrm{p}}
\left\{
\sum_{i=1}^N 
\psi_A(V_i;\mu_g,\tau_0)
\mid \F_N
\right\}
\right] \\
&&&+
nN^{-2}\var_\zeta 
\left[
\E_{\mathrm{p}}
\left\{
\sum_{i=1}^N 
\psi_A(V_i;\mu_g,\tau_0)
\mid \F_N
\right\}
\right],\\
 V_B &=
&&nN^{-2}\E_\zeta 
\left[
\var_{\mathrm{p}\text{-}\mathrm{np}}
\left\{
\sum_{i=1}^N 
\psi_B(V_i;\mu_g,\tau_0)
\mid \F_N
\right\}
\right] \\
&&&+
nN^{-2}\var_\zeta 
\left[
\E_{\mathrm{p}\text{-}\mathrm{np}}
\left\{
\sum_{i=1}^N 
\psi_B(V_i;\mu_g,\tau_0)
\mid \F_N
\right\}
\right],\\
\Gamma &=&&
nN^{-2}\E_\zeta 
\left[
\cov_{\mathrm{p}\text{-}\mathrm{np}}
\left\{
\sum_{i=1}^N 
\psi_A(V_i;\mu_g,\tau_0),
\sum_{i=1}^N 
\psi_B(V_i;\mu_g,\tau_0)
\mid \F_N
\right\}
\right] \\ 
&&&+
nN^{-2}\var_\zeta 
\left[
\E_{\mathrm{p}}
\left\{
\sum_{i=1}^N 
\psi_A(V_i;\mu_g,\tau_0)\mid \F_N
\right\},
\E_{\mathrm{p}\text{-}\mathrm{np}}
\left\{\sum_{i=1}^N 
\psi_B(V_i;\mu_g,\tau_0)
\mid \F_N
\right\}
\right],
\end{alignat*}
where the first term is attributed to the randomness of probability (and non-probability) sample designs, and the second term is attributed to the randomness of the superpopulation model. The rest of the proof is summarized in Lemma \ref{lem:mu_eff}.

\begin{lemma}\label{lem:mu_eff} Under the Assumption \ref{asmp:regularity} and
the asymptotic joint distribution for $\widehat{\mu}_{A}$ and $\widehat{\mu}_{B}$
in Lemma \ref{lemma:mu_A_B_Hn}, the form of $\widehat{\mu}_{\eff}$ which
maximizes the variance reduction under $H_{0}$ would be
\begin{align*}
n^{1/2}(\widehat{\mu}_{{\eff}}-\mu_{0}) & \cong n^{1/2}\{\omega_{A}(\Lambda_\eff)(\widehat{\mu}_{A}-\mu_{g})+\text{\ensuremath{\omega_{B}}}(\Lambda_\eff)(\widehat{\mu}_{B}-\mu_{g})\},
\end{align*}
where the weight functions are
\begin{align}
\omega_{A}(\Lambda) & = \E\left\{\dot{\Phi}_{A,B,n}(\Lambda,\mu_{g,0},\tau_0)\right\}^{-1}\E\left\{ \dot{\Phi}_{A}(V_{i},\delta_{A,i};\mu_{g,0},\tau_{0})\right\}\label{eq:W_A_func} ,\\
\omega_{B}(\Lambda) & =\E\left\{\dot{\Phi}_{A,B,n}(\Lambda,\mu_{g,0},\tau_0)\right\}^{-1}\Lambda
\E\left\{\dot{\Phi}_{B}(V_{i},\delta_{A,i},\delta_{B,i};\mu_{g,0},\tau_{0})\right\}\label{eq:W_B_func},
\end{align}
where $\dot{\Phi}_{A,B,n}(\Lambda,\mu_{g,0},\tau_0) = 
\dot{\Phi}_{A}(V_{i},\delta_{A,i};\mu_{g,0},\tau_{0}) +\Lambda 
\dot{\Phi}_{B}(V_{i},\delta_{A,i},\delta_{B,i};\mu_{g,0},\tau_{0})$. The most efficient estimator $\widehat{\mu}_\eff$ with
\begin{equation}
    \Lambda_{\eff}=\E\left\{\dot{\Phi}_{A}(V_{i};\mu_{g,0},\tau_{0})\right\}(V_{A}-\Gamma)(V_{B}-\Gamma^{\T})^{-1}\E\left\{\dot{\Phi}_{B}(V_{i};\mu_{g,0},\tau_{0})\right\}^{-1}
    \label{eq:lambda_eff}
\end{equation}
has the asymptotic distribution under $H_{a,n}$ as 
\begin{align*}
n^{1/2}(\widehat{\mu}_{{\eff}}-\mu_{g}) & \rightarrow\N\{b_{{\rm eff}}(\eta),V_{\eff}\},
\end{align*}
where $b_{{\rm eff}}(\eta)=-f_{B}^{-1/2}\omega_{B}(\Lambda_\eff)\left\{ \E\partial \Phi_{B}(\mu_{g,0},\tau_0)/\partial\mu\right\} ^{-1}\eta$
and 
\[
V_{{\rm eff}}=\left(\begin{array}{c}
\omega_{A}^{\T}(\Lambda_\eff)\\
\omega_{B}^{\T}(\Lambda_\eff)
\end{array}\right)^{\T}\left(\begin{array}{cc}
V_{A} & \Gamma\\
\Gamma^{\T} & V_{B}
\end{array}\right)\left(\begin{array}{c}
\omega_{A}(\Lambda_\eff)\\
\omega_{B}(\Lambda_\eff)
\end{array}\right).
\]
When $\mu_{A}$ and $\mu_{B}$ are both scalar, $V_{\eff}$ would reduce
to 
\begin{align*}
V_{\eff} & =(V_{A}V_{B}-\Gamma^{2})(V_{A}+V_{B}-2\Gamma)^{-1}=V_{A}-V_{\Delta},
\end{align*}
where $V_{\Delta}=(V_{A}-\Gamma)^{2}(V_{A}+V_{B}-2\Gamma)^{-1}$.
\end{lemma}

\subsection{Proof of Lemma \ref{lemma:test}}
By applying the Taylor expansion with Lagrange forms of remainder
to the asymptotic distribution for $n_{B}^{1/2}N^{-1}\sum_{i=1}^{N}\Phi_{B}(V_{i},\delta_{A,i},\delta_{B,i};\widehat{\mu}_{A},\widehat{\tau})$
in (\ref{eq:testT}) could be shown as 
\begin{align*}
 & n_{B}^{1/2}N^{-1}\sum_{i=1}^{N}\Phi_{B}(V_{i},\delta_{A,i},\delta_{B,i};\widehat{\mu}_{A},\widehat{\tau})=n_{B}^{1/2}N^{-1}\sum_{i=1}^{N}\Phi_{B}(V_{i},\delta_{A,i},\delta_{B,i};\mu_{g},\tau_{0})\\
 & +n_{B}^{1/2}N^{-1}\sum_{i=1}^{N}\left(\begin{array}{cc}
\frac{\partial \Phi_{B}(V_{i},\delta_{A,i},\delta_{B,i};\widehat{\mu}_{A}^*,\tau^*)}{\partial\mu} & \frac{\partial \Phi_{B}(V_{i},\delta_{A,i},\delta_{B,i};\widehat{\mu}_{A}^*,\tau^*)}{\partial\tau}\end{array}\right)\left(\begin{array}{c}
\widehat{\mu}_{A}-\mu_{g}\\
\widehat{\tau}-\tau_{0}
\end{array}\right)
\end{align*}
where $(\widehat{\mu}_{A}^*\quad \tau^*)^{\intercal}$ is the neighborhood of $(\mu_{g,0},\tau_{0})^{\intercal}$ as $\plim\widehat{\mu}_A = \mu_{g,0}$ and $\plim\widehat{\tau}=\tau_0$. Under the Assumption \ref{asmp:regularity} e), we have
\begin{align}
&n_{B}^{1/2}N^{-1}\sum_{i=1}^{N}\Phi_{B} (V_{i},\delta_{A,i},\delta_{B,i};\widehat{\mu}_{A},\widehat{\tau})\nonumber \\
&=n_{B}^{1/2}N^{-1}\sum_{i=1}^{N}\Phi_{B}(V_{i},\delta_{A,i},\delta_{B,i};\mu_{g},\tau_{0})\\
&+n_{B}^{1/2}N^{-1}\sum_{i=1}^{N}\frac{\partial \Phi_{B,j}(V_{i},\delta_{A,i},\delta_{B,i};\widehat{\mu}_{A}^*,\tau^*)}{\partial\mu}(\widehat{\mu}_{A}-\mu_{g})\nonumber \\
 & +n_{B}^{1/2}N^{-1}\sum_{i=1}^{N}\frac{\partial \Phi_{B}(V_{i},\delta_{A,i},\delta_{B,i};\widehat{\mu}_{A}^*,\tau^*)}{\partial\tau}(\widehat{\tau}-\tau_{0})\nonumber \\
 & =n_{B}^{1/2}N^{-1}\sum_{i=1}^{N}\Phi_{B}(V_{i},\delta_{A,i},\delta_{B,i};\mu_{g},\tau_{0})\nonumber\\
 &+\E\left\{
 \frac{\partial \Phi_B(V,\delta_{A},\delta_{B};\mu_{g,0},\tau_0)}{\partial \tau}
 \right\}
 n_{B}^{1/2}(\widehat{\tau}-\tau_{0})
\label{eq:test_SM}\\
 & +\E\left\{
 \frac{\partial \Phi_B(V,\delta_{A},\delta_{B};\mu_{g,0},\tau_0)}{\partial \mu}
 \right\}
 n_{B}^{1/2}
 (\widehat{\mu}_{A}-\mu_{g})+o_{\zeta\text{-}\mathrm{p}\text{-}\mathrm{np}}(1). \nonumber
\end{align}
Next, by replacing the first two term in Equation (\ref{eq:test_SM}) with
Equation (\ref{eq:mu_B-mu_0}), we have 
\begin{align*}
&n_{B}^{1/2}N^{-1}\sum_{i=1}^{N}\Phi_{B}(V_{i},\delta_{A,i},\delta_{B,i};\widehat{\mu}_{A},\widehat{\tau})=-\E\left\{
 \frac{\partial \Phi_B(V,\delta_{A},\delta_{B};\mu_{g,0},\tau_0)}{\partial \mu}
 \right\}
 n_{B}^{1/2}(\widehat{\mu}_{B}-\mu_{g})\\
 & +
 \E\left\{
 \frac{\partial \Phi_B(V,\delta_{A},\delta_{B};\mu_{g,0},\tau_0)}{\partial \mu}
 \right\}n_{B}^{1/2}(\widehat{\mu}_{A}-\mu_{g})+o_{\zeta\text{-}\mathrm{p}\text{-}\mathrm{np}}(1)\\
 & =-(n_{B}/n)^{1/2}\cdot \E\dot{\Phi}_{B}(V_{i};\mu_{g,0},\tau_{0})\cdot  n^{1/2}(\widehat{\mu}_{B}-\mu_{g})\\
 & +(n_{B}/n)^{1/2}\cdot \E\dot{\Phi}_{B}(V_{i};\mu_{g,0},\tau_{0}) \cdot n^{1/2}(\widehat{\mu}_{A}-\mu_{g})+o_{\zeta\text{-}\mathrm{p}\text{-}\mathrm{np}}(1),
\end{align*}
provided by WLLN under Assumptions 
\ref{asmp:Sampling-design}, \ref{asmp:MAR} (iii) and Assumption \ref{asmp:regularity}. By the joint distribution of $\widehat{\mu}_{A}$ and $\widehat{\mu}_{B}$
in Lemma \ref{lemma:mu_A_B_Hn}, the variance of $\Phi_{B,n}(V_{i},\delta_{A,i},\delta_{B,i};\widehat{\mu}_{A},\widehat{\tau})$
would be 
\[
\Sigma_{T}=f_{B}\left\{ \E\dot{\Phi}_{B}(V_{i};\mu_{g,0},\tau)\right\} \left(V_{A}+V_{B}-\Gamma^{\intercal}-\Gamma\right)\left\{ \E\dot{\Phi}_{B}(V_{i};\mu_{g,0},\tau)\right\} ^{\intercal}.
\]
Thus, the asymptotic distribution for $\Phi_{B,n}(V_{i},\delta_{A,i},\delta_{B,i};\widehat{\mu}_{A},\widehat{\tau})$
would be
$$
\begin{aligned}
    &n_{B}^{1/2}N^{-1}\sum_{i=1}^{N}\Phi_{B}(V_{i},\delta_{A,i},\delta_{B,i};\widehat{\mu}_{A},\widehat{\tau})\\
    &{\rightarrow}\N\left\{ \eta,f_{B}\left\{ \E\dot{\Phi}_{B}(V_{i};\mu_{g,0},\tau)\right\} \left(V_{A}+V_{B}-\Gamma^{\intercal}-\Gamma\right)\left\{ \E\dot{\Phi}_{B}(V_{i};\mu_{g,0},\tau)\right\} ^{\intercal}\right\} .
\end{aligned}
$$

\subsection{Proof of Theorem \ref{thm:mu_TAP}}
From Lemma \ref{lemma:mu_A_B_H0} and \ref{lemma:mu_A_B_Hn}, we know
that the asymptotic joint distribution for $\widehat{\mu}_{A}$ and
$\widehat{\mu}_{B}$ would be 
\begin{align*}
&n^{1/2}\left(\begin{array}{c}
\widehat{\mu}_{A}-\mu_{g}\\
\widehat{\mu}_{B}-\mu_{g}
\end{array}\right) \\
&{\rightarrow}\N\left\{ \left(\begin{array}{c}
\boldsymbol{0}_{l\times1}\\
-f_{B}^{-1/2}\left[\E\left\{ \partial \Phi_{B}(\mu_{g,0},\tau_{0})/\partial\mu\right\}\right]^{-1}\eta
\end{array}\right),\left(\begin{array}{cc}
V_{A} & \Gamma\\
\Gamma^{\intercal} & V_{B}
\end{array}\right)\right\} .
\end{align*}
For simplicity, we let $n^{1/2}(\widehat{\mu}_{A}-\mu_{g})$ and $n^{1/2}(\widehat{\mu}_{B}-\mu_{g})$ be asymptotically distributed as $Z_{1}$ and $Z_{2}$, respectively. Then, $n_{B}^{1/2}N^{-1}\sum_{i=1}^{N}\Phi_{B}(V_{i},\delta_{A,i},\delta_{B,i};\widehat{\mu}_{A},\widehat{\tau})$
could be expressed as 
\begin{alignat*}{2}
&&&n_{B}^{1/2}N^{-1}\sum_{i=1}^{N}\Phi_{B}(V_{i},\delta_{A,i},\delta_{B,i};\widehat{\mu}_{A},\widehat{\tau}) 
\\
&\cong&& -n_{B}^{1/2}N^{-1}\sum_{i=1}^{N}
\dot{\Phi}_B(V_i,\delta_{A,i},\delta_{B,i};\mu_{g,0},\tau_0)
(\widehat{\mu}_{B}-\mu_{g})\\
 & &&+n_{B}^{1/2}N^{-1}\sum_{i=1}^{N}\dot{\Phi}_B(V_i,\delta_{A,i},\delta_{B,i};\mu_{g,0},\tau_0)
 (\widehat{\mu}_{A}-\mu_{g})\\
 &{\rightarrow} && f_{B}^{1/2}\left\{ \E\dot{\Phi}_B(V,\delta_{A},\delta_{B};\mu_{g,0},\tau_0)\right\} (Z_{1}-Z_{2}).
\end{alignat*}
Let $U_{2}=f_{B}^{1/2}\left\{ \E\dot{\Phi}_B(V,\delta_{A},\delta_{B};\mu_{g,0},\tau_0)\right\} (Z_{1}-Z_{2})$.
Next step, we attempt to find another linear combination of $Z_{1}$
and $Z_{2}$ which is orthogonal to $U_{2}$. Observed that when $U_{1}=f_{B}^{1/2}\{(\Gamma^{\intercal}-V_{B})(\Gamma-V_{A})^{-1}Z_{1}+Z_{2}\}$,
it is easy to verify that the covariance of $U_{1}$ and $U_{2}$
is zero under $H_{0}$. 
\begin{align*}
\text{cov}(U_{2},U_{1}) & =f_{B}\left\{ \E\dot{\Phi}_B(V,\delta_{A},\delta_{B};\mu_{g,0},\tau_0)\right\} \\
&\times \left(\begin{array}{c}
I_{l\times l}\\
-I_{l\times l}
\end{array}\right)^{\intercal}\times\left(\begin{array}{cc}
V_{A} & \Gamma\\
\Gamma^{\intercal} & V_{B}
\end{array}\right)\times\left(\begin{array}{c}
(\Gamma^{\intercal}-V_{A})^{-1}(\Gamma-V_{B})\\
I_{l\times l}
\end{array}\right)\\
 & =f_{B}\left\{ \E\dot{\Phi}_B(V,\delta_{A},\delta_{B};\mu_{g,0},\tau_0)\right\}\\
 &\times (\begin{array}{cc}
V_{A}-\Gamma^{\intercal} & \Gamma-V_{B}\end{array})\times\left(\begin{array}{c}
(\Gamma^{\intercal}-V_{A})^{-1}(\Gamma-V_{B})\\
I_{l\times l}
\end{array}\right)\\
 & ={0}_{l\times l}.
\end{align*}
Also, since $U_{1}$and $U_{2}$ are both asymptotically normal distributions,
which implies that zero covariance leads to independency. After a
few standardization procedures, we have $W_{1}$ and $W_{2}$ as $W_{1}=\Sigma_{S}^{-1/2}U_{1}$,$W_{2}=\Sigma_{T}^{-1/2}U_{2}$ with $\Sigma_{S}$ and $\Sigma_{T}$ defined as 
\begin{align}
\Sigma_{S} & =\var(U_{1})=f_{B}\var\{(\Gamma^{\intercal}-V_{B})(\Gamma-V_{A})^{-1}Z_{1}+Z_{2}\},\label{eq:sigma_S}\\
\Sigma_{T} & =f_{B}\left\{ \E\dot{\Phi}_B(\mu_{g,0},\tau_0)\right\}\left(V_{A}+V_{B}-\Gamma^{\intercal}-\Gamma\right)\left\{ \E\dot{\Phi}_B(\mu_{g,0},\tau_0)\right\}^{\intercal}.\label{eq:sigma_T} 
\end{align}
Therefore, we have the form
for the standardized random variables $W_{1}$ and $W_{2}$ as 
\begin{align*}
W_{1} & =\Sigma_{S}^{-1/2}U_{1}=f_{B}^{1/2}\Sigma_{S}^{-1/2}\{(\Gamma^{\intercal}-V_{B})(\Gamma-V_{A})^{-1}Z_{1}+Z_{2}\},\\
W_{2} & =-\Sigma_{T}^{-1/2}U_{2}=-(V_{A}+V_{B}-\Gamma^{\intercal}-\Gamma)^{-1/2}(Z_{1}-Z_{2}).
\end{align*}
Here we use $-\Sigma_{T}^{-1/2}$ to standardize $U_{2}$ for the
sake of convenience later. Therefore, under the local alternative
$H_{a,n}:\E\{\Phi_{B}(V,\delta_A,\delta_B;\mu_{g,0},\tau_0)\}=n_{B}^{-1/2}\eta$, we have that
$\E(Z_{1})=0,\E(Z_{2})=-f_{B}^{-1/2}\left\{ \E\dot{\Phi}_B(\mu_{g,0},\tau_0)\right\} ^{-1}\eta$.
Combining the above leads to 
\begin{align*}
W_{1} & \sim N(\mu_{1},I_{l\times l}),~W_{2}\sim N(\mu_{2},I_{l\times l}),
\end{align*}
where
\begin{align*}
\mu_{1} & =\E(W_{1})=-\Sigma_{S}^{-1/2}\left\{ \E\dot{\Phi}_B(\mu_{g,0},\tau_0)\right\} ^{-1}\eta,\\
\mu_{2} & =\E(W_{2})=-f_{B}^{-1/2}(V_{A}+V_{B}-\Gamma^{\intercal}-\Gamma)^{-1/2}\left\{ \E\dot{\Phi}_B(\mu_{g,0},\tau_0)\right\}^{-1}\eta=-\Sigma_{T}^{-1/2}\eta,
\end{align*}
and since $W_{1}\perp W_{2}$, we could project out TAP estimator
$\widehat{\mu}_{\tap}$ with the optimal tuning parameter $(\Lambda^{*},c_{\gamma^{*}})$
onto these two basis respectively. First, on the condition that 
\[
T>c_{\gamma^{*}}=\left\{ {\Phi}_{B,n}(\widehat{\mu}_A,\widehat{\tau})\right\} ^{\intercal}\widehat{\Sigma}_{T}^{-1}\left\{ {\Phi}_{B,n}(\widehat{\mu}_A,\widehat{\tau})\right\}>c_{\gamma^{*}}
{\rightarrow}
W_{2}^{\intercal}W_{2}>c_{\gamma^{*}},
\]
we have
\begin{align*}
&n^{1/2}(\widehat{\mu}_{\tap}  -\mu_{g})\mid T>c_{\gamma^{*}}=n^{1/2}(\widehat{\mu}_{A}-\mu_{g})\mid T>c_{\gamma^{*}}\\
 & {\rightarrow} Z_{1}|W_{2}^{\intercal}W_{2}>c_{\gamma^{*}}\\
 & {\rightarrow} -f_{B}^{-1/2}(\Gamma-V_{A})(V_{A}+V_{B}-\Gamma^{\intercal}-\Gamma)^{-1}U_{1}\\
 & +f_{B}^{-1/2}(\Gamma-V_{A})(V_{A}+V_{B}-\Gamma^{\intercal}-\Gamma)^{-1}\left\{ \E\dot{\Phi}_{B}(\mu_{g,0},\tau_0)\right\} ^{-1}U_{2}|W_{2}^{\intercal}W_{2}>c_{\gamma^{*}}\\
 & {\rightarrow} -f_{B}^{-1/2}(V_{A}+V_{B}-\Gamma^{\intercal}-\Gamma)^{-1}\Sigma_{S}^{1/2}W_{1}\\
 & +(\Gamma-V_{A})(V_{A}+V_{B}-\Gamma-\Gamma^{\intercal})^{-1/2}W_{2}|W_{2}^{\intercal}W_{2}>c_{\gamma^{*}}\\
 & {\rightarrow} -V_{\eff}^{1/2}W_{1}+V_{\text{A-eff}}^{1/2}W_{2}|W_{2}^{\intercal}W_{2}>c_{\gamma^{*}}.
\end{align*}
Next, on the condition $T=W_{2}^{\intercal}W_{2}\leq c_{\gamma^{*}},$
we have 
\begin{align*}
&n^{1/2}(\widehat{\mu}_{\tap}-\mu_{g})  {\rightarrow}
\omega_{A}^{*}Z_{1}+\omega_{B}^{*}Z_{2}|W_{2}^{\intercal}W_{2}\leq c_{\gamma^{*}}\\
 & {\rightarrow}
 -f_{B}^{-1/2}(\Gamma-V_{A})(V_{A}+V_{B}-\Gamma^{\intercal}-\Gamma)^{-1}U_{1}\\
 & +f_{B}^{-1/2}\omega_{A}^{*}(\Gamma-V_{A})(V_{A}+V_{B}-\Gamma^{\intercal}-\Gamma)^{-1}\left\{ \E\dot{\Phi}_{B}(\mu_{g,0},\tau_0)\right\}^{-1}U_{2}\mid W_{2}^{\intercal}W_{2}\le c_{\gamma^{*}}\\
 & -f_{B}^{-1/2}\omega_{B}^{*}(\Gamma^{\intercal}-V_{B})(V_{A}+V_{B}-\Gamma^{\intercal}-\Gamma)^{-1}\left\{ \E\dot{\Phi}_{B}(\mu_{g,0},\tau_0)\right\}^{-1}U_{2}\mid W_{2}^{\intercal}W_{2}\le c_{\gamma^{*}}\\
 & {\rightarrow} -f_{B}^{-1/2}(V_{A}+V_{B}-\Gamma^{\intercal}-\Gamma)^{-1}\Sigma_{S}^{1/2}W_{1}\\
 & +f_{B}^{-1/2}\omega_{A}^{*}(\Gamma-V_{A})(V_{A}+V_{B}-\Gamma-\Gamma^{\intercal})^{-1/2}W_{2}\mid W_{2}^{\intercal}W_{2}\le c_{\gamma^{*}}\\
 & -f_{B}^{-1/2}\omega_{B}^{*}(\Gamma^{\intercal}-V_{B})(V_{A}+V_{B}-\Gamma^{\intercal}-\Gamma)^{-1/2}W_{2}\mid W_{2}^{\intercal}W_{2}\leq c_{\gamma^{*}}\\
 & {\rightarrow} -V_{\eff}^{1/2}W_{1}+(\omega_{A}V_{\text{A-eff}}^{1/2}-\omega_{B}V_{\text{B-eff}}^{1/2})W_{2}|W_{2}^{\intercal}W_{2}\leq c_{\gamma^{*}},
\end{align*}
where $W_{2}^{t}=W_{2}|W_{2}^{\intercal}W_{2}\leq c_{\gamma}$, and
$\omega_{A}^{*},\omega_{B}^{*}$ are the new tuned weighted functions
defined in (\ref{eq:W_A_func}) and (\ref{eq:W_B_func}) with $\Lambda=\Lambda^{*}$. In this way, we could fully characterize
the asymptotic distribution for the TAP estimator $\widehat{\mu}_{\tap}$ under
the optimal tuning parameter as, 
$$
n^{1/2}({\widehat{\mu}_{\tap}}-\mu_{g})
{\rightarrow}
\begin{cases}
-V_{\eff}^{1/2}W_{1}+(\omega_{A}V_{\text{A-eff}}^{1/2}-\omega_{B}V_{\text{B-eff}}^{1/2})W_{[0,c_{\gamma}]}^{t} & w.p.\ \xi,\\
-V_{\eff}^{1/2}W_{1}+V_{\text{A-eff}}^{1/2}W_{[c_{\gamma^{*}},\infty]}^{t} & w.p.\ 1-\xi,
\end{cases}
$$
where $\xi=\pr(W_{2}^{\intercal}W_{2}<c_{\gamma^{*}})$.

\subsection{Proof of the bias and mean squared error of {$n^{1/2}(\widehat{\mu}_{\tap}-\mu_g)$}{the test-and-pool estimator}}
For general case, given $W_{2}\sim N_{p}(\mu_{2},I_{p\times p})$, the
MGF of truncated normal distribution $W_{2}|a\leq W_{2}^{\intercal}W_{2}\leq b$ is \citep{tallis1963elliptical} 
\begin{align*}
    \alpha m(t)	&=\E\{\exp(t^{\intercal}W_{2})\}\\
	&=(2\pi)^{-p/2}\int_{\mathbb{C}}\exp(t^{\intercal}W_{2})\exp\left\{ -\frac{1}{2}(W_{2}-\mu_{2})^{\intercal}(W_{2}-\mu_{2})\right\} dW_{2}\\
	&=(2\pi)^{-p/2}\exp(\frac{1}{2}t^{\intercal}t+\mu_{2}^{\intercal}t)\int_{\mathbb{C}}\exp\left\{ -\frac{1}{2}(W_{2}-\mu_{2}-t)^{\intercal}(W_{2}-\mu_{2}-t)\right\} dW_{2}\\
	&=\exp(-\frac{1}{2}\mu_{2}^{\intercal}\mu_{2})\sum_{k=0}^{\infty}\{F_{p+2k}(b)-F_{p+2k}(a)\}\{(\mu_{2}+t)^{\intercal}(\mu_{2}+t)/2\}^{k}/k!,
\end{align*}
where $\alpha=F_{p}(b;\mu_{2}^{\intercal}\mu_{2}/2)-F_{p}(a;\mu_{2}^{\intercal}\mu_{2}/2)$
is the normalization constant and $F_{p}(a;\mu_{2}^{\intercal}\mu_{2}/2)$
is CDF of chi-square distribution at value $a$ with non-central parameter
$\mu_{2}^{\intercal}\mu_{2}/2$. The second and the third equality above are justified by
\begin{align*}
	    &(2\pi)^{-p/2}\int_{\mathbb{C}}\exp\left\{ -\frac{1}{2}(W_{2}-\mu_{2}-t)^{\intercal}(W_{2}-\mu_{2}-t)\right\} \\
	    &=\pr\{a\leq W_{2}^{\intercal}W_{2}\leq b\mid W_{2}\sim\mathcal{N}(\mu_{2}+t,I_{p\times p})\}
	\\
	&=F\{b;k=p,\lambda=(\mu_{2}+t)^{\intercal}(\mu_{2}+t)\}-F\{a;k=p,\lambda=(\mu_{2}+t)^{\intercal}(\mu_{2}+t)\}
	\\
	&=\exp\{-\frac{1}{2}(\mu_{2}+t)^{\intercal}(\mu_{2}+t)\}\sum_{k=0}^{\infty}\{F_{p+2k}(b)-F_{p+2k}(a)\}\{(\mu_{2}+t)^{\intercal}(\mu_{2}+t)/2\}^{k}/k!.
	\end{align*} 
To compute the first and second moment
of this truncated normal distribution, we take derivative of the MGF
and evaluate the function at $t=0$ 
\begin{align*}
\alpha\frac{dm(t)}{dt^{\intercal}}\bigg|_{t=0} & =(\mu_{2}+t)\exp(-\frac{1}{2}\mu_{2}^{\intercal}\mu_{2})\\
&\times\sum_{k=0}^{\infty}\{F_{p+2k+2}(b)-F_{p+2k+2}(a)\}\{(\mu_{2}+t)^{\intercal}(\mu_{2}+t)\}/k!|_{t=0}\\
 & =\mu_{2}\exp(-\frac{1}{2}\mu_{2}^{\intercal}\mu_{2})\sum_{k=0}^{\infty}\{F_{p+2k+2}(b)-F_{p+2k+2}(a)\}\{\mu_{2}^{\intercal}\mu_{2}/2\}^{k}/k!\\
 & =\mu_{2}\left\{ F_{p+2}(b;\mu_{2}^{\intercal}\mu_{2}/2)-F_{p+2}(a;\mu_{2}^{\intercal}\mu_{2}/2)\right\} .
\end{align*}
By the nature of MGF, we obtain the expectation of the first moment
of $W_{2}$ 
\[
\E(W_{2}|a\leq W_{2}^{\intercal}W_{2}\leq b)=\mu_{2}\cdot\frac{F_{p+2}(b;\mu_{2}^{\intercal}\mu_{2}/2)-F_{p+2}(a;\mu_{2}^{\intercal}\mu_{2}/2)}{F_{p}(b;\mu_{2}^{\intercal}\mu_{2}/2)-F_{p}(a;\mu_{2}^{\intercal}\mu_{2}/2)}.
\]
Then, taking the second derivative of the MGF follows by 
\begin{align*}
 &\alpha\frac{d^{2}m(t)}{dtdt^{\intercal}} \bigg|_{t=0}\\
 &=\exp(-\frac{1}{2}\mu_{2}^{\intercal}\mu_{2})\left[\sum_{k=0}^{\infty}\{F_{p+2k+2}(b)-F_{p+2k+2}(a)\}\{(\mu_{2}+t)^{\intercal}(\mu_{2}+t)/2\}^{k}/k!|_{t=0}\right.\\
 & \left.+(\mu_{2}+t)(\mu_{2}+t)^{\intercal}\sum_{k=0}^{\infty}\{F_{p+2k+4}(b)-F_{p+2k+4}(a)\}\{(\mu_{2}+t)^{\intercal}(\mu_{2}+t)/2\}^{k}/k!|_{t=0}\right]\\
 & =\exp(-\frac{1}{2}\mu_{2}^{\intercal}\mu_{2})\left[\sum_{k=0}^{\infty}\{F_{p+2k+2}(b)-F_{p+2k+2}(a)\}\{\mu_{2}^{\intercal}\mu_{2}/2\}^{k}/k!\right.\\
 & \left.+\mu_{2}\mu_{2}^{\intercal}\sum_{k=0}^{\infty}\{F_{p+2k+4}(b)-F_{p+2k+4}(a)\}\{\mu_{2}^{\intercal}\mu_{2}/2\}^{k}/k!\right]\\
 & =I_{p\times p}(F_{p+2}(b;\mu_{2}^{\intercal}\mu_{2}/2)-F_{p+2}(a;\mu_{2}^{\intercal}\mu_{2}/2))\\
 &+\mu_{2}\mu_{2}^{\intercal}(F_{p+4}(b;\mu_{2}^{\intercal}\mu_{2}/2)-F_{p+4}(a;\mu_{2}^{\intercal}\mu_{2}/2)),
\end{align*}
which leads to 
\begin{align*}
\E(W_{2}W_{2}^{T}|a\leq W_{2}^{T}W_{2}\leq b) & =I_{p\times p}\frac{F_{p+2}(b;\mu_{2}^{T}\mu_{2}/2)-F_{p+2}(a;\mu_{2}^{T}\mu_{2}/2)}{F_{p}(b;\mu_{2}^{T}\mu_{2}/2)-F_{p}(a;\mu_{2}^{T}\mu_{2}/2)}\\
 & +\mu_{2}\mu_{2}^{\intercal}\frac{F_{p+4}(b;\mu_{2}^{T}\mu_{2}/2)-F_{p+4}(a;\mu_{2}^{T}\mu_{2}/2)}{F_{p}(b;\mu_{2}^{T}\mu_{2}/2)-F_{p}(a;\mu_{2}^{T}\mu_{2}/2)}.
\end{align*}
In our case, 
\[
p=l,\quad \mu_{1}=-\Sigma_{S}^{-1/2}\left[\E\left\{ \partial \Phi_{B}(\mu_{g,0},\tau_{0})/\partial\mu\right\} \right]^{-1}\eta, \quad \mu_{2}=-\Sigma_{T}^{-1/2}\eta.
\]
Recall, for $T\leq c_{\gamma}$, we have $n^{1/2}(\widehat{\mu}_{\tap}-\mu_{g})
{\rightarrow}
-V_{\eff}^{1/2}W_{1}+(\omega_{A}V_{\text{A-eff}}^{1/2}-\omega_{B}V_{\text{B-eff}}^{1/2})W_{2}\allowbreak|W_{2}^{\intercal}W_{2}\leq c_{\gamma}$
with probability $\xi=F_{l}(c_{\gamma};\mu_{2}^{\intercal}\mu_{2})$
, the bias would be 
\begin{align*}
\text{bias}(\lambda,c_{\gamma};\eta)_{T\leq c_{\gamma}} & =-V_{\eff}^{1/2}\mu_{1}+(\omega_{A}V_{\text{A-eff}}^{1/2}-\omega_{B}V_{\text{B-eff}}^{1/2})\cdot\E(W_{2}|W_{2}^{\intercal}W_{2}\leq c_{\gamma})\\
 & =-V_{\eff}^{1/2}\mu_{1}+(\omega_{A}V_{\text{A-eff}}^{1/2}-\omega_{B}V_{\text{B-eff}}^{1/2})\cdot\frac{F_{l+2}(c_{\gamma};\mu_{2}^{T}\mu_{2}/2)\mu_{2}}{F_{l}(c_{\gamma};\mu_{2}^{T}\mu_{2}/2)}.
\end{align*}
The MSE can be derived based on the known formula $\text{mse}(X+Y)=\text{var}(X+Y)+\{\E(X+Y)\}^{\otimes2}=\{\text{var}(X)+\mu_{X}^{\otimes2}\}+\{\text{var}(Y)+\mu_{Y}^{\otimes2}\}+2\mu_{X}\mu_{Y}^{\intercal}$
\begin{align*}
&\text{mse}(\lambda,c_{\gamma};\eta)_{T\leq c_{\gamma}}  =V_{\eff}^{1/2}(\mu_{1}\mu_{1}^{\intercal}+I_{l\times l})V_{\eff}^{1/2}+(\omega_{A}V_{\text{A-eff}}^{1/2}-\omega_{B}V_{\text{B-eff}}^{1/2})\\
 & \times E(W_{2}W_{2}^{T}|W_{2}^{\intercal}W_{2}\leq c_{\gamma})(\omega_{A}V_{\text{A-eff}}^{1/2}-\omega_{B}V_{\text{B-eff}}^{1/2})\\
 & -2V_{\eff}^{1/2}\mu_{1}E(W_{2}^{\intercal}|W_{2}^{\intercal}W_{2}\leq c_{\gamma})(\omega_{A}V_{\text{A-eff}}^{1/2}-\omega_{B}V_{\text{B-eff}}^{1/2})\\
 & =V_{\eff}^{1/2}(\mu_{1}\mu_{1}^{\intercal}+I_{l\times l})V_{\eff}^{1/2}+(\omega_{A}V_{\text{A-eff}}^{1/2}-\omega_{B}V_{\text{B-eff}}^{1/2})\\
 & \times\left\{ \frac{F_{l+2}(c_{\gamma};\mu_{2}^{T}\mu_{2}/2)}{F_{l}(c_{\gamma};\mu_{2}^{T}\mu_{2}/2)}I_{l\times l}+\frac{F_{l+4}(c_{\gamma};\mu_{2}^{T}\mu_{2}/2)}{F_{l}(c_{\gamma};\mu_{2}^{T}\mu_{2}/2)}\mu_{2}\mu_{2}^{\intercal}\right\} (\omega_{A}V_{\text{A-eff}}^{1/2}-\omega_{B}V_{\text{B-eff}}^{1/2})\\
 & -\frac{2F_{l+2}(c_{\gamma};\mu_{2}^{T}\mu_{2}/2)}{F_{l}(c_{\gamma};\mu_{2}^{T}\mu_{2}/2)}V_{\eff}^{1/2}\mu_{1}\mu_{2}^{\intercal}(\omega_{A}V_{\text{A-eff}}^{1/2}-\omega_{B}V_{\text{B-eff}}^{1/2}).
\end{align*}
For $T>c_{\gamma},$we have $n^{1/2}(\widehat{\mu}_{\tap}-\mu_{g})
{\rightarrow}
-V_{\eff}^{1/2}W_{1}+V_{\text{A-eff}}^{1/2}W_{2}|W_{2}^{\intercal}W_{2}>c_{\gamma}$
with probability $1-\xi=1-F_{l}(c_{\gamma};\mu_{2}^{\intercal}\mu_{2})$,
the corresponding bias and MSE would be 
\begin{align*}
\text{bias}(\lambda,c_{\gamma};\eta)_{T>c_{\gamma}} & =-V_{\eff}^{1/2}\mu_{1}+V_{\text{A-eff}}^{1/2}\cdot E(W_{2}|W_{2}^{\intercal}W_{2}>c_{\gamma})\\
 & =-V_{\eff}^{1/2}\mu_{1}+V_{\text{A-eff}}^{1/2}\cdot\frac{1-F_{l+2}(c_{\gamma};\mu_{2}^{T}\mu_{2}/2)\mu_{2}}{1-F_{l}(c_{\gamma};\mu_{2}^{T}\mu_{2}/2)},
\end{align*}
and
\begin{align*}
&\text{mse}(\lambda,c_{\gamma};\eta)_{T>c_{\gamma}}  =V_{\eff}^{1/2}(\mu_{1}\mu_{1}^{\intercal}+I_{l\times l})V_{\eff}^{1/2}+V_{\text{A-eff}}^{1/2}E(W_{2}W_{2}^{T}|W_{2}^{\intercal}W_{2}>c_{\gamma})V_{\text{A-eff}}^{1/2}\\
 & -2V_{\eff}^{1/2}\mu_{1}E(W_{2}^{\intercal}|W_{2}^{\intercal}W_{2}>c_{\gamma})V_{\text{A-eff}}^{1/2}\\
 & =V_{\eff}^{1/2}(\mu_{1}\mu_{1}^{\intercal}+I_{l\times l})V_{\eff}^{1/2}+V_{\text{A-eff}}^{1/2}\\
 & \times\left\{ \frac{1-F_{l+2}(c_{\gamma};\mu_{2}^{T}\mu_{2}/2)}{1-F_{l}(c_{\gamma};\mu_{2}^{T}\mu_{2}/2)}I_{l\times l}+\frac{1-F_{l+4}(c_{\gamma};\mu_{2}^{T}\mu_{2}/2)}{1-F_{l}(c_{\gamma};\mu_{2}^{T}\mu_{2}/2)}\mu_{2}\mu_{2}^{\intercal}\right\} V_{\text{A-eff}}^{1/2}\\
 & -2\left\{ \frac{1-F_{3}(c_{\gamma};\mu_{2}^{T}\mu_{2}/2)}{1-F_{1}(c_{\gamma};\mu_{2}^{T}\mu_{2}/2)}\right\} V_{\eff}^{1/2}\mu_{1}\mu_{2}^{\intercal}V_{\text{A-eff}}^{1/2}.
\end{align*}
Overall, the bias and mean squared error for $n^{1/2}({\widehat{\mu}_{\tap}}-\mu_{g})$
can be characterized as 
\[
\text{bias}(\lambda,c_{\gamma};\eta)=\xi\cdot\text{bias}(\lambda,c_{\gamma};\eta)_{T\leq c_{\gamma}}+(1-\xi)\cdot\text{bias}(\lambda,c_{\gamma};\eta)_{T>c_{\gamma}},
\]
\[
\text{mse}(\lambda,c_{\gamma};\eta)=\xi\cdot\text{mse}(\lambda,c_{\gamma};\eta)_{T\leq c_{\gamma}}+(1-\xi)\cdot\text{mse}(\lambda,c_{\gamma};\eta)_{T>c_{\gamma}}.
\]

\subsection{Proof of the asymptotic distribution for {${U}(a)$}{U(a)}\label{subsec:asym_u_a}}

Throughout the proof, we assume that the regularity conditions in
Lemma \ref{lemma:mu_A_B_H0} and assumptions in Theorem \ref{thm:ACI} hold, we prove
that the coverage probability for the adaptive projection sets is
guaranteed to be larger than $1-\alpha$, which is 
\[
\pr\left\{a^{\T}\mu_{g}\in
\mathbb{C}_{\mu_g,1-\alpha}^{\baci}(a)\right\}\geq1-\alpha+o(1),
\]
where $\mathbb{C}_{\mu_g,1-\alpha}^{\baci}(a)=\left[a^{\T}\widehat{\mu}_{\tap}-{\widehat{U}}_{1-\alpha/2}(a)/\surd{n},a^{\T}\widehat{\mu}_{\tap}-{\widehat{L}}_{\alpha/2}(a)/\surd{n}\right]$.
As we already know that 
\[
a^{\T}n^{1/2}(\widehat{\mu}_{\tap}-\mu_{g})\leq{U}(a),\quad a^{\T}n^{1/2}(\widehat{\mu}_{\tap}-\mu_{g})\geq{L}(a),
\]
it is needed to show that $\text{\ensuremath{{\widehat{U}}}}(a)$
obtained by bootstrapping converges to the same asymptotic distribution
as ${U}(a)$. Let $D_{p\times p}$ denotes the space of $p\times p$
symmetric positive-definite matrices equipped with the spectral norm.
We can rewrite $\text{\ensuremath{{U}}}(a)$ as 
\begin{align*}
&\text{\ensuremath{{U}}}(a)  =-a^{\T}V_{\eff}^{1/2}W_{1}\{\Sigma_{S},n^{1/2}(\widehat{\mu}_{A}-\mu_{g}),n^{1/2}(\widehat{\mu}_{B}-\mu_{g}),\tau\}\\
 & +a^{\T}(\omega_{A}V_{\text{A-eff}}^{1/2}-\omega_{B}V_{\text{B-eff}}^{1/2})W_{2}\{\Sigma_{T},n^{1/2}(\widehat{\mu}_{A}-\mu_{g}),n^{1/2}(\widehat{\mu}_{B}-\mu_{g}),\tau\}\\
 & +a^{\T}\omega_{B}(V_{\text{B-eff}}^{1/2}+V_{\text{A-eff}}^{1/2})\mu_{[c_{\gamma},\infty)}^{t}\\
 & +a^{\T}\omega_{B}(V_{\text{B-eff}}^{1/2}+V_{\text{A-eff}}^{1/2})\\
 &\times\left[ W_{2}\{\Sigma_{T},n^{1/2}(\widehat{\mu}_{A}-\mu_{g}),n^{1/2}(\widehat{\mu}_{B}-\mu_{g}),\tau\}_{[c_{\gamma},\infty)}-\mu_{[c_{\gamma},\infty)}^{t}\right] \bone_{T\geq\upsilon_{n}}\\
 & +a^{\T}\omega_{B}(V_{\text{B-eff}}^{1/2}+V_{\text{A-eff}}^{1/2})\\
 &\times \sup_{\mu_{2}\in\R^{l}}\left[ W_{2}\{\Sigma_{T},n^{1/2}(\widehat{\mu}_{A}-\mu_{g}),n^{1/2}(\widehat{\mu}_{B}-\mu_{g}),\tau\}{}_{[c_{\gamma},\infty)}-\mu_{[c_{\gamma},\infty)}^{t}\right] \bone_{T<\upsilon_{n}}.
\end{align*}
Next, we adopt the notation for the bootstrapping to express the upper
bound $\widehat{{U}}(a)=\text{\ensuremath{{U}^{(b)}}}(a)$
as 
\begin{align*}
&\text{\ensuremath{{U}^{(b)}}}(a)  =-a^{\T}\widehat{V}_{\eff}^{1/2}W_{1}\{\widehat{\Sigma}_{S},n^{1/2}(\widehat{\mu}_{A}^{(b)}-\widehat{\mu}_{A}),n^{1/2}(\widehat{\mu}_{B}^{(b)}-\widehat{\mu}_{A}),\widehat{\tau}\}\\
 & +a^{\T}(\omega_{A}\widehat{V}_{\text{A-eff}}^{1/2}-\omega_{B}\widehat{V}_{\text{B-eff}}^{1/2})W_{2}\{\widehat{\Sigma}_{T},n^{1/2}(\widehat{\mu}_{A}^{(b)}-\widehat{\mu}_{A}),n^{1/2}(\widehat{\mu}_{B}^{(b)}-\widehat{\mu}_{A}),\widehat{\tau}\}\\
 & +a^{\T}\omega_{B}(\widehat{V}_{\text{B-eff}}^{1/2}+\widehat{V}_{\text{A-eff}}^{1/2})\bar{W}_{2}^{(b)}{}_{[c_{\gamma},\infty)}\\
 & +a^{\T}\omega_{B}(\widehat{V}_{\text{B-eff}}^{1/2}+\widehat{V}_{\text{A-eff}}^{1/2})\\
 &\times \left[ W_{2}\{\widehat{\Sigma}_{T},n^{1/2}(\widehat{\mu}_{A}^{(b)}-\widehat{\mu}_{A}),n^{1/2}(\widehat{\mu}_{B}^{(b)}-\widehat{\mu}_{A}),\widehat{\tau}\}_{[c_{\gamma},\infty)}-\bar{W}_{2}^{(b)}{}_{[c_{\gamma},\infty)}\right] \bone_{T\geq\upsilon_{n}}\\
 & +a^{\T}\omega_{B}(\widehat{V}_{\text{B-eff}}^{1/2}+\widehat{V}_{\text{A-eff}}^{1/2})\\
 &\times \sup_{\mu_{2}\in\R^{l}}\left[ W_{2}\{\widehat{\Sigma}_{T},n^{1/2}(\widehat{\mu}_{A}^{(b)}-\widehat{\mu}_{A}),n^{1/2}(\widehat{\mu}_{B}^{(b)}-\widehat{\mu}_{A}),\widehat{\tau}\}_{[c_{\gamma},\infty)}-\mu_{[c_{\gamma},\infty)}^{t}\right] \bone_{T<\upsilon_{n}},
\end{align*}
where $\bar{W}_{2}^{(b)}=(1/K)\sum_{b=1}^{K}W_{2}\{\widehat{\Sigma}_{T},n^{1/2}(\widehat{\mu}_{A}^{(b)}-\widehat{\mu}_{A}),n^{1/2}(\widehat{\mu}_{B}^{(b)}-\widehat{\mu}_{A}),\widehat{\tau}\}$.
Next, we define some functions to proceed our proof. $w_{11}:D_{l\times l}\times D_{l\times l}\times\R^{l}\times\R^{l}\times\R^{d}\times\R\rightarrow\R$,
$w_{12}:D_{l\times l}\times\R^{l}\times\R^{l}\times\R^{d}\times\R^{l}\rightarrow\R$
and $\rho:D_{2l\times2l}\times D_{l\times l}\times\R^{l}\times\R^{l}\times\R^{d}\times\R\times\R^{l}\rightarrow\R$
are functions defined as below 
\begin{alignat*}{2}
&w_{11}(\Sigma_{T},\Sigma_{S},&&\mathbb{G}_{A},\mathbb{G}_{B},\tau,\mu_{2})  =-a^{\T}V_{\eff}^{1/2}W_{1}(\Sigma_{S},\mathbb{G}_{A},\mathbb{G}_{B},\tau)\\
 & &&+a^{\T}(\omega_{A}V_{\text{A-eff}}^{1/2}-\omega_{B}V_{\text{B-eff}}^{1/2})W_{2}(\Sigma_{T},\mathbb{G}_{A},\mathbb{G}_{B},\tau)\\
 & &&+a^{\T}\omega_{B}(V_{\text{B-eff}}^{1/2}+V_{\text{A-eff}}^{1/2})\mu_{[c_{\gamma},\infty)}^{t}\\
 & &&+a^{\T}\omega_{B}(V_{\text{B-eff}}^{1/2}+V_{\text{A-eff}}^{1/2})\\
 & &&\times\left\{ W_{2}(\Sigma_{T},\mathbb{G}_{A},\mathbb{G}_{B},\tau)_{[c_{\gamma},\infty)}-\mu_{[c_{\gamma},\infty)}^{t}\right\} \bone_{\mu_{2}^{\T}\mu_{2}\in\mathbb{B}^{\complement},}\\
 &w_{12}(\Sigma_{T},\mathbb{G}_{A},&&\mathbb{G}_{B},\mu_{2}) =a^{\T}\omega_{B}(V_{\text{B-eff}}^{1/2}+V_{\text{A-eff}}^{1/2})\\
&&&\times\left\{ W_{2}(\Sigma_{T},\mathbb{G}_{A},\mathbb{G}_{B},\tau){}_{[c_{\gamma},\infty)}-\mu_{[c_{\gamma},\infty)}^{t}\right\} \bone_{\mu_{2}^{\T}\mu_{2}\in\mathbb{B},}\\
&\rho_{11}(\Sigma_{T},\mathbb{G}_{A},&&\mathbb{G}_{B},\mu_{2})  =a^{\T}\omega_{B}(V_{\text{B-eff}}^{1/2}+V_{\text{A-eff}}^{1/2})\\
 & &&\times\left\{ W_{2}(\Sigma_{T},\mathbb{G}_{A},\mathbb{G}_{B},\tau)_{[c_{\gamma},\infty)}-\mu_{[c_{\gamma},\infty)}^{t}\right\} (\bone_{T\geq\upsilon_{n}}-\bone_{\mu_{2}^{\T}\mu_{2}\in\mathbb{B}^{\complement}}),\\
&\rho_{12}(\Sigma_{T},\mathbb{G}_{A},&&\mathbb{G}_{B},\mu_{2})  =a^{\T}\omega_{B}(V_{\text{B-eff}}^{1/2}+V_{\text{A-eff}}^{1/2})\\
 & &&\times\left\{ W_{2}(\Sigma_{T},\mathbb{G}_{A},\mathbb{G}_{B},\tau)_{[c_{\gamma},\infty)}-\mu_{[c_{\gamma},\infty)}^{t}\right\} (\bone_{T<\upsilon_{n}}-\bone_{\mu_{2}^{\T}\mu_{2}\in\mathbb{B}}),
\end{alignat*}
where $\mathbb{G}_{A}=n^{1/2}(\widehat{\mu}_{A}-\mu_{g})$
and $\mathbb{G}_{B}=n^{1/2}(\widehat{\mu}_{B}-\mu_{g})$.
Using the functions we have defined, we could re-express the upper
bound $\text{\ensuremath{{U}}}(a)$ in terms of 
\begin{align*}
\text{\ensuremath{{U}}}(a) & =w_{11}(\Sigma_{T},\Sigma_{S},\mathbb{G}_{A},\mathbb{G}_{B},\tau,\mu_{2})+\rho_{11}(\Sigma_{T},\mathbb{G}_{A},\mathbb{G}_{B},\mu_{2})\\
 & +\sup_{\mu_{2}\in\R^{l}}\left\{ w_{12}(\Sigma_{T},\mathbb{G}_{A},\mathbb{G}_{B},\mu_{2})+\rho_{12}(\Sigma_{T},\mathbb{G}_{A},\mathbb{G}_{B},\mu_{2})\right\} .
\end{align*}
Assume the conditions in Theorem \ref{thm:ACI}, we can show that 
\begin{enumerate}
    \item $w_{11}$ is continuous at points in $(\Sigma_T, \Sigma_S, \R^l, \R^l, \R^d,\mu_2)$ and $w_{12}$ is continuous at points in $(\Sigma_T, \R^l, \R^l,\mu_2)$ uniformly in $\mu_2$. That is, for any $\widehat{\Sigma}_T \rightarrow \Sigma_T$, $\widehat{\Sigma}_S \rightarrow \Sigma_S$, $\mathbb{G}_A^{(b)}=n^{1/2}(\widehat{\mu}_A^{(b)}-\widehat{\mu}_A) \rightarrow Z_1$,  $\mathbb{G}_B^{(b)}=n^{1/2}(\widehat{\mu}_B^{(b)}-\widehat{\mu}_A) \rightarrow Z_2$ and $\widehat{\tau}\rightarrow \tau$, we have
    \begin{equation}
    \begin{aligned}
    &\sup_{\mu_2 \in \R^l}|
        w_{11}(\widehat{\Sigma}_T, \widehat{\Sigma}_S, \mathbb{G}_A^{(b)},
        \mathbb{G}_B^{(b)},\widehat{\tau},\mu_2
        )-
        w_{11}({\Sigma}_T, {\Sigma}_S, Z_1,
       Z_2,\tau, \mu_2
        )
        |\rightarrow 0,\\
    &\sup_{\mu_2 \in \R^l}|
        w_{12}(\widehat{\Sigma}_T, \mathbb{G}_A^{(b)},
        \mathbb{G}_B^{(b)},\mu_2
        )-
        w_{12}({\Sigma}_T, Z_1,
       Z_2,\mu_2
        )
        |\rightarrow 0.
    \end{aligned}
    \label{eq:cont:1}
    \end{equation}
    \item $\rho_{11}(\widehat{\Sigma}_T, \mathbb{G}_A^{(b)}, \mathbb{G}_B^{(b)}, \mu_2)$ and $\rho_{12}(\widehat{\Sigma}_T, \mathbb{G}_A^{(b)}, \mathbb{G}_B^{(b)}, \mu_2)$ converge to zeros with probability one as $n\rightarrow \infty$ uniformly in $\mu_2$. That is,
    \begin{equation}
        \begin{aligned}
            \sup_{\mu_2 \in \R^l}|\rho_{11}(\widehat{\Sigma}_T, \mathbb{G}_A^{(b)}, \mathbb{G}_B^{(b)}, \mu_2)| \rightarrow 0,
            \quad \max_{\mu_2 \in \R^l}|\rho_{12}(\widehat{\Sigma}_T, \mathbb{G}_A^{(b)}, \mathbb{G}_B^{(b)}, \mu_2)| 
            \rightarrow 0.
        \end{aligned}
        \label{eq:cont:2}
    \end{equation}
    See Lemma B.9. and Lemma B.11. in \cite{laber2014dynamic} for details.
\end{enumerate}
By far, combine (\ref{eq:cont:1}) and (\ref{eq:cont:2}), ${U}(a)$
is guaranteed to be continuous, and the continuity of ${L}(a)$
can be derived in the same way. Based on continuous mapping theorem
and Theorem 4.2 in \cite{laber2014dynamic}, we can state that 
\[
\sup_{M}|\E\{{L}(a),{U}(a)\}-\E_{M}\{{L}^{(b)}(a),{U}^{(b)}(a)\}|
\]
converges to zero in probability, where $\E_{M}(\cdot)$ denotes the
expectation taken with respect to the bootstrap weights.

\subsection{Proof of Theorem \ref{thm:ACI}}
Based on the established consistency of the bootstrapping bounds in Section \ref{subsec:asym_u_a}, the proof can be decomposed into two parts. One part is for 
\begin{align*}
\pr\{a^{\T}\surd{n}(\widehat{\mu}_{\tap}-\mu_{g})\leq\widehat{U}_{1-\alpha/2}(a)\} & \geq \pr\{{U}(a)\leq\widehat{U}_{1-\alpha/2}(a)\}\\
 & =G_{{U}}\{\widehat{U}_{1-\alpha/2}(a)\}-\widehat{G}_{{U}}\{\widehat{U}_{1-\alpha/2}(a)\}\\
 & +\widehat{G}_{{U}}\{\widehat{U}_{1-\alpha/2}(a)\}\\
 & =o(1)+1-\alpha/2,
\end{align*}
where $G_{{U}}(\cdot)$ is the cumulative distribution function for ${U}(a)$. Let $\widehat{G}_{{U}}(\cdot)$
be the empirical cumulative distribution function $\widehat{{U}}(a)$ estimated by bootstrapping.
Similarly, we can show that the other part of our proof as 
\begin{align*}
\pr\{a^{\T}\surd{n}(\widehat{\mu}_{\tap}-\mu_{g})\leq\widehat{L}_{\alpha/2}(a)\} & \leq \pr\{{L}(a)\leq\widehat{L}_{\alpha/2}(a)\}\\
 & =G_{{L}}\{\widehat{L}_{\alpha/2}(a)\}-\widehat{G}_{{L}}\{\widehat{L}_{\alpha/2}(a)\}\\
 & +\widehat{G}_{{L}}\{\widehat{L}_{\alpha/2}(a)\}\\
 & =o(1)+\alpha/2,
\end{align*}
where $G_{{L}}(\cdot)$ is the cumulative distribution function for ${L}(a)$. Combine the results we have above, we can obtain that 
\begin{align*}
 &\pr(\widehat{L}_{\alpha/2}(a)\leq a^{\T}\surd{n}(\widehat{\mu}_{\tap}-\mu_{g})\leq\widehat{U}_{1-\alpha/2}(a)) \\
 &= \pr\{a^{\T}\surd{n}(\widehat{\mu}_{\tap}-\mu_{g})\leq\widehat{U}_{1-\alpha/2}(a)\}\\
&- \pr\{a^{\T}\surd{n}(\widehat{\mu}_{\tap}-\mu_{g})\leq\widehat{L}_{\alpha/2}(a)\} \\
 & \geq1-\alpha/2+o(1)-\alpha/2+o(1)=1-\alpha. 
\end{align*}
Thus, the proof is completed.

\subsection{Proof of Remark \ref{rem:1}}
In this section, we construct a data-adaptive confidence interval
based on the projection sets proposed in \cite{robins2004optimal}.
Starting from the common projection sets, we re-express the test-and-pool estimator
\begin{align*}
a^{\T}n^{1/2}(\widehat{\mu}_{\tap}-\mu_{g}) & =-a^{\T}V_{\eff}^{1/2}W_{1}+a^{\T}(\omega_{A}V_{\text{A-eff}}^{1/2}-\omega_{B}V_{\text{B-eff}}^{1/2})W_{2}\\
 & +a^{\T}\omega_{B}(V_{\text{B-eff}}^{1/2}+V_{\text{A-eff}}^{1/2})W_{[c_{\gamma},\infty)}^{t}.
\end{align*}
For given $\mu_{2}$, we know that 
\begin{align*}
n^{1/2}\{\widehat{\mu}_{\tap}(\mu_{2})-\mu_{g}\} & =-a^{\T}V_{\eff}^{1/2}W_{1}+a^{\T}(\omega_{A}V_{\text{A-eff}}^{1/2}-\omega_{B}V_{\text{B-eff}}^{1/2})W_{2}(\mu_{2})\\
 & +a^{\T}\omega_{B}(V_{\text{B-eff}}^{1/2}+V_{\text{A-eff}}^{1/2})W_{[c_{\gamma},\infty)}^{t}(\mu_{2}),
\end{align*}
where the right hand side can be approximated by empirical sample distribution
as $\widehat{Q}_{n}(\mu;a)$ and we could construct a $(1-\tilde{\alpha}_{1})\times100\%$
confidence interval $\mathbb{B}_{\mu_{g},1-\tilde{\alpha}_{1}}(a;\mu_{2})$
of $\mu_{g}$ given $\mu_{2}$ by the empirical quantile confidence
interval as 
\begin{align*}
& \mathbb{B}_{\mu_{g},1-\tilde{\alpha}_{1}}(a;\mu_{2})\\
&=\left\{ \mu_{g}\in\R^{l}:\widehat{\mu}_{\tap}(\mu_{2})-\frac{\widehat{Q}_{n}^{-1}(1-\alpha/2;a)}{\surd{n}}\leq\mu_{g}\leq\widehat{\mu}_{\tap}(\mu_{2})-\frac{\widehat{Q}_{n}^{-1}(\alpha/2;a)}{\surd{n}}\right\} ,
\end{align*}
where $\widehat{Q}_{n}^{-1}(d;a)$ is the $d$-th sample quantiles
based on our empirical distribution.

However, the value of $\mu_{2}$ is unknown, a useful approach is
to form a $(1-\tilde{\alpha}_{2})\times100\%$ confidence region $\mathbb{B}_{\mu_{2,}1-\tilde{\alpha}_{2}}$
for $\mu_{2}$, and thus the projection confidence interval for $\mu_{g}$
is the union of $\mathbb{B}_{\mu_{g},1-\tilde{\alpha}_{1}}(a;\mu_{2})$
over all $\mu_{2}\in\mathbb{B}_{\mu_{2,}1-\tilde{\alpha}_{2}}$. Here,
the confidence bounds for $\mu_{2}$ can be constructed as $\mathbb{B}_{\mu_{2}1-\tilde{\alpha}_{2}}=\widehat{\mu}_{2}\pm\Phi^{-1}(1-\tilde{\alpha}_{2}/2)$
where 
$$\widehat{\mu}_{2}=n^{1/2}f_{B}^{1/2}\Sigma_{T}^{-1/2}\left\{ N^{-1}\sum_{i=1}^{N}\dot{\Phi}_{B}(V_{i},\delta_{A,i},\delta_{B,i};\widehat{\mu}_{A},\widehat{\tau})\right\} ^{-1}(\widehat{\mu}_{A}-\widehat{\mu}_{B}),$$
$\Phi^{-1}(\cdot)$ is the inverse cdf for a standard normal
distribution. Thus, let $\alpha=\tilde{\alpha}_{1}+\tilde{\alpha}_{2}$
and the union would be the data-adaptive projection $(1-\alpha)\times100\%$
confidence interval for $\mu_{g}$ 
\begin{equation}
\mathbb{C}_{\mu_g,1-\alpha}^{\pci}(a)=\cup_{\mu_{2}\in\mathbb{B}_{\mu_{2,}1-\tilde{\alpha}_{2}}}\mathbb{B}_{\mu_{g},1-\tilde{\alpha}_{1}}(a;\mu_{2}).\label{eq:PACI_form}
\end{equation}
To limit conservatism, a pretest procedure is carried out while we
construct the projection adaptive confidence intervals $\mathbb{C}_{\mu_g,1-\alpha}^{\paci}(a)$, and
we would use the $\mathbb{C}_{\mu_g,1-\alpha}^{\pci}(a)$ if we cannot reject the $H_{0}:\mu^{\T}\mu\in\mathbb{B}$.
To prove the coverage for the projection adaptive confidence interval,
denote for $\alpha\in(0,1)$, we have that 
\begin{align*}
&\pr\left(a^{\T}\mu_{g}\notin \mathbb{C}_{\mu_g,1-\alpha}^{\paci}(a)\right)  =\pr\left(a^{\T}\mu_{g}\notin \mathbb{C}_{\mu_g,1-\alpha}^{\paci}(a)\mid T\leq v_{n}\right)\pr(T\leq v_{n})\\
 & +\pr\left\{ a^{\T}\mu_{g}\notin\mathbb{B}_{\mu_{g},1-\alpha}(a;\widehat{\mu}_{2})|T>v_{n}\right\} \pr(T>v_{n})\\
 & =\pr(a^{\T}\mu_{g}\notin \mathbb{C}_{\mu_g,1-\alpha}^{\pci}(a),\mu_{2}\in\mathbb{B}_{\mu_{2,}1-\tilde{\alpha}_{2}}\mid T\leq v_{n})\pr(T\leq v_{n})\\
 & +\pr(a^{\T}\mu_{g}\notin \mathbb{C}_{\mu_g,1-\alpha}^{\pci}(a),\mu_{2}\notin\mathbb{B}_{\mu_{2,}1-\tilde{\alpha}_{2}}\mid T\leq v_{n})\pr(T\leq v_{n})\\
 & +\{\tilde{\alpha}_{1}+o(1)\}\pr(T>v_{n})\\
 & \leq \pr\{a^{\T}\mu_{g}\notin\mathbb{B}_{\mu_{g},1-\tilde{\alpha}_{1}}(a;\mu_{2}),\mu_{2}\in\mathbb{B}_{\mu_{2,}1-\tilde{\alpha}_{2}}\mid T\leq v_{n}\}\pr(T\leq v_{n})\\
 & +\pr(\mu_{2}\notin\mathbb{B}_{\mu_{2,}1-\tilde{\alpha}_{2}}\mid T\leq v_{n})\pr(T\leq v_{n})+\alpha \pr(T>v_{n})\\
 & \leq(\tilde{\alpha}_{1}+\tilde{\alpha}_{2})\pr(T\leq v_{n})+\alpha \pr(T>v_{n})\\
 & =\alpha,
\end{align*}
where we know that $\pr\{a^{\T}\mu_{g}\notin\mathbb{B}_{\mu_{g},1-\tilde{\alpha}_{1}}(a;\mu_{2}),\mu_{2}\in\mathbb{B}_{\mu_{2,}1-\tilde{\alpha}_{2}}\}\leq\tilde{\alpha}_{1}$
holds for any value $\mu_{2}$.

\subsection{Proof of Lemma \ref{lemm:uncondition}}
Following the similar arguments in \cite{schenker1988asymptotic}, let $F(\cdot)$ and $G(\cdot)$ be the cumulative distribution function (c.d.f.) of $\mathcal{N}(\mu_g,V_1)$ and $\mathcal{N}(-\mu_{g,0},V_2)$. Let $\Phi(t)$ be the convolution of $G(\cdot)$ and $F(\cdot)$ as $\Phi(\cdot)=(G*F)(\cdot)$, then we have
\begin{align*}
&|\pr\{(\widehat{\mu}_{g}-\mu_g)\leq t\}-\Phi(t)|\\
&\leq 
\left|\mathbb{E}_\zeta\left\{\sup_x \pr(\widehat{\mu}_{g}
\leq x\mid \mathcal{F}_N)-
F(x)\right\}\right|+
|\mathbb{E}_\zeta\left\{
F(s) - \Phi(t)
\right\}|,
\end{align*}
where $s=t+\mu_g=t-(-\mu_g)$. By Lemma 3.2 in \cite{rao1962relations}, $|\pr(\widehat{\mu}_{g}\leq x\mid \mathcal{F}_N)-F(x)|$ converges to $0$ uniformly in $x$. For the first term, we have
\begin{align*}
\lim_{N\rightarrow\infty}
&\left|\mathbb{E}_\zeta\left\{\sup_x \pr(\widehat{\mu}_{g}\leq x\mid \mathcal{F}_N)-
F(x)\right\}\right|
\\
&\leq 
\mathbb{E}_\zeta\left\{
\lim_{N\rightarrow\infty}
\left|\sup_x \pr(\widehat{\mu}_{g}\leq x\mid \mathcal{F}_N)-
F(x)\right|\right\}\rightarrow 0.
\end{align*}
Since $F(\cdot)$ and $G(\cdot)$ are both bounded and continuous, by the dominated convergence theorem, the second term is
$$
\lim_{N\rightarrow \infty}\mathbb{E}_\zeta\{\F(s)\}-\Phi(t)=
\mathbb{E}_\zeta\left\{
\lim_{N\rightarrow \infty}
\F(t-(-\mu_g))
\right\}-\Phi(t)=
\int_{x} G(x)F(t-x)dx-\Phi(t),
$$
which also converges to $0$ \citep[~Lemma 1]{schenker1988asymptotic}. Hence, the asymptotic c.d.f of $\widehat{\mu}_{g}-\mu_g$ is $\Phi(\cdot)$ and the result follows as the convolution of Gaussians is still Gaussian \citep{abramowitz1988handbook, boas2006mathematical}.
\subsection{Proof of Lemma \ref{lemma:Phi_B_expectation}}
Under Assumptions 
\ref{asmp:Sampling-design}, \ref{asmp:MAR} (iii) and Assumption \ref{asmp:regularity} f), we have 
\begin{alignat*}{2}
    0&=&&N^{-1}\sum_{i=1}^N
    \E_{\mathrm{np}\text{-}\mathrm{p}}\left\{\Phi_B(V_i,\delta_{A,i},\delta_{B,i};\mu_{B,0},\tau_0)
    \mid \F_N
    \right\} \\
    &= &&
N^{-1}\sum_{i=1}^N \E_{\mathrm{np}\text{-}\mathrm{p}}\left\{\Phi_B(V_i,\delta_{A,i},\delta_{B,i};\mu_{g,0},\tau_0)
    \mid \F_N
    \right\} \\
    &&&+ 
N^{-1}\sum_{i=1}^N
\E_{\mathrm{np}\text{-}\mathrm{p}}\left\{
\dot{\Phi}_B(V_i,\delta_{A,i},\delta_{B,i};\mu_{B}^{*},\tau_0)\mid \F_N\right\}
(\mu_{B,0}-\mu_{g,0})\\
&= &&
\E\{\Phi_B(V_i,\delta_{A,i},\delta_{B,i};\mu_{g,0},\tau_0)\}
\\
&&&+ 
\E\left\{\dot{\Phi}_B(V_i;\mu_{B}^{*},\tau_0)
\right\}
(\mu_{B,0}-\mu_{g,0}) + 
O_{\mathrm{np}\text{-}\mathrm{p}\text{-}\zeta}(n^{-1/2}),
\end{alignat*}
for some $\mu_B^{*}$ between $\mu_{B,0}$ and $\mu_{g,0}$, where
\begin{align}
&N^{-1}\sum_{i=1}^N \E_{\mathrm{np}\text{-}\mathrm{p}}\left\{\Phi_B(V_i,\delta_{A,i},\delta_{B,i};\mu_{g,0},\tau_0)
    \mid \F_N
    \right\}\nonumber\\
    &=
\E_\zeta 
\left[
\E_{\mathrm{np}\text{-}\mathrm{p}}
\left\{
\Phi_{B}(V_{i},\delta_{A,i},\delta_{B,i};\mu_{g},\tau_{0})\mid \F_N
\right\}
\right] + O_{\mathrm{np}\text{-}\mathrm{p}\text{-}\zeta}(n^{-1/2})\label{eq:first_approx}\\
&=\E\{\Phi_{B}(V_{i},\delta_{A,i},\delta_{B,i};\mu_{g,0},\tau_{0})\}  + O_{\mathrm{np}\text{-}\mathrm{p}\text{-}\zeta}(N^{-1/2})  + O_{\mathrm{np}\text{-}\mathrm{p}\text{-}\zeta}(n^{-1/2}),
\label{eq:second_approx}
\end{align}
where for (\ref{eq:first_approx}),  the first approximation $\E_{\mathrm{np}\text{-}\mathrm{p}}(\cdot\mid\F_N)$ is based on the design consistency and the non-probability sample-based Weak Law of Large
Numbers under Assumption \ref{asmp:MAR} (iii), and the second approximation $\E_{\zeta}(\cdot)$ is justified under Assumption \ref{asmp:regularity} f); For (\ref{eq:second_approx}), it can be obtained by continuous mapping theorem as $\mu_g = \mu_{g,0}+O_\zeta(N^{-1/2})$ under Assumption \ref{asmp:regularity} f). By rearranging the terms under the local alternative, it follows that
\begin{align*}
  &\mu_{B,0}-\mu_{g,0} \\
  &= \left[
\E\left\{\dot{\Phi}_B(V;\mu_{B}^{*},\tau_0)
\right\}\right]^{-1}
\E\{\Phi_B(V_i,\delta_{A,i},\delta_{B,i};\mu_{g,0},\tau_0)\} + O_{\mathrm{np}\text{-}\mathrm{p}\text{-}\zeta}(n^{-1/2})
\\
&= 
O(1)\times n_B^{-1/2}\eta +O_{\mathrm{np}\text{-}\mathrm{p}\text{-}\zeta}(n^{-1/2})=o_{\mathrm{np}\text{-}\mathrm{p}\text{-}\zeta}(1).  
\end{align*}

\subsection{Proof of Lemma \ref{lem:mu_eff}}
First, we show that the composite estimator $\widehat{\mu}_{\pool}$ is
essentially the solution to 
$$
\sum_{i=1}^{N}\{\Phi_{A}(V_{i},\delta_{A,i};\mu,\tau)+\Lambda \Phi_{B}(V_{i},\delta_{A,i},\delta_{B,i};\mu,\tau)\}=0.
$$
Next, under the Assumption \ref{asmp:regularity} a)-d), we apply
the Taylor expansion at point $(\mu_{g},\tau_{0})$ which leads to
\begin{align*}
0 & =\sum_{i=1}^{N}\{\Phi_{A}(V_{i},\delta_{A,i};\widehat{\mu}_{\pool},\widehat{\tau})+\Lambda \Phi_{B}(V_{i},\delta_{A,i},\delta_{B,i};\widehat{\mu}_{\pool},\widehat{\tau})\}\\
 & =\sum_{i=1}^{N}\{\Phi_{A}(V_{i},\delta_{A,i};\mu_{g},\tau_{0})+\Lambda \Phi_{B}(V_{i},\delta_{A,i},\delta_{B,i};\mu_{g},\tau_{0})\}\\
 & +\sum_{i=1}^{N}\left\{ \frac{\partial \Phi_{A}(V_{i},\delta_{A,i};\widehat{\mu}_{\pool}^*,\widehat{\tau}^*)}{\partial\mu}+\Lambda\frac{\partial \Phi_{B}(V_{i},\delta_{A,i},\delta_{B,i};\widehat{\mu}_{\pool}^*,\widehat{\tau}^*)}{\partial\mu}\right\} (\widehat{\mu}_{\pool}-\mu_{g})\\
 & +\sum_{i=1}^{N}\left\{ \frac{\partial \Phi_{A}(V_{i},\delta_{A,i};\widehat{\mu}_{\pool}^*,\widehat{\tau}^*)}{\partial\tau}+\Lambda\frac{\partial \Phi_{B}(V_{i},\delta_{A,i},\delta_{B,i};\widehat{\mu}_{\pool}^*,\widehat{\tau}^*)}{\partial\tau}\right\} (\widehat{\tau}-\tau_{0}),
\end{align*}
for some $(\widehat{\mu}_{\pool}^*,\widehat{\tau}^*)$ between $(\widehat{\mu}_{\pool},\widehat{\tau})$ and $(\mu_g, \tau_0)$. Given the asymptotic joint
distribution for $\widehat{\mu}_{A}$ and $\widehat{\mu}_{B}$ in
Lemma \ref{lemma:mu_A_B_Hn}, we obtain 
\begin{align}
 & n^{1/2}(\widehat{\mu}_{\pool}-\mu_{g})\nonumber\\
 &=-n^{1/2}\left[ \sum_{i=1}^{N}\left\{\dot{\Phi}_{A}(V_i,\delta_{A,i};\widehat{\mu}_{\pool}^*,\widehat{\tau}^*)+\Lambda\dot{\Phi}_{B}(V_i,\delta_{A,i},\delta_{B,i};\widehat{\mu}_{\pool}^*,\widehat{\tau}^*)\right\}
 \right] ^{-1}\nonumber \\
 &\times\left[\sum_{i=1}^{N}\left\{\Phi_{A}(V_{i},\delta_{A,i};\mu_{g},\tau_{0})+\Lambda \Phi_{B}(V_{i},\delta_{A,i},\delta_{B,i};\mu_{g},\tau_{0})\right\}\right.\nonumber \\
 &\left.+\sum_{i=1}^{N}\left(\partial \Phi_{A}(V_i,\delta_{A,i};\widehat{\mu}_{\pool}^*,\widehat{\tau}^*)/\partial\tau+\Lambda\partial \Phi_{B}(V_i,\delta_{A,i},\delta_{B,i};\widehat{\mu}_{\pool}^*,\widehat{\tau}^*)/\partial\tau\right)(\widehat{\tau}-\tau_{0})\right]\nonumber \\
 & \cong\E\left\{ 
 \dot{\Phi}_{A,B,n}(\Lambda,{\mu}_{\pool}^*,{\tau}_0)
 \right\} ^{-1}\nonumber \\
 & \times\left[ 
 \E\left\{\dot{\Phi}_A(V;\mu_{g,0},\tau_0)
 \right\}
 \cdot n^{1/2}(\widehat{\mu}_{A}-\mu_{g})+
 \Lambda\E\left\{\dot{\Phi}_B(V;\mu_B^*,\tau_0)
 \right\}\cdot n^{1/2}(\widehat{\mu}_{B}-\mu_{g})\right],\label{eq:step1}
\end{align}
for some intermittent value ${\mu}_{\pool}^*$ between $\plim\widehat{\mu}_\pool$ and $\mu_{g,0}$, where Equation (\ref{eq:step1}) is obtained by using Equation (\ref{eq:mu_A-mu_0})
and (\ref{eq:mu_B-mu_0}) collectively. By Assumptions \ref{asmp:Sampling-design}, \ref{asmp:MAR} (iii) and suitable moments condition in Assumption \ref{asmp:regularity}, under the local alternative, $n^{1/2}(\widehat{\mu}_{\pool}-\mu_{g})$
would follow the normal distribution with mean and variance as 
\begin{align*}
&\E\left\{ n^{1/2}(\widehat{\mu}_{\pool}-\mu_{g})\right\}   =-f_{B}^{-1/2}\E\left\{
\dot{\Phi}_{A,B,n}(\Lambda,{\mu}_{g,0},\tau_0)
\right\} ^{-1}\Lambda\eta,\\
&\var\left\{ n^{1/2}(\widehat{\mu}_{\pool}-\mu_{g})\right\}   =\E\left\{
\dot{\Phi}_{A,B,n}(\Lambda,{\mu}_{g,0},\tau_0)
\right\}^{-1}\\
 & \times\left\{ 
 \left(\begin{array}{c}
\E\dot{\Phi}_{A}(V;\mu_{g,0},\tau_0)\\
\Lambda\E\dot{\Phi}_{B}(V;\mu_{g,0},\tau_0)
\end{array}\right)\left(\begin{array}{cc}
V_{A} & \Gamma\\
\Gamma^{\T} & V_{B}
\end{array}\right)\left(\begin{array}{c}
\E\dot{\Phi}_{A}(V;\mu_{g,0},\tau_0)\\
\Lambda\E\dot{\Phi}_{B}(V;\mu_{g,0},\tau_0)
\end{array}\right)^{\T}\right\}  \\
&\times \left[\E\left\{
\dot{\Phi}_{A,B,n}(\Lambda,{\mu}_{g,0},\tau_0)
\right\}^{-1}\right] ^{\T},
\end{align*}
obtained by the similar arguments in (\ref{eq:mu_B2mu_0}). Plugging (\ref{eq:lambda_eff}) into Equation (\ref{eq:step1}), the asymptotic distribution of the most efficient estimator $\widehat{\mu}_\eff$ follows
\begin{align*}
&n^{1/2}(\widehat{\mu}_{\text{eff}}-\mu_{g})\cong  \E\left\{
\dot{\Phi}_{A,B,n}(\Lambda_\eff,{\mu}_{g,0},\tau_0)
\right\}^{-1}\times\\
 & \left\{ 
 \E\dot{\Phi}_{A}(V;\mu_{g,0},\tau_0)
 \cdot n^{1/2}(\widehat{\mu}_{A}-\mu_{g})+\Lambda_{\eff}\E\dot{\Phi}_{B}(V;\mu_{g,0},\tau_0) \cdot n^{1/2}(\widehat{\mu}_{B}-\mu_{g})\right\}\\
 & \cong n^{1/2}\left\{ \omega_{A}(\Lambda_\eff)(\widehat{\mu}_{A}-\mu_{g})+\omega_{B}(\Lambda_\eff)(\widehat{\mu}_{B}-\mu_{g})\right\}.
\end{align*}
It yields a similar efficient estimator as derived in \cite{yang2020combining}
\begin{align}
n^{1/2}(\widehat{\mu}_{\text{eff}}-\mu_{g}) & \cong n^{1/2}\left\{ \omega_{A}(\Lambda_\eff)\widehat{\mu}_{A}+\omega_{B}(\Lambda_\eff)\widehat{\mu}_{B}-\mu_{g}\right\} ,\label{eq:mu_eff}
\end{align}
with 
\begin{align*}
\omega_{A}(\Lambda) & = \E\left\{\dot{\Phi}_{A,B,n}(\Lambda,\mu_{g,0},\tau_0)\right\}^{-1}\E\left\{ \dot{\Phi}_{A}(V_{i},\delta_{A,i};\mu_{g,0},\tau_{0})\right\},\\
\omega_{B}(\Lambda) & =\E\left\{\dot{\Phi}_{A,B,n}(\Lambda,\mu_{g,0},\tau_0)\right\}^{-1}\Lambda
\E\left\{\dot{\Phi}_{B}(V_{i},\delta_{A,i},\delta_{B,i};\mu_{g,0},\tau_{0})\right\},
\end{align*}
where it is easy to show that $\omega_A+\omega_B=I_{l\times l}$. So that the asymptotic variance $V_\eff$ of this efficient estimator will become
$$
V_{{\rm eff}}=\left(\begin{array}{c}
\omega_{A}^{\T}(\Lambda_\eff)\\
\omega_{B}^{\T}(\Lambda_\eff)
\end{array}\right)^{\T}\left(\begin{array}{cc}
V_{A} & \Gamma\\
\Gamma^{\T} & V_{B}
\end{array}\right)\left(\begin{array}{c}
\omega_{A}(\Lambda_\eff)\\
\omega_{B}(\Lambda_\eff)
\end{array}\right).
$$
The expression
of $V_{\eff}$ can be complicated when the dimension of the parameters
of interest is greater than $1$. Here, we provide the form of $V_{\eff}$
when estimating equations are (\ref{eq:S_A_1}) and (\ref{eq:S_B_1}):
\begin{eqnarray*}
\omega_{A}(\Lambda_\eff)&=&
\E\left\{\dot{\Phi}_{A,B,n}(\Lambda_\eff,\mu_{g,0},\tau_0)\right\}^{-1}
\E\left\{\dot{\Phi}_A(V,\delta_A;\mu_{g,0},\tau_0)\right\},\\
 &=&\{I_{l\times l}+(V_A-\Gamma)(V_B-\Gamma^\T)^{-1}\}^{-1}\\
 &=&(V_B-\Gamma^\T)(V_A+V_B-\Gamma-\Gamma^\T)^{-1},\\
\omega_{B}(\Lambda_\eff)&=&\E\left\{\dot{\Phi}_{A,B,n}(\Lambda_\eff,\mu_{g,0},\tau_0)\right\}^{-1}
\Lambda_\eff
\E\left\{\dot{\Phi}_B(V,\delta_A,\delta_B;\mu_{g,0},\tau_0)\right\},\\
  &=&\{I_{l\times l}+(V_A-\Gamma)(V_B-\Gamma^\T)^{-1}\}^{-1}(V_A-\Gamma)(V_B-\Gamma^\T)^{-1}\\
 &=&(V_B-\Gamma^\T)(V_A+V_B-\Gamma-\Gamma^\T)^{-1}(V_A-\Gamma)(V_B-\Gamma^\T)^{-1},
\end{eqnarray*}
and
\begin{align*}
V_{\eff} & =\left\{ \left(\begin{array}{c}
1\\
-1
\end{array}\right)\left(\begin{array}{cc}
V_{A} & \Gamma\\
\Gamma^{\T} & V_{B}
\end{array}\right)\left(\begin{array}{c}
1\\
-1
\end{array}\right)\right\} ^{-2}\\
&\times\left(\begin{array}{c}
V_{B}-\Gamma^{\T}\\
V_{A}-\Gamma
\end{array}\right)^{\T}\left(\begin{array}{cc}
V_{A} & \Gamma\\
\Gamma^{\T} & V_{B}
\end{array}\right)\left(\begin{array}{c}
V_{B}-\Gamma^{\T}\\
V_{A}-\Gamma
\end{array}\right)\\
 & =(V_{A}+V_{B}-\Gamma^{\T}-\Gamma)^{-2}\{(V_{B}-\Gamma^{\T})^{2}V_{A}+(V_{A}-\Gamma)^{2}V_{B}\\
 & +\Gamma(V_{B}-\Gamma^{\T})(V_{A}-\Gamma^{\T})+\Gamma^{\T}(V_{A}-\Gamma)(V_{B}-\Gamma)\}\\
 & =(V_{A}V_{B}-\Gamma^{2})(V_{A}+V_{B}-2\Gamma)^{-1}\\
 & =V_{A}-V_{\Delta},
\end{align*}
with $V_{\Delta}=(V_{A}-\Gamma)^{2}(V_{A}+V_{B}-2\Gamma)^{-1}$ guaranteed to be non-negative definite, i.e., non-negative quantity. By Cauchy-Schwarz inequality, we have
\begin{align*}
   \surd{\E\{(\widehat{\mu}_A-\mu_g)^2\} \times 
\E\{(\widehat{\mu}_B-\mu_g)^2\}} \geq 
\E\{(\widehat{\mu}_A-\mu_g)(\widehat{\mu}_B-\mu_g)\},
\end{align*}
which leads to $\surd{V_{A}V_{B}}\geq\Gamma$, and therefore
\[
V_{A}+V_{B}-2\Gamma\geq
2\{|V_{A}V_{B}|^{1/2}-\Gamma\}\geq0,
\]
where the two sides are equal if and only if $V_{A}=V_{B}=\Gamma$.
The asymptotic variance of the efficient estimator for other
multi-dimensional estimating equations can be obtained in an analogous
way but with much heavier notations.

\global\long\def\thetable{B.\arabic{table}}%
\setcounter{table}{0}

\global\long\def\thefigure{B.\arabic{figure}}%
\setcounter{figure}{0}

\makeatletter
\renewcommand{\thealgocf}{B.\@arabic\c@algocf}
\renewcommand{\fnum@algocf}{\AlCapSty{\algorithmcfname\nobreakspace\AlCapFnt\thealgocf}}%
\makeatother
\SetAlgorithmName{Algorithm}{}{}

\section{Simulation}

\subsection{A detailed illustration of simulation}
Here, we will provide detailed proof for estimating the finite-population
parameter $\mu_y=\mu_g=N^{-1}\sum_{i=1}^N Y_i$ and $\mu_0 = \mathbb{E}_\zeta(Y)$. First, we know the following expectation that 
$$
 \E_{\mathrm{np}}(\delta_{B,i}\mid X_{i},Y_{i})=\pi_B(X_{i},Y_i),\quad\E_{\mathrm{np}}(Y_{i}\mid X_{i})=m(X_{i}).
$$
To obtain the asymptotic joint
distribution $\widehat{\mu}_{A}$ and $\widehat{\mu}_{B}$, the stacked estimating equation system $\Phi(V,\delta_A,\delta_B;\theta)$ is constructed with $\theta = (\mu_A^\T,\mu_B^\T,\tau^\T)^\T$ where
\begin{equation}
\Phi(V,\delta_A,\delta_B; \theta)=\left\{\Phi_{A}(V,\delta_A; \mu_A)^\T,
\Phi_{B}(V,\delta_A,\delta_B; \mu_B,\tau)^\T, \Phi_{\tau}(V,\delta_A,\delta_B; \tau)^\T\right\}^\T,\label{eq:Phi}
\end{equation}
where we use $\mu_A$ and $\mu_B$ to distinguish between estimators yielded by $\Phi_A(V,\delta_A;\allowbreak\mu_A)$ and $\Phi_B(V,\delta_A,\delta_B;\mu_B,\tau)$. By positing a logistic regression model $\pi_B(X_i;\alpha) = \exp(X_{i}^{\T}\alpha)/\{1+\exp(X_{i}^{\T}\alpha)\}$ and a linear model $m(X_i;\beta) = X_{i}^{\T}\beta$, one common choices for $\Phi_A(V,\delta_A;\mu_A)$ and $\Phi_B(V,\delta_A,\delta_B;\mu_B,\tau)$ are 
\begin{align*}
&\Phi_A(V,\delta_A;\mu_A) = \delta_{A}\pi_A^{-1}(Y-\mu_{A}),\\
&\Phi_B(V,\delta_A,\delta_B;\mu_B,\tau)=\frac{\delta_{B}}{\pi_{B}\left(X ; \alpha\right)}\left\{Y-m\left(X ; \beta\right)\right\}+\frac{\delta_{A}}{\pi_{A}} m\left(X;\beta\right)-\mu_B,
\end{align*}
where $\tau = (\alpha,\beta)$ and $\pi_A$ is the known sample weights under probability samples accounting
for sample design. There are various ways to construct the estimating functions $\Phi_\tau(V_i;\alpha,\beta)$ for $(\alpha,\beta)$. One standard approach is to use the pseudo maximum likelihood estimator $\widehat{\alpha}$ and the ordinary least square estimator $\widehat{\beta}$ \citep{scharfstein1999adjusting, haziza2006nonresponse}. In usual, the maximum likelihood estimator of $\alpha$ can be computed by maximizing the log-likelihood function $l(\alpha)$
\begin{align*}
\widehat{\alpha} & = \arg\max_\alpha 
\sum_{i=1}^N\left[\delta_{B,i} \log\pi_B(X_i;\alpha) + 
(1-\delta_{B,i}) \log\{1-\pi_B(X_i;\alpha)\}\right]\\
& = \arg\max_\alpha 
\sum_{i=1}^N\delta_{B,i} 
\log \left\{
\frac{\pi_B(X_i;\alpha)}{1-\pi_B(X_i;\alpha)}
\right\}+
\sum_{i=1}^N 
\log\{1-\pi_B(X_i;\alpha)\}.
\end{align*}
Since we do not have the $X_i$ for all units in the finite population, we then instead construct the following pseudo log-likelihood function $l^*(\alpha)$
\begin{align*}
    l^*(\alpha)& =
\sum_{i=1}^N\delta_{B,i} 
\log \left\{
\frac{\pi_B(X_i;\alpha)}{1-\pi_B(X_i;\alpha)}
\right\}+
\sum_{i=1}^N \delta_{A,i}\pi_{A,i}^{-1}
\log\{1-\pi_B(X_i;\alpha)\}\\
& = 
\sum_{i=1}^N\left[ \delta_{B,i}X_i^\T \alpha -
\delta_{A,i}\pi_{A,i}^{-1}
\log\{1+\exp(X_i^\T \alpha)\}\right],
\end{align*}
where the second equality is derived under the logistic regression model for $\pi_B(X_i;\alpha)$. By taking derivative of $l^*(\alpha)$ with respect to $\alpha$, the estimating functions for $(\alpha,\beta)$ can be constructed as follows:
\begin{align}
&\Phi_{\tau,1}(V,\delta_A,\delta_B;\alpha,\beta)= \delta_{B}X-\delta_{A}\pi_{A}^{-1}\pi_B(X;\alpha)X, \label{eq:Phi_tau_alpha_1}\\
&\Phi_{\tau,2}(V,\delta_A,\delta_B;\alpha,\beta)=\delta_{B}X\{Y-m(X;\beta)\},
\label{eq:Phi_tau_beta_1}
\end{align}
with $\Phi_\tau(V,\delta_A,\delta_B;\alpha,\beta) = (\Phi_{\tau, 1}(V,\delta_A,\delta_B;\alpha,\beta)^\T\quad \Phi_{\tau, 2}(V,\delta_A,\delta_B;\alpha,\beta)^\T)^\T$. Under our setup, both Sample A and Sample B provide information on $X$ and $Y$, thus we can also consider the estimating equation based on the combined samples for $\beta$:
\begin{align}
&\Phi_{\tau,1}(V,\delta_A,\delta_B;\alpha,\beta)= \delta_{B}X-\delta_{A}\pi_{A}^{-1}\pi_B(X;\alpha)X,\nonumber \\
&\Phi_{\tau,2}^*(V,\delta_A,\delta_B;\alpha,\beta)=(\delta_{A}+\delta_{B})X\{Y-m(X;\beta)\}.\label{eq:Phi_tau_beta_2}
\end{align}
In addition, \cite{kim2014doubly} propose a new set of estimating functions, in which $(\widehat{\alpha},\widehat{\beta})$ are obtained by jointly solve the following estimating functions:
\begin{align}
   &\Phi_{\tau,1}^{\rm KH}(V,\delta_A,\delta_B;\alpha,\beta) = \left\{\delta_{B}
   \pi^{-1}_B(X;\alpha)-\delta_{A}\pi_{A}^{-1}\right\}X, \label{eq:Phi_tau_alpha_KH}\\
   &\Phi_{\tau,2}^{\rm KH}(V,\delta_A,\delta_B;\alpha,\beta) = \delta_{B}\{\pi^{-1}_B(X;\alpha)-1\}X\{Y-m(X;\beta)\}. \label{eq:Phi_tau_beta_KH}
\end{align}
Denote the solution to $\sum_{i=1}^N \Phi(V_i,\delta_{A,i},\delta_{B,i}; \theta)=0$ as $\widehat{\theta}=(\widehat{\mu}_A,\widehat{\mu}_B,\widehat{\tau}^\T)^\T$. Under Assumption \ref{asmp:regularity} a)-e), we could apply the Taylor expansion to 
around $\theta_{y}=(\mu_y,\mu_y,\tau_0^\T)^\T$ and obtain
\begin{align}
  &0 = \sum_{i=1}^N \Phi(V_i,\delta_{A,i},\delta_{B,i}; \widehat{\theta})\nonumber\\
  &= 
\sum_{i=1}^N \Phi(V_i,\delta_{A,i},\delta_{B,i}; \theta_y) + 
\left\{
\sum_{i=1}^N \frac{\partial \Phi(V_i,\delta_{A,i},\delta_{B,i};\widehat{\theta}^*)}{\partial \theta^\T}
\right\}
(\widehat{\theta}-\theta_y),
\label{eq:Taylor_theta}
\end{align}
for some $\widehat{\theta}^* = (\widehat{\mu}_A^*,
\widehat{\mu}_B^*,
\widehat{\tau}^{*\T})^\T$ lying between $\widehat{\theta}$ and $\theta_y$. Under Assumption \ref{asmp:Sampling-design}, the consistency of $\widehat{\mu}_A$ for $\mu_y$ can be established, i.e., $\widehat{\mu}_A = \mu_y + O_{\mathrm{p}}(n^{-1/2})$. Moreover, under Assumption \ref{asmp:regularity} f), we have $\mu_y=\mu_0+O_\zeta(N^{-1/2})$ and hence $\plim\widehat{\mu}_A^* = \mu_0$, i.e., $\widehat{\mu}_A^*$ converges to $\mu_0$ in probability. Under Assumption \ref{asmp:regularity} b), $\widehat{\mu}_B$ is consistent to $\mu_{B,0}$, and $\mu_{B,0} = \mu_0 + O_{\zeta\text{-}\mathrm{p}\text{-}\mathrm{np}}(n^{-1/2})$ under the local alternative. Denote $\theta_0=(\mu_0,\mu_0,\tau_0^\T)^\T$, and the following uniform convergence can be established under Assumption \ref{asmp:regularity} (a)-(c) and (e)
\begin{align*}
    &N^{-1}\sum_{i=1}^N \frac{\partial \Phi(V_i,\delta_{A,i},\delta_{B,i};\widehat{\theta}^*)}{\partial \theta^\T}\\
    &= 
\mathbb{E}
\left\{
\frac{\partial \Phi(V_i,\delta_{A,i},\delta_{B,i};{\theta}_0)}{\partial \theta^\T}
\right\} +
O_{\zeta\text{-}\mathrm{p}\text{-}\mathrm{np}}(n^{-1/2}) + 
O_\zeta(N^{-1/2}),
\end{align*}

and by Assumption \ref{asmp:regularity} (d), we have
$$
\left\{N^{-1}\sum_{i=1}^N \frac{\partial \Phi(V_i,\delta_{A,i},\delta_{B,i};\widehat{\theta}^*)}{\partial \theta^\T}\right\}^{-1} =
\left[
\mathbb{E}
\left\{
\frac{\partial \Phi(V_i,\delta_{A,i},\delta_{B,i};{\theta}_0)}{\partial \theta^\T}
\right\}
\right]+
o_{\zeta\text{-}\mathrm{p}\text{-}\mathrm{np}}(1).
$$
Rearrange the terms of (\ref{eq:Taylor_theta}), we then have
\begin{align*}
n^{1/2}(\widehat{\theta}-\theta_{y}) & =\left\{ -N^{-1}\sum_{i=1}^{N}\phi(\widehat{\theta}^*)\right\} ^{-1}\left\{ n^{1/2}N^{-1}\sum_{i=1}^{N}\Phi(V_i,\delta_{A,i},\delta_{B,i};\theta_y)\right\} +o_{\zeta\text{-}\mathrm{p}\text{-}\mathrm{np}}(1)\\
 & =-\left\{ \E\phi(\theta_0)\right\} ^{-1}\left\{ n^{1/2}N^{-1}\sum_{i=1}^{N}\Phi(V_i,\delta_{A,i},\delta_{B,i};\theta_y)\right\} +o_{\zeta\text{-}\mathrm{p}\text{-}\mathrm{np}}(1),
\end{align*}
where $\phi(\theta)=\partial\Phi(V,\delta_A,\delta_B;\theta)/\partial\theta^\T$. For the simplicity of notation, we denote $\pi_B(X_i;\alpha)=\pi_{B,i}, m(X_i;\beta)=m_i, \dot{m}_{i}=\partial m\left(X_i;\beta\right)/\partial\beta$, and its expectation is given by
\begin{align}
&\E\left\{ \phi(\theta)\right\}  \nonumber\\
 & =-\left(\begin{array}{cccc}
\mathbb{E}(\delta_{A,i}d_i) & 0 & 0 & 0\\
0 & 1 & \E\left\{
\frac{\delta_{B,i}(1-\pi_{B,i})(Y_{i}-m_{i})X_{i}^{\T}}{\pi_{B, i}}\right\}  & \E\left\{\left( \delta_{B,i}\pi_{B, i}^{-1}-\delta_{A,i}d_{i}\right)X_{i}^\T\right\} \\
0 & 0 & \E\left\{ \delta_{A,i}d_{i}\pi_{B, i}(1-\pi_{B, i})X_{i}X_{i}^{\T}\right\}  & 0\\
0 & 0 & 0 & \E\left\{ (\delta_{B,i} + \Omega \delta_{A,i})X_{i}X_{i}^{\T} \right\} 
\end{array}\right)\nonumber \\
&= -\text{diag}\left[\begin{array}{cccc}
    1 & 1 & \pi_{B}(X_i;\alpha)\{1-\pi_{B}(X_i;\alpha)\}X_iX_i^\T& (\pi_{B,i}^* + \Omega d_i^{-1}) X_iX_i^\T 
\end{array}\right],\label{eq:phi} 
\end{align}
where $\Omega = 0$ if $\Phi_\tau(V,\delta_A,\delta_B;\alpha,\beta)$ is constructed by (\ref{eq:Phi_tau_alpha_1}) and (\ref{eq:Phi_tau_beta_1}), and $\Omega = 1$ if $\Phi_\tau(V,\delta_A,\delta_B;\alpha,\beta)$ is constructed by (\ref{eq:Phi_tau_alpha_1}) and (\ref{eq:Phi_tau_beta_2}); $\pi_{B,i}^* = \pr(\delta_{B,i}=1\mid X_i)$ is the true probability. In addition, if (\ref{eq:Phi_tau_alpha_KH}) and (\ref{eq:Phi_tau_beta_KH}) are used to estimate $\tau$, it gives us
\begin{align}
&\E\left\{ \phi_{{\rm KH}}(\theta)\right\} \nonumber \\
& =-\left(\begin{array}{cccc}
\mathbb{E}(\delta_{A,i}d_i) & 0 & 0 & 0\\
0 & 1 & 0 & 0\\
0 & 0 & \E\left\{ \frac{\delta_{B,i}(1-\pi_{B,i})X_{i}X_{i}^{\T}}{\pi_{B,i}}\right\}  & 0\\
0 & 0 & \E\left\{ \frac{\delta_{B,i}(1-\pi_{B,i})(Y_{i}-m_i)X_{i}X_{i}^{\T}}{\pi_{B,i}}\right\}  & \E\left\{
\frac{\delta_{B,i}(1-\pi_{B,i})X_iX_i^\T}{\pi_{B,i}}\right\} 
\end{array}\right)\nonumber\\
& = -\text{diag}\left\{
    1\quad 1 \quad (1-\pi_{B,i}^*)X_{i}X_{i}^{\T} \quad (1-\pi_{B,i}^*)X_{i}X_{i}^{\T}\right\}.\label{eq:phi-KH} 
\end{align}
Below, we focus on the asymptotic properties of $n^{1/2}(\widehat{\theta}-\theta_y)$ under (\ref{eq:phi}), and
the asymptotics under under (\ref{eq:phi-KH}) can be obtained in
an analogous way. First, the inverse of $\E\left\{ \phi(\theta)\right\}$ is
\begin{align*}
&\left[\E\left\{ \phi(\theta)\right\} \right]^{-1} \\
&
= -\text{diag}\left[
    1 \quad 1 \quad \pi_{B}(X_i;\alpha)\{1-\pi_{B}(X_i;\alpha)\}X_iX_i^\T \quad (\pi_{B,i}^* + \Omega d_i^{-1}) X_iX_i^\T \right]^{-1}.
\end{align*}
As shown in \cite{chen2019doubly} under Assumption \ref{asmp:regularity}
g), the asymptotic variance of $\widehat{\mu}_{B}$ will not be affected
by the estimated $\widehat{\beta}$. Let $\pi_{B,i,0}=\pi_B(X_{i};\alpha_{0})$ and $m_{i,0}=m(X_{i};\beta_{0})$
be the correct working model evaluated the true parameter value $(\alpha_0,\beta_{0})$.
Therefore, the $\sum_{i=1}^{N}\Phi(V_i,\delta_{A,i},\delta_{B,i};\theta_y)$ can be found by using
the decomposition 
\begin{align*}
&\sum_{i=1}^{N}\Phi(V_i,\delta_{A,i},\delta_{B,i};\theta_y) \\
& =\left(\begin{array}{c}
0\\
N\left(h_{N}-\mu_{y}\right)+\sum_{i=1}^{N}\delta_{B,i}\left\{ \pi_{B,i,0}^{-1}\left(Y_{i}-m_{i,0}-h_{N}\right)-b^{\T}X_{i}\right\} \\
\sum_{i=1}^{N}\delta_{B,i}X_{i}-\sum_{i=1}^{N}\pi_{B,i,0}X_{i}\\
\sum_{i=1}^{N}\delta_{B,i}(Y_{i}-X_{i}^{\T}\beta_{0})X_{i}
\end{array}\right)\\
 & +\left(\begin{array}{c}
\sum_{i=1}^N\delta_{A,i}d_{i}(Y_{i}-\mu_y)\\
\sum_{i=1}^N\delta_{A,i}d_{i}t_{i}\\
\sum_{i=1}^{N}\pi_{B,i,0}X_{i}-\sum_{i=1}^N\delta_{A,i}d_{i}\pi_{B,i,0}X_{i}\\
0
\end{array}\right),
\end{align*}
where 
$$
\begin{aligned}
    &h_{N}=N^{-1}\sum_{i=1}^{N}\left(Y_{i}-m_{i,0}\right),\\ &b^{\T}=\left[(1-\pi_{B,i,0})\{Y_{i}-m_{i,0}-h_{N}\}X_{i}^{\T}\right]\{N^{-1}\sum_{i=1}^{N}\pi_{B,i,0}(1-\pi_{B,i,0})X_{i}X_{i}^{\T}\}^{-1},\\
    &t_{i}=\pi_{A,i}X_{i}^{\T}b+m_{i,0}-N^{-1}\sum_{i=1}^{N}m_{i,0}.
\end{aligned}
$$
Since the probability sample is assumed to be independent of the non-probability sample \citep{chen2019doubly}, we could express the variance
for $\sum_{i=1}^{N}\Phi(V_i,\delta_{A,i},\delta_{B,i};\theta_y)$ as two components $\mathcal{V}_{1}$
and $\mathcal{V}_{2}$ under Assumption \ref{asmp:Sampling-design} and \ref{asmp:MAR} (iii)
\begin{align}
 & \var\left\{
 n^{1/2}N^{-1}\sum_{i=1}^{N}\Phi(V_i,\delta_{A,i},\delta_{B,i};\theta_y)\right\} =\mathcal{V}_{1}+\mathcal{V}_{2}\nonumber\\
 & =nN^{-2}\sum_{i=1}^{N} \pi_{B,i,0}(1-\pi_{B,i,0})\nonumber\\
&\times \mathbb{E}_\zeta\left\{
 \left(\begin{array}{cccc}
0 & 0 & 0 & 0\\
0 &\Delta^{2} & \Delta X_{i}^{\T} & \Delta Y_{i}X_{i}^{\T}\\
0 & \Delta X_{i} & X_{i}X_{i}^{\T} & Y_{i}X_{i}X_{i}^{\T}\\
0 & \Delta Y_{i}X_{i} & Y_{i}X_{i}X_{i}^{\T} & (Y_{i}-X_{i}^{\T}\beta_{0})^{2}X_{i}X_{i}^{\T}
\end{array}\right)
\right\} \label{eq:var_1}\\
 & +
 nN^{-2}\mathbb{E}_\zeta 
 \left\{
 \left(\begin{array}{cccc}
\mathcal{D}_{11} & \mathcal{D}_{12} & \mathcal{D}_{13} & 0\\
\mathcal{D}_{12}^{\T} & \mathcal{D}_{22} & \mathcal{D}_{23} & 0\\
\mathcal{D}_{13}^{\T} & \mathcal{D}_{23}^{\T} & \mathcal{D}_{33} & 0\\
0 & 0 & 0 & 0
\end{array}\right)
\right\}+ o(1),\label{eq:var_2}
\end{align}
where $\mathcal{V}_1=
\var_{\zeta\text{-}\mathrm{np}}\left\{\sum_{i=1}^N \Phi(V_i,\delta_{A,i},\delta_{B,i};\theta_y)\right\}$, $\mathcal{V}_2=\var_{\zeta\text{-}\mathrm{p}}\left\{\sum_{i=1}^N \Phi(V_i,\delta_{A,i},\delta_{B,i};\theta_y)\right\}$ and $\Delta=\pi_{B,i,0}^{-1}\left\{ y_{i}-m_{i,0}-h_{N}\right\} -b^{\T}x_{i}$. By the law of total variance, we have
$$
\begin{aligned}
   \var_{\zeta\text{-}\mathrm{np}}\left\{\sum_{i=1}^N \Phi(V_i,\delta_{A,i},\delta_{B,i};\theta_y)\right\} & =
\E_\zeta 
\left[
\var_\mathrm{np}
\left\{
\sum_{i=1}^N \Phi(V_i,\delta_{A,i},\delta_{B,i};\theta_y)
\mid \F_N
\right\}
\right] \\
&+\var_\zeta 
\left[
\E_\mathrm{np}
\left\{
\sum_{i=1}^N \Phi(V_i,\delta_{A,i},\delta_{B,i};\theta_y)
\mid \F_N
\right\}
\right],
\end{aligned}
$$
where the second term will be negligible under Assumption \ref{asmp:regularity} g) and h). Similar arguments hold for $\var_{\zeta\text{-}\mathrm{p}}\left\{\sum_{i=1}^N \Phi(V_i,\delta_{A,i},\delta_{B,i};\theta_y)\right\}$, therefore, (\ref{eq:var_1}) and (\ref{eq:var_2}) follow. The sub-matrices $\mathcal{D}_{kl}, k=1,\cdots,3, l=1,\cdots,3$ are all design-based variance-covariance matrices under the probability sampling design, and can be obtained using standard plug-in approach. 

Alternatively, a with-replacement bootstrap variance estimation can also be used here \citep{rao1992some}. To
illustrate, we consider a single-stage probability proportional to size sampling with negligible sampling
ratios. Following \cite{shao2012jackknife}, 
the bootstrap procedures in Algorithm \ref{alg:replication} are conducted.
\begin{algorithm}[tb]
\caption{\label{alg:replication} Replication-based method for estimating variance of $\widehat{\mu}_A$ and $\widehat{\mu}_B$}
\textbf{Input}: the probability sample $\{(V_i,\delta_{A,i}): i\in\mathcal{A}\}$, the non-probability sample $\{(V_i,\delta_{B,i}): i\in\mathcal{B}\}$ and the number of bootstrap $K$.\\
\For {$b=1,\cdots,K$}
{Sample $n_{A}$ units from the probability sample with replacement
as $\mathcal{A}^{(b)}$.\\
Sample $n_{B}$ units from the non-probability sample with
replacement as $\mathcal{B}^{(b)}$.\\
Compute the bootstrap replicates $\widehat{\mu}_{A}^{(b)}$
and $\widehat{\mu}_{B}^{(b)}$ by solving 
$$
\sum_{i\in\mathcal{A}^{(b)}}\Phi_{A}(V_{i},\delta_{A,i};\mu)=0,\quad \sum_{i\in\mathcal{A}^{(b)}\cup\mathcal{B}^{(b)}}\Phi_{B}(V_{i},\delta_{A,i},\delta_{B,i};\mu,\widehat{\tau})=0.
$$\\
}
Calculate the variance estimator $\widehat{V}_{A},\widehat{\Gamma}$
and $\widehat{V}_{B}$
\begin{eqnarray*}
 & \widehat{\Gamma} & =n(K-1)^{-1}\sum_{b=1}^{K}(\widehat{\mu}_{A}^{(b)}-\overline{\widehat{\mu}}_{A})(\widehat{\mu}_{B}^{(b)}-\overline{\widehat{\mu}}_{B})^{\T},\\
 & \widehat{V}_{D} & =n(K-1)^{-1}\sum_{b=1}^{K}(\widehat{\mu}_{D}^{(b)}-\overline{\widehat{\mu}}_{D})(\widehat{\mu}_{D}^{(b)}-\overline{\widehat{\mu}}_{D})^{\T},\ \ \ D=A,B,
\end{eqnarray*}
where $\overline{\widehat{\mu}}_{D}=K^{-1}\sum_{b=1}^{K}\widehat{\mu}_{D}^{(b)}$
for $D=A,B$. 
\end{algorithm}

Under Assumptions \ref{asmp:Sampling-design} and \ref{asmp:regularity}, $\widehat{\theta}-\theta_y \mid \F_N$ and $\theta_y$ are both approximately normal, which leads to the asymptotic normality of the unconditional distribution over all the finite
populations by Lemma \ref{lemm:uncondition}:
$$
\begin{aligned}
    &n^{1/2}(\widehat{\theta}-\theta_y)\rightarrow\\
    &\N
    \left(\theta^*,
    \left\{
    \E\phi(\theta_0)
    \right\}^{-1}
    \var\left\{
 n^{1/2}N^{-1}\sum_{i=1}^{N}\Phi(V_i,\delta_{A,i},\delta_{B,i};\theta_y)\right\}
 \left\{
    \E\phi(\theta_0)^\T
    \right\}^{-1}
 \right),
\end{aligned}
$$
where $\theta^* = (0\quad  -f_B^{-1/2} \left[\mathbb{E}\{\partial \Phi_B(\mu_0,\tau_0)/\partial \mu\}\right]^{-1}\eta \quad 0)^\T$. Thus, the
asymptotic variance for the joint distribution $n^{1/2}(\widehat{\mu}_{A}-\mu_{y},\widehat{\mu}_{B}-\mu_{y})^{\T}$
is obtain by the $2\times2$ submatrix corresponding as 
\begin{align*}
&\var\{n^{1/2}(\widehat{\mu}_{A}-\mu_{y},\widehat{\mu}_{B}-\mu_{y})^{\T}\} \\
& = nN^{-2}\left(\begin{array}{cc}
\mathcal{D}_{11} & \mathcal{D}_{12}\\
\mathcal{D}_{21} & \sum_{i=1}^{N}(1-\pi_{B,i,0})\pi_{B,i,0}\Delta^{2}+\mathcal{D}_{22}
\end{array}\right)+o(1)\\
 & =\left(\begin{array}{cc}
V_{A} & \Gamma\\
\Gamma^{\T} & V_{B}
\end{array}\right) +o(1),
\end{align*}
and
\begin{align*}
n^{1/2}\left(\begin{array}{c}
\widehat{\mu}_{A}-\mu_{y}\\
\widehat{\mu}_{B}-\mu_{y}
\end{array}\right) & \rightarrow\N\left\{ \left(\begin{array}{c}
0\\
-f_B^{-1/2} \left[\mathbb{E}\{\partial \Phi_B(\mu_0,\tau_0)/\partial \mu\}\right]^{-1}\eta
\end{array}\right),\left(\begin{array}{cc}
V_{A} & \Gamma\\
\Gamma^{\T} & V_{B}
\end{array}\right)\right\}\\
&\rightarrow\N\left\{ \left(\begin{array}{c}
0\\
f_B^{-1/2}\eta
\end{array}\right),\left(\begin{array}{cc}
V_{A} & \Gamma\\
\Gamma^{\T} & V_{B}
\end{array}\right)\right\},
\end{align*}
where $\mathbb{E}\{\partial \Phi_B(\mu_0,\tau_0)/\partial \mu\}=-1$.

\subsection{A detailed illustration of bias and mean squared error}
\begin{figure}[tb]
    \centering
    \includegraphics[width=.85\linewidth]{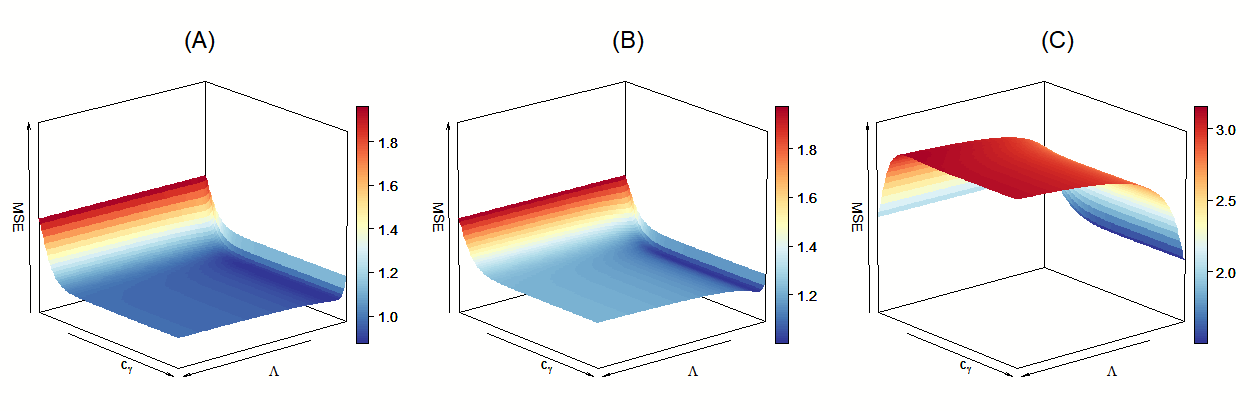}
    \caption{\label{fig:mse_mu_tap}The plots for the mean squared errors in a synthetic example.
Leftmost (A) plots the mean square error $\text{mse}(\Lambda,c_{\gamma};\eta)$
of $n^{1/2}(\widehat{\mu}_{\tap}-\mu_{g})$ as function of $\Lambda$
and $c_{\gamma}$ when the null hypothesis $H_{0}$ holds true $(\eta=0)$;
Middle (B) plots $\text{mse}(\Lambda,c_{\gamma};\eta)$ when the null
hypothesis $H_{0}$ is weakly violated $(\eta=0.5)$; Rightmost (C)
plots $\text{mse}(\Lambda,c_{\gamma};\eta)$ when the null hypothesis
$H_{0}$ is strongly violated $(\eta=1.5)$.}
\end{figure}

Here, we take $\Phi_{A}(V,\delta_A;\mu)$ as Equation (\ref{eq:S_A_1}) and $\Phi_B(V,\delta_A,\delta_B;\mu,\tau)$ as Equation
(\ref{eq:S_B_1}) for an illustration. For $T\leq c_{\gamma}$,
we have 
\[
n^{1/2}(\widehat{\mu}_{\tap}-\mu_{g})=-\left(\frac{V_{A}V_{B}-\Gamma^{2}}{V_{A}+V_{B}-2\Gamma}\right)^{1/2}W_{1}+\frac{(\Gamma-V_{A})-\lambda(\Gamma-V_{B})}{(1+\lambda)(V_{A}+V_{B}-2\Gamma)^{1/2}}W_{2}|W_{2}^{2}\leq c_{\gamma},
\]
with probability $\xi=F_{1}(c_{\gamma};\mu_{2}^{2})$, which leads
to 
\begin{align*}
&\text{bias}(\lambda,c_{\gamma};\eta)_{T\leq c_{\gamma}}  =-\left(\frac{V_{A}V_{B}-\Gamma^{2}}{V_{A}+V_{B}-2\Gamma}\right)^{1/2}\mu_{1}\\
&+\frac{(\Gamma-V_{A})-\lambda(\Gamma-V_{B})}{(1+\lambda)(V_{A}+V_{B}-2\Gamma)^{1/2}}\cdot E(W_{2}|W_{2}^{2}\leq c_{\gamma})\\
 & =-\left(\frac{V_{A}V_{B}-\Gamma^{2}}{V_{A}+V_{B}-2\Gamma}\right)^{1/2}\mu_{1}\\
 &+\frac{(\Gamma-V_{A})-\lambda(\Gamma-V_{B})}{(1+\lambda)(V_{A}+V_{B}-2\Gamma)^{1/2}}\cdot\mu_{2}\frac{F_{3}(c_{\gamma};\mu_{2}^{T}\mu_{2}/2)}{F_{1}(c_{\gamma};\mu_{2}^{T}\mu_{2}/2)}\\
 & =\frac{-\eta f_{B}^{-1/2}(\Gamma-V_{A})}{V_{A}+V_{B}-2\Gamma}+\frac{\eta f_{B}^{-1/2}\{(\Gamma-V_{A})-\lambda(\Gamma-V_{B})\}}{(1+\lambda)(V_{A}+V_{B}-2\Gamma)}\frac{F_{3}(c_{\gamma};\mu_{2}^{T}\mu_{2}/2)}{F_{1}(c_{\gamma};\mu_{2}^{T}\mu_{2}/2)},
\end{align*}
and 
\begin{align*}
&\text{mse}(\lambda,c_{\gamma};\eta)_{T\leq c_{\gamma}}  =\frac{V_{A}V_{B}-\Gamma^{2}}{V_{A}+V_{B}-2\Gamma}\cdot(\mu_{1}^{2}+1)+\left\{ \frac{(\Gamma-V_{A})-\lambda(\Gamma-V_{B})}{(1+\lambda)(V_{A}+V_{B}-2\Gamma)^{1/2}}\right\} ^{2}\\
 & \times E(W_{2}^{2}|W_{2}^{2}\leq c_{\gamma})\\
 & -2\frac{\left(V_{A}V_{B}-\Gamma^{2}\right)^{1/2}\left\{ (\Gamma-V_{A})-\lambda(\Gamma-V_{B})\right\} }{(1+\lambda)(V_{A}+V_{B}-2\Gamma)}\mu_{1}\cdot\mu_{2}\frac{F_{3}(c_{\gamma};\mu_{2}^{T}\mu_{2}/2)}{F_{1}(c_{\gamma};\mu_{2}^{T}\mu_{2}/2)}\\
 & =\frac{V_{A}V_{B}-\Gamma^{2}}{V_{A}+V_{B}-2\Gamma}\cdot(\mu_{1}^{2}+1)+\left\{ \frac{\lambda(\Gamma-V_{B})-(\Gamma-V_{A})}{(1+\lambda)(V_{A}+V_{B}-2\Gamma)^{1/2}}\right\} ^{2}\\
 & \times\left\{ \frac{F_{3}(c_{\gamma};\mu_{2}^{2}/2)}{F_{1}(c_{\gamma};\mu_{2}^{2}/2)}+\mu_{2}^{2}\frac{F_{5}(c_{\gamma};\mu_{2}^{2}/2)}{F_{1}(c_{\gamma};\mu_{2}^{2}/2)}\right\} \\
 & -2\frac{\left(V_{A}V_{B}-\Gamma^{2}\right)^{1/2}\left\{ (\Gamma-V_{A})-\lambda(\Gamma-V_{B})\right\} }{(1+\lambda)(V_{A}+V_{B}-2\Gamma)}\mu_{1}\cdot\mu_{2}\frac{F_{3}(c_{\gamma};\mu_{2}^{T}\mu_{2}/2)}{F_{1}(c_{\gamma};\mu_{2}^{T}\mu_{2}/2)}.
\end{align*}
For $T>c_{\gamma}$, we have 
\[
n^{1/2}(\widehat{\mu}_{\tap}-\mu_{g})=-\left(\frac{V_{A}V_{B}-\Gamma^{2}}{V_{A}+V_{B}-2\Gamma}\right)^{1/2}W_{1}+\frac{-(\Gamma-V_{A})}{(V_{A}+V_{B}-2\Gamma)^{1/2}}W_{2}|W_{2}^{2}>c_{\gamma},
\]
with probability $1-\xi=1-F_{1}(c_{\gamma};\mu_{2}^{2})$, the corresponding bias and mean squared error would be 
\begin{alignat*}{2}
\text{bias}(\lambda,c_{\gamma};\eta)_{T>c_{\gamma}}  &=&&-\left(\frac{V_{A}V_{B}-\Gamma^{2}}{V_{A}+V_{B}-2\Gamma}\right)^{1/2}\mu_{1}\\
&&&+\frac{(\Gamma-V_{A})}{(V_{A}+V_{B}-2\Gamma)^{1/2}}\cdot\mu_{2}\frac{1-F_{3}(c_{\gamma};\mu_{2}^{T}\mu_{2}/2)}{1-F_{1}(c_{\gamma};\mu_{2}^{T}\mu_{2}/2)}\\
  &=&&\frac{-\eta f_{B}^{-1/2}(\Gamma-V_{A})}{V_{A}+V_{B}-2\Gamma}+\frac{\eta f_{B}^{-1/2}(\Gamma-V_{A})}{V_{A}+V_{B}-2\Gamma}\frac{1-F_{3}(c_{\gamma};\mu_{2}^{T}\mu_{2}/2)}{1-F_{1}(c_{\gamma};\mu_{2}^{T}\mu_{2}/2)},
\end{alignat*}
and 
\begin{align*}
&\text{mse}(\lambda,c_{\gamma};\eta)_{T>c_{\gamma}}  =\frac{V_{A}V_{B}-\Gamma^{2}}{V_{A}+V_{B}-2\Gamma}\cdot(\mu_{1}^{2}+1)+\frac{(\Gamma-V_{A})^{2}}{V_{A}+V_{B}-2\Gamma}\\
 & \times E(W_{2}^{2}|W_{2}^{2}>c_{\gamma})\\
 & -2\frac{\left(V_{A}V_{B}-\Gamma^{2}\right)^{1/2}(\Gamma-V_{A})}{V_{A}+V_{B}-2\Gamma}\mu_{1}\cdot\mu_{2}\frac{1-F_{3}(c_{\gamma};\mu_{2}^{T}\mu_{2}/2)}{1-F_{1}(c_{\gamma};\mu_{2}^{T}\mu_{2}/2)}\\
 & =\frac{V_{A}V_{B}-\Gamma^{2}}{V_{A}+V_{B}-2\Gamma}+\frac{(\Gamma-V_{A})^{2}}{V+V_{B}-2\Gamma}\\
 & \times\left\{ \frac{1-F_{3}(c_{\gamma};\mu_{2}^{2}/2)}{1-F_{1}(c_{\gamma};\mu_{2}^{2}/2)}+\mu_{2}^{2}\frac{1-F_{5}(c_{\gamma};\mu_{2}^{2}/2)}{1-F_{1}(c_{\gamma};\mu_{2}^{2}/2)}\right\} \\
 & -2\frac{\left(V_{A}V_{B}-\Gamma^{2}\right)^{1/2}(\Gamma-V_{A})}{V_{A}+V_{B}-2\Gamma}\mu_{1}\cdot\mu_{2}\frac{1-F_{3}(c_{\gamma};\mu_{2}^{T}\mu_{2}/2)}{1-F_{1}(c_{\gamma};\mu_{2}^{T}\mu_{2}/2)}.
\end{align*}
Then, the bias and mean squared error for $n^{1/2}(\widehat{\mu}_{\tap}-\mu_{g})$
would be 
\begin{align}
&\text{bias}(\lambda,c_{\gamma};\eta)  =\text{bias}(\lambda,c_{\gamma};\eta)_{T\leq c_{\gamma}}\cdot\xi+\text{bias}(\lambda,c_{\gamma};\eta)_{T>c_{\gamma}}\cdot(1-\xi)\nonumber\\
 & =\frac{-\eta f_{B}^{-1/2}(\Gamma-V_{A})}{V_{A}+V_{B}-2\Gamma}+\frac{\eta f_{B}^{-1/2}\{-\lambda(\Gamma-V_{B})+(\Gamma-V_{A})\}}{(1+\lambda)(V_{A}+V_{B}-2\Gamma)}F_{3}(c_{\gamma};\mu_{2}^{T}\mu_{2}/2)\nonumber\\
 & +\frac{\eta f_{B}^{-1/2}(\Gamma-V_{A})}{V_{A}+V_{B}-2\Gamma}\left\{ 1-F_{3}(c_{\gamma};\mu_{2}^{T}\mu_{2}/2)\right\} \nonumber\\
 & =\frac{-\lambda\eta f_{B}^{-1/2}}{1+\lambda}\left(\frac{\Gamma-V_{B}}{V_{A}+V_{B}-2\Gamma}+\frac{\Gamma-V_{A}}{V_{A}+V_{B}-2\Gamma}\right)F_{3}(c_{\gamma};\mu_{2}^{T}\mu_{2}/2)\nonumber\\
 & =\eta d_{0} \label{eq:bias_general},
\end{align}
with 
$$
d_{0}=-\lambda f_{B}^{-1/2}(1+\lambda)^{-1}(\omega_{A}+\omega_{B})F_{3}(c_{\gamma};\mu_{2}^{T}\mu_{2}/2),
$$
and 
\begin{align}
&\text{mse}(\lambda,c_{\gamma};\eta)  =\frac{V_{A}V_{B}-\Gamma^{2}}{V_{A}+V_{B}-2\Gamma}\cdot(\mu_{1}^{2}+1)\nonumber\\
&+\frac{\left\{ \lambda(\Gamma-V_{B})-(\Gamma-V_{A})\right\} ^{2}}{(1+\lambda)^{2}(V_{A}+V_{B}-2\Gamma)}\times\left\{ F_{3}(c_{\gamma};\mu_{2}^{2}/2)+\mu_{2}^{2}F_{5}(c_{\gamma};\mu_{2}^{2}/2)\right\} \nonumber \\
 & +\frac{(\Gamma-V_{A})^{2}}{V_{A}+V_{B}-2\Gamma}\times\left\{ 1-F_{3}(c_{\gamma};\mu_{2}^{2}/2)+\mu_{2}^{2}-\mu_{2}^{2}F_{5}(c_{\gamma};\mu_{2}^{2}/2)\right\} \nonumber \\
 & -2\frac{\left(V_{A}V_{B}-\Gamma^{2}\right)^{1/2}(\Gamma-V_{A})}{V_{A}+V_{B}-2\Gamma}\left\{ 1-F_{3}(c_{\gamma};\mu_{2}^{T}\mu_{2}/2)\right\} \mu_{1}\mu_{2}\nonumber \\
 & -2\frac{\left(V_{A}V_{B}-\Gamma^{2}\right)^{1/2}\left\{ (\Gamma-V_{A})-\lambda(\Gamma-V_{B})\right\} }{(1+\lambda)(V_{A}+V_{B}-2\Gamma)}F_{3}(c_{\gamma};\mu_{2}^{T}\mu_{2}/2)\mu_{1}\mu_{2}\nonumber \\
 & =V_{\eff}d_{1}+V_{\text{B-eff}}d_{2}+V_{\text{A-eff}}d_{3}+V_{\eff}^{1/2}(V_{\text{B-eff}}^{1/2}d_{4}+V_{\text{A-eff}}^{1/2}d_{5}),\label{eq:mse_general}
\end{align}
with 
\begin{eqnarray*}
d_{1}&=&\mu_{1}^{2}+1,\\
d_{2}&=&\lambda(1+\lambda)^{-2}\left\{ F_{3}(c_{\gamma};\mu_{2}^{2}/2)+\mu_{2}^{2}F_{5}(c_{\gamma};\mu_{2}^{2}/2)\right\} \left\{ \lambda-2\omega_{B}/\omega_{A}\right\},\\
d_{3}&=&1-F_{3}(c_{\gamma};\mu_{2}^{2}/2)+\mu_{2}^{2}\left\{ 1-F_{5}(c_{\gamma};\mu_{2}^{2}/2)\right\} \\
&&+(1+\lambda)^{-2}\left\{ F_{3}(c_{\gamma};\mu_{2}^{2}/2)+\mu_{2}^{2}F_{5}(c_{\gamma};\mu_{2}^{2}/2)\right\},\\
d_{4}&=&2\lambda(1+\lambda)^{-1}\mu_{1}\mu_{2}F_{3}(c_{\gamma};\mu_{2}^{2}/2),\\
d_{5}&=&-2\mu_{1}\mu_{2}\left\{ 1-F_{3}(c_{\gamma};\mu_{2}^{2}/2)+F_{3}(c_{\gamma};\mu_{2}^{2}/2)(1+\lambda)^{-1}\right\}.
\end{eqnarray*}
Let $V_{A}=2,V_{B}=1,\Gamma=0.5$, and $\eta=0,0.5$
and $1.5$ (encoding zero, weak, and strong violation of $H_{0}$) in \eqref{eq:bias_general} and \eqref{eq:mse_general}. Figure \ref{fig:mse_mu_tap} shows three mean squared error surfaces as functions of $(\Lambda,c_{\gamma})$ with three values of $\eta$. 

\begin{itemize}
\item [a)] In the leftmost plot, where $H_{0}$ holds, for a given $\Lambda$,
the mean squared error decreases drastically and then flattens out as $c_{\gamma}$
increases. Moreover, for a given $c_{\gamma}$, there exists a minimizer
$\Lambda^{*}$ such that the mean squared error achieves the minimum. These observations
justify our strategy by viewing $\Lambda$ and $c_{\gamma}$ jointly
as tuning parameters since both of them are playing important roles
when searching for the minimum value of mean squared error. 

\item [b)]  In the middle plot, where $H_{0}$ is weakly violated, the pattern
of the mean squared error retains the similar features for $c_{\gamma}$ as shown in (A). In
addition, the optimal choice $\Lambda^{*}$ leads to a sharp decline
of the mean squared error compared to other choices of $\Lambda$. These findings
imply that despite the bias due to accepting the non-probability sample, the
impact would be less compared to the increased variance due to rejecting
the non-probability sample. But care is needed to determine the amount of information
borrowed from the non-probability sample since a small deviation from the optimal
value $\Lambda^{*}$ can lead to a non-ignorable increase of the mean squared error.
Once the optimal mean squared error is reached at $(\Lambda^{*},c^*_{\gamma})$, the
further increment of $c_{\gamma}$ will not be influential.

\item [c)] In the rightmost plot, where $H_{0}$ is strongly violated, the
mean squared error behaves differently as in (A) and (B). It is advisable to choose
both $\Lambda$ and $c_{\gamma}$ close to zero (the low probability
of combining the non-probability sample with the probability sample) to minimize the mean squared error. As
above, keeping increasing $c_{\gamma}$ after the mean squared error flattens out is of
no importance. 
\end{itemize}

\subsection{Additional simulation results}

\begin{figure}[!tb]
    \centering
    \includegraphics[width=.85\linewidth]{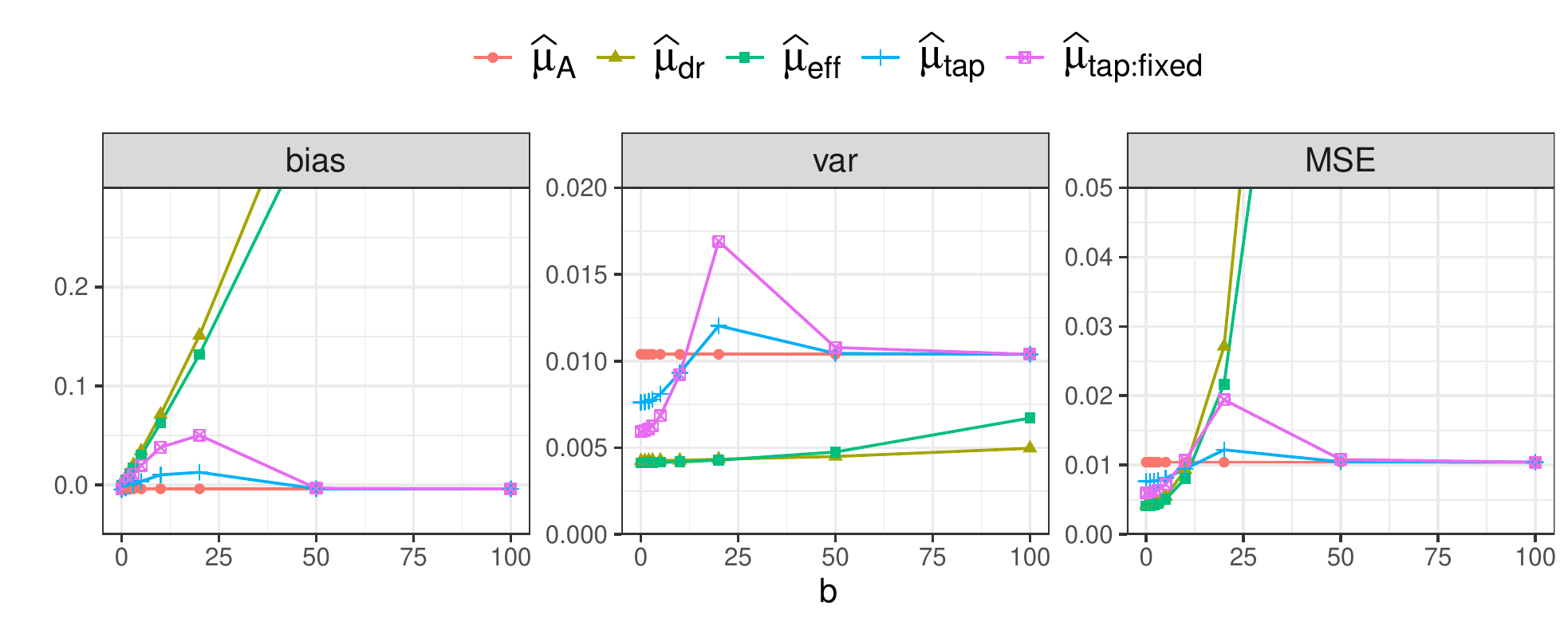}
    \caption{\label{fig:with_fixed}Summary statistics plots of estimators of $\mu_y$ with respect to the strength of violation, labeled by $b$. Each column of the plots corresponds to a different metric: “bias” for bias, “var” for variance, “MSE” for mean square error.}
\end{figure}

\begin{table}[!tb]
 \caption{\label{tab:nuisance}Simulation results of Monte Carlo averages of the tuning parameters $(\Lambda, c_\gamma)$ and the proportion $\pr(\text{comb})$ of combining the probability and non-probability samples}
 \vspace{0.15cm}
 \centering
 \resizebox{.8\textwidth}{!}{
    \begin{tabular}{llcccccc}
    \toprule
    \multicolumn{1}{l}{$H_0$} &       & \multicolumn{2}{c}{$\Lambda$} & \multicolumn{2}{c}{$c_\gamma$} & \multicolumn{2}{c}{$\pr(\text{comb})$} \\       &       & \multicolumn{1}{l}{\textsc{est}} & \multicolumn{1}{l}{\textsc{se}} & \multicolumn{1}{l}{\textsc{est}} & \multicolumn{1}{l}{\textsc{se}} & \multicolumn{1}{l}{\textsc{est}} & \multicolumn{1}{l}{\textsc{se}} \\
    \midrule
    \multicolumn{1}{l}{holds} & $\widehat{\mu}_{\tap}$   & 3.02  & 4.26  & 35.06  & 9.45  & 0.95  & 0.22  \\
          & $\widehat{\mu}_{\tap:B}$  & 3.05  & 4.62 & 35.06  & 9.44  & 0.95  & 0.22  \\
          & $\widehat{\mu}_{\tap:\rm KH}$     & 3.06  & 4.66  & 35.06  & 9.44  & 0.95  & 0.22  \\
    \multicolumn{1}{l}{slightly violated} & $\widehat{\mu}_{\tap}$    & 2.21  & 3.39  & 31.60  & 13.76  & 0.86  & 0.35  \\
          & $\widehat{\mu}_{\tap:B}$  & 2.22  & 3.47  & 31.60  & 13.75  & 0.86  & 0.35  \\
          & $\widehat{\mu}_{\tap:\rm KH}$     & 2.23  & 3.60  & 31.60  & 13.75  & 0.86  & 0.35  \\
    \multicolumn{1}{l}{strongly violated} & $\widehat{\mu}_{\tap}$    & 0.16  & 0.28  & 1.40  & 1.97  & 0.00  & 0.06  \\
          & $\widehat{\mu}_{\tap:B}$  & 0.16  & 0.28  & 1.40  & 1.97  & 0.00  & 0.06   \\
          & $\widehat{\mu}_{\tap:\rm KH}$     & 0.16  & 0.28  & 1.40  & 1.98  & 0.00  & 0.06   \\
    \bottomrule
    \end{tabular}%
    }
\end{table}%
Table \ref{tab:nuisance} provides the Monte Carlo averages and standard errors of the data-adaptive tuned parameters $(\Lambda, c_\gamma)$ and the Monte Carlo proportion of combining the probability and non-probability samples.
Figure \ref{fig:with_fixed} presents the plots of Monte Carlo biases, variances and mean squared errors of the $\widehat{\mu}_A$, $\widehat{\mu}_\dr$, $\widehat{\mu}_\eff$, $\widehat{\mu}_\tap$ and $\widehat{\mu}_{\tap:\text{fix}}$ based on $2000$ replicated datasets. For the fixed threshold strategy $\widehat{\mu}_{\tap:\text{fix}}$, the threshold $c_\gamma$ is held fixed to be the 95th quantile of a $\chi_1^2$ distribution (i.e., $3.84$) and the tuning parameter $\Lambda$ is selected by minimizing the asymptotic mean square error at the fixed $c_\gamma$.

In Table \ref{tab:nuisance}, we find that the adaptive procedure tends to select smaller values of $\Lambda$ and $c_\gamma$ as $b$ increases. As a result, the Monte Carlo proportions of combining the probability and non-probability samples together are decreasing, which is desired for down-weighting the biased non-probability sample. Moreover, we compare the adaptive tuning strategy of $c_\gamma$ with a fixed thresholding strategy, and Figure \ref{fig:with_fixed} shows that the strategy with pre-defined cutoff cannot satisfactorily control the mean squared error when $H_0$ is slightly or strongly violated.

\subsection{Double-bootstrap procedure for
{$v_{n}$}{v\_n} selection\label{subsec:double-bootstrap}}

Following the algorithm mentioned by \cite{chakraborty2013inference},
where optimal $v_{n}$ is selected to ensure the coverage probability,
we need to retain the $K$ bootstrapped samples, called ${V}^{(1)}$, ${V}^{(2)}$, $\cdots$, ${V}^{(K)}$
where ${V}^{(b)}=\{V_{i}=(X_{i}^{(b)\T},Y_{i}^{(b)})^{\T}:i\in1,\cdots,n\},b=1,\cdots,K$ with $n=n_A+n_B$.
The reason it is called double bootstrap is that each bootstrap
sample spawns itself to a set of $K'$ second-order bootstrap samples.
Next, we set up the candidates for $v_{n}$. Under the assumption
(A2), we let $v_{n}$ be the form of $\kappa\log\log n$ with $\kappa\in\{2, 4, 10,20,30\}$,
and construct the bound-based adaptive confidence intervals for each given $\kappa$ at $1-\alpha$ confidence
level, denoted as $\mathbb{C}_{\mu_g,1-\alpha}^{\paci,\kappa}(a)$. Given each $\kappa$, we compute the coverage probability for
the associated adaptive confidence intervals regarding these $K'$ second-ordered simulated
datasets. Then, choose the smallest $\kappa$ that ensures the actual
coverage probability larger than $1-\alpha$. Specifically, we use
the estimator $\widehat{\mu}_{A}^{(b)}$ for $\mu_{A}$ in each bootstrapped
dataset as the ground truth and count the number of datasets in which
the adaptive confidence interval covers the ground truth, say $c(\kappa)=\sum_{b=1}^{K'}\bone\{\widehat{\mu}_{A}^{(b)}\in\mathbb{C}_{\mu_g,1-\alpha}^{\paci,\kappa, (b)}(a)\}$
and therefore the $v_{n}$ can be determined by using $v_{n}=\inf\{\kappa:c(\kappa)/K'>1-\alpha\}\times\log\log n$. In our simulation, $K'$ is set to be $100$.

\subsection{Details of the Bayesian method}\label{sec:Bayesian}
In this section, we provide the details of the Bayesian approaches proposed by \cite{sakshaug2019supplementing} to combine the probability and non-probability samples as follows.
\begin{enumerate}
    \item Solve the score function for $\beta$ by using the non-probability sample:
    $$
    \widehat{\beta}_{\text{NPR}}=\arg\min_\beta
    \sum_{i=1}^N \delta_{B,i}X_i(Y_i-X_i^\intercal \beta)=0.
    $$
    \item Construct the informative prior with three choices:
    \begin{itemize}
        \item[Prior 1:] Choose a weakly informative parameterization of the prior as
        $$
        \beta \sim \mathcal{N}(0, 10^6),
        $$
        which can be treated as a reference for comparison.
        \item[Prior 2:] Let $\widehat{\beta}_{\text{PR}}$ be the solution to the score function based on the probability sample
        $$ \widehat{\beta}_{\text{PR}}=\arg\min_\beta
    \sum_{i=1}^N \delta_{A,i}X_i(Y_i-X_i^\intercal \beta)=0.
        $$
        Then consider the squared Euclidean distance between $\widehat{\beta}_{\text{PR}}$ and $\widehat{\beta}_{\text{NPR}}$ as the hyper-parameter $\sigma_\beta^2$ for the variance of $\beta$:
        $$
        \beta \sim  \mathcal{N}\left\{\widehat{\beta}_{\text{NPR}}, \text{diag}(\|\widehat{\beta}_{\text{PR}}-\widehat{\beta}_{\text{NPR}}\|_2^2)\right\}.
        $$
        \item[Prior 3:] In lieu of using the squared distance to extract information on $\sigma_\beta^2$, a nonparametric with-replacement bootstrap procedure can be implemented ($B=1000$). After estimating the coefficient in each of them, denoted by $\widehat{\beta}^{(i)}_{\text{NPR}}$, one replication-based variance estimator can be obtained, $\widehat{\sigma}_{\beta_{\text{NPR}}}^2=\sum_{i=1}^B(\widehat{\beta}^{(i)}_{\text{NPR}}-\bar{\widehat{\beta}}_{\text{NPR}})^2/(B-1)$ with $\bar{\widehat{\beta}}_{\text{NPR}}=1/B
        \sum_{i=1}^B \widehat{\beta}^{(i)}_{\text{NPR}}
        $. Then, the informative prior can be constructed 
        $$
        \beta \sim \mathcal{N}(\widehat{\beta}_{\text{NPR}}, I_{p\times p}\cdot \widehat{\sigma}_{\beta_{\text{NPR}}}^2).
        $$
    \end{itemize}
    \item Assume that the model for the observed probability sample is
    $$
    Y_i\mid \delta_{A,i} = 1 \sim \mathcal{N}( X_i^\intercal \beta, \sigma^2 ).
    $$
    By imposing an informative non-probability-based prior, the resulting posterior estimates are expected to be more efficient. Specifically, these priors are:
    \begin{align*}
        &\beta \sim \mathcal{N}(\beta_0,
        \sigma_\beta^2),\quad \sigma^{-2}\sim \Gamma(r,m), \quad r=m=10^{-3},
    \end{align*}
    where
    \begin{align*}
        &\text{Prior 1: }\beta_0 = 0,\quad
        \sigma_\beta^2 = 10^6,\\
        &\text{Prior 2: }\beta_0 = \widehat{\beta}_{\text{NPR}},\quad
        \sigma_\beta^2 = \text{diag}(\|\widehat{\beta}_{\text{PR}}-\widehat{\beta}_{\text{NPR}}\|_2^2),\\
        &\text{Prior 3: }\beta_0 = \widehat{\beta}_{\text{NPR}},\quad
        \sigma_\beta^2 = I_{p\times p}\cdot \widehat{\sigma}_{\beta_{\text{NPR}}}^2.
    \end{align*}
\end{enumerate}
The posterior Markov chain Monte Carlo (MCMC) samples of ${\beta}$ and $Y_i$ are obtained by drawing 2000 samples from the posterior distributions and discarding the first 500 samples as the burn-in procedures. The Bayesian estimator is
$$
\widehat{\mu}_{\text{Bayes}}=
    1/\widehat{N}
    \sum_{i=1}^{n_A}d_i \bar{Y}_i
    \text{ with } \widehat{N} = \sum_{i=1}^{n_A} d_i,
$$
where $\bar{Y}_i$ is the posterior mean calculated by $\bar{Y}_i={1}/(2000-500)\sum_{k=501}^{2000}{Y}_{i,k}$. Borrowed from Bayes' Theorem, its variance and 95\% highest posterior density intervals can be estimated via the MCMC posterior samples. Denote $\widehat{\mu}_{\text{Bayes},k}=1/\widehat{N}\sum_{i=1}^{n_A}d_i Y_{i,k},k=501,\cdots,2000$. Then, we have
\begin{align*}
    &\text{var}(\widehat{\mu}_{\text{Bayes}})=
\frac{1}{2000-500-1}
\sum_{k=501}^{2000}
(
\widehat{\mu}_{\text{Bayes},k}-
\widehat{\mu}_{\text{Bayes}}
)^2,\\
&\text{HPDI}=\left\{
Q(\widehat{\mu}_{\text{Bayes},k};\alpha/2),
Q(\widehat{\mu}_{\text{Bayes},k};1-\alpha/2)
\right\},
\end{align*}
where $Q(\widehat{\mu}_{\text{Bayes},k};\alpha_0)$ represents the $\alpha_0$-th sample quantile of the posterior samples $\widehat{\mu}_{\text{Bayes},k}$, $k=501, \cdots,2000$ after burn-in. 
\end{document}